\documentclass[aps, prd, twocolumn,floats,floatfix,showpacs,prd,superscriptaddress,nofootinbib]{revtex4-1}
\usepackage{graphicx}
\usepackage[export]{adjustbox}
\usepackage{caption}
\usepackage{color}
\usepackage{url}
\usepackage{bm}  
\usepackage{times}
\usepackage{amsmath}
\usepackage{amssymb}
\usepackage{float}
\usepackage[colorlinks=true,allcolors=blue]{hyperref}
\usepackage{subcaption}
\usepackage[normalem]{ulem}
\usepackage{todonotes}
\usepackage{subcaption} 
\usepackage[export]{adjustbox}
\usepackage{soul}
\usepackage{varioref}
\usepackage{hyperref}
\usepackage{cleveref}
\usepackage{cancel}

\usepackage{tikz,xcolor,hyperref}

\definecolor{lime}{HTML}{A6CE39}
\DeclareRobustCommand{\orcidicon}{%
	\begin{tikzpicture}
	\draw[lime, fill=lime] (0,0) 
	circle [radius=0.16] 
	node[white] {{\fontfamily{qag}\selectfont \tiny ID}};
	\draw[white, fill=white] (-0.0625,0.095) 
	circle [radius=0.007];
	\end{tikzpicture}
	\hspace{-2mm}
}
\foreach \x in {A, ..., Z}{%
	\expandafter\xdef\csname orcid\x\endcsname{\noexpand\href{https://orcid.org/\csname orcidauthor\x\endcsname}{\noexpand\orcidicon}}
}

\begin{document}

\title{Dipolar tidal effects in gravitational waves from scalarized black hole binary inspirals in quadratic gravity}

\author{Iris van Gemeren\orcidA{}}
\affiliation{
Institute for Theoretical Physics,
Utrecht University, Princetonplein 5, 3584 CC Utrecht, The Netherlands}
\author{Banafsheh Shiralilou\orcidB{}}
\affiliation{GRAPPA, Anton Pannekoek Institute for Astronomy and Institute of High-Energy Physics, University of Amsterdam, Science Park 904, 1098 XH Amsterdam, The Netherlands}

\author{Tanja Hinderer\orcidC{}}
\affiliation{
Institute for Theoretical Physics,
Utrecht University, Princetonplein 5, 3584 CC Utrecht, The Netherlands}

\begin{abstract}
Gravitational waves (GWs) from merging binary black holes (BHs) enable unprecedented tests of gravitational theories beyond Einstein's General Relativity (GR) in highly nonlinear, dynamical regimes.
Such GW measurements require an accurate 
description of GW signatures that may arise in alternative gravitational models. 
In this work, we focus on a class of higher-curvature extensions of GR, the scalar-Gauss-Bonnet theories, where BHs can develop scalar hair. In an inspiraling binary system, this leads to scalar-induced tidal effects in the dynamics and radiation. We calculate the dominant adiabatic dipolar tidal effects via an approximation scheme based on expansions in post-Newtonian, higher-curvature, and tidal corrections. The tidal effects depend on a characteristic scalar Love number, which we compute using BH perturbation theory, and have the same scaling with GW frequency 
as the higher-curvature corrections. We perform case studies to characterize the net size and parameter dependencies of these effects, showing that at low frequencies, tidal effects dominate over the higher-curvature contributions for small couplings within current bounds, regardless of the total BH mass, while at high frequencies they are subdominant. We further consider prospects observing both of these regimes, which would be interesting for breaking parameter degeneracies, with multiband detections of LISA and ground-based detectors or the Einstein Telescope alone. We also assess the frequency range of the transition between these regimes by numerically solving the energy balance law. Our results highlight the importance of the dipolar scalar tidal effects for BHs with scalar hair, which arise in several beyond-GR paradigms, and provide ready-to-use inputs for improved GW constraints on Gauss-Bonnet theories.   
\end{abstract}
\maketitle

\section{introduction}

With the advent of Gravitational-Wave (GW) astronomy~\cite{LIGOScientific:2016aoc}, compact objects as sources of GWs have become unique laboratories for testing our understanding of General Relativity (GR) in unexplored strong-field, nonlinear, and dynamical regimes of gravity -- a  regime inaccessible to Solar System tests~\cite{Will:2014kxa}, binary pulsars~\cite{Wex:2014nva}, as well as observations around the Galactic center~\cite{GRAVITY:2018ofz,EventHorizonTelescope:2019ths}.
Extracting the information on the fundamental source physics from the GW data relies on cross-correlating the detector output with theoretical models. 
With the GW detectors such as advanced LIGO, Virgo and KAGRA poised to yield much larger and more precise datasets in the coming years~\cite{KAGRA:2013rdx}, and the next-generation detectors such as Einstein Telescope (ET)~\cite{Maggiore:2019uih}, Cosmic Explorer~\cite{Reitze:2019iox} and LISA~\cite{eLISA:2013xep} planned, accurate theoretical modeling of the gravitational waveforms is of high interest and essential for precision tests of gravity.
While current analysis pipelines for theory-agnostic null tests of GR are successfully established~\cite{LIGOScientific:2016lio, LIGOScientific:2019fpa, LIGOScientific:2020tif, Yunes:2009ke}, the interpretation of the results to constrain classes of alternative theories is limited and faces subtleties (c.f.,~\cite{Chua:2020oxn,Johnson-McDaniel:2021yge}).
It is thus important to complement such tests with full inspiral-merger-ringdown waveform models in the well-motivated beyond-GR theories. This is needed for
systematic searches of predicted beyond-GR signatures in the data and for setting stronger theoretical constraints. In beyond-GR theories the GW propagation, generation, and properties of compact objects may be altered. 
This affects the entire binary coalescence process, which begins with the inspiral regime amenable to perturbative methods.

In this work, we focus on the derivation of the so-called \textit{scalar-induced dipolar tidal effects} during a binary inspiral in scalar-Gauss-Bonnet (sGB) gravity, to push ahead on the ongoing efforts of computing analytical waveforms in this class of theories~\cite{Yagi:2011xp,Julie:2019sab,Shiralilou:2020gah,Shiralilou:2021mfl,Julie:2022qux}. These theories have a dynamical scalar field $\varphi$ that non-minimally couples to the Gauss-Bonnet (GB) topological invariant, which is a quadratic combination of curvature quantities. They represent higher-order curvature extensions to GR involving a specific combination of curvature terms that guarantees ghost-free second-order equations of motion~\cite{Nojiri:2018ouv}. The sGB gravity arises from the low energy limit of several quantum gravity paradigms \cite{Zwiebach:1985uq, Gross:1986mw, Boulware:1985wk} and agrees with GR in the weak-curvature regime~\cite{Sotiriou:2006pq}.  Hence, it provides a useful effective theory of gravity for astrophysical tests. Moreover, recent work~\cite{Kovacs:2020pns,Kovacs:2020ywu,R:2022tqa} proved the mathematical well-posedness of the theory, paving the way for the first set of fully-nonlinear and non-iterative numerical computations of GWs~\citep{Corman:2022xqg,East:2022rqi,East:2021bqk,East:2020hgw, Ripley:2022cdh,Evstafyeva:2022rve,R:2022hlf}.

Black holes, in particular, attain interesting features in sGB gravity, such as having non-trivial scalar-hair profiles~\cite{Kanti:1995vq,Pani:2009wy, Sotiriou:2013qea,Benkel:2016rlz, Antoniou:2017hxj,Mignemi:1992nt,Kanti:1995vq, Antoniou:2017acq,Papageorgiou:2022umj,Prabhu:2018aun,Saravani:2019xwx,Ayzenberg:2014aka,R:2022tqa} or exhibiting spontaneous (de-)scalarization~\cite{Silva:2017uqg, Doneva:2017bvd,Dima:2020yac,Herdeiro:2020wei,Berti:2020kgk,Collodel:2019kkx, Doneva:2020nbb}.
These features are related to the properties of the coupling function $f(\varphi)$ between the scalar field $\varphi$ and the GB quadratic curvature invariant (see~\cite{Herdeiro:2015waa} for a general review). Coupling functions having a non-vanishing derivative when evaluated at a zero field configuration
lead to BHs with non-trivial scalar hair. Examples of such couplings include $f(\varphi) = 2\varphi$  and $f(\varphi)= e^{2\varphi}$ corresponding to shift symmetric~\cite{Sotiriou:2013qea} and dilatonic~\cite{Pani:2009wy, Ripley:2019irj} GB theories respectively. For couplings with vanishing derivatives, the (de-)scalarization can happen spontaneously. Coupling functions of this type are, for example, $f(\varphi)= \varphi^2$~\cite{Berti:2020kgk,Collodel:2019kkx,Silva:2017uqg} and $f(\varphi)= e^{2\varphi^2}$~\cite{Doneva:2017bvd}.
Thus, BHs in sGB gravity can evade the “no-hair” theorems~\cite{Bekenstein:1995un,Kanti:1995vq,Antoniou:2017acq, Papageorgiou:2022umj} and have a monopole scalar charge. In an inspiraling binary system, this moving charge sources dipolar scalar radiation, which affects the dynamics and the morphology of GWs.

Another novel effect for BHs with scalar hair, which we consider here for the first time in the sGB context, are scalar-induced tidal effects. They arise from the interaction of the scalar condensate configuration surrounding a BH with the scalar field sourced by the companion, whose gradients across the condensate produce scalar tidal fields. In general, couplings with tensor-tidal effects also play a role and give rise to a rich phenomenology, however, in this paper we focus on the dominant dipolar effect involving tidal interactions only in the scalar sector. The induced scalar dipole moment contributes to the scalar radiation and changes the energetics of the system. 

More precisely, the tidal response of a compact object is characterized by its tidal Love numbers. These characteristic parameters quantify the ratio of the induced multipole moments to the perturbing tidal field in the limit of a static tidal environment and encode information about the nature and interior structure of the object~\cite{poisson_will_2014,1999ssd..book.....M}. Within GR, the Love numbers of BHs vanish~\cite{Binnington:2009bb,PhysRevLett.114.151102,Hui:2020xxx,PhysRevD.91.104018,Chia:2020yla,Katagiri:2022vyz, LeTiec:2020spy}, while they are non-zero for other objects such as exotica~\cite{Cardoso:2017cfl, Herdeiro:2020kba} or neutron stars~\cite{Hinderer:2007mb,Hinderer:2009ca,Damour:2009vw}, for which recent GW discoveries enabled the first constraints~\cite{LIGOScientific:2018cki}. 

Tidal effects have been incorporated into the waveforms of inspiraling binaries in GR by exploiting the hierarchy of scales: the size of the objects is small compared to the orbital scale, which in turn is small compared to the wavelength of GWs. This enables meshing different approximation schemes adapted to different scales: a fully relativistic description of the region near the compact objects, a post-Newtonian (PN) approximation for the dynamics, and a multipolar post-Minkowski approximation for the asymptotic radiation~\cite{Flanagan:2007ix,   Damour:2009vw,Vines:2011ud,  Banihashemi:2018xfb,Henry:2020ski}. The PN approximation applies at large orbital separations, where the gravitational fields are weak and the motion is slow. In this regime, the dynamics and GWs are dominated by point-mass effects, with the finite size of the objects contributing small but relatively clean and cumulative corrections. This perturbative information on finite-size effects has also been incorporated in complete waveform models~\cite{Damour:2009wj,Bernuzzi:2014owa,Dietrich:2017aum,Dietrich:2019kaq,Hinderer:2016eia,Steinhoff:2016rfi}. Similar to the tensor tidal Love numbers, a scalarized compact object can also have scalar Love numbers, as well as mixed scalar-tensor ones. Scalar dipolar tidal effects have previously been computed for neutron stars in scalar-tensor theories for the dynamics~\cite{PhysRevLett.70.2220,PhysRevD.57.4802} and radiation~\cite{Bernard:2019yfz}.

In this work, we compute, for the first time, the scalar tidal love number for BHs with scalar hair in sGB gravity, focusing on nonspinning BHs. We use relativistic perturbation theory to compute the scalar dipolar Love numbers 
and obtain closed-form expressions under the assumption that the sGB corrections are small.
Building on the computation of the two-body Lagrangian~\cite{Julie:2019sab} and the scalar and tensor waveforms~\cite{Shiralilou:2020gah, Shiralilou:2021mfl} 
to linear order in the GB coupling and to relative 1PN order, we calculate the tidal contributions to the BH dynamics and radiation from the tidally-induced scalar dipole moment and obtain explicit expressions for the corresponding signatures in the Fourier-domain GW phasing. 
We perform case studies to analyze the parameter dependencies and features of the tidal and higher curvature corrections to the GW phase evolution. We also quantify the regimes of validity of our analytic results for the GW phase expressions, which distinguish between a low (high) frequency regime dominated by the scalar dipole (tensor) radiation by comparing with numerical solutions of the energy balance condition. 

This paper is organised as follows. In Sec.~\ref{sec:sGB} we outline the sGB theory and a general treatment of the matter action including scalar-induced dipolar finite size effects. We also explain the multiple expansion parameters used in the calculations and the different approximation schemes. Sec.~\ref{sec:Binarydyn} contains our derivation of the tidal contributions to the two body Lagrangian and binding energy, and  Sec.~\ref{sec:WaveformPhasing} the tidal contributions to the scalar and tensor waveforms and the GW phasing. Independently in Sec.~\ref{sec:Lovenumber} we derive the scalar tidal Love number using BH perturbation theory and a small coupling approximation.    
Collecting our results we show different case studies of BH systems to analyse the dependencies of the tidal and higher curvature corrections to the GW phasing in Sec.~\ref{sec:results} and conclude in Sec.~\ref{sec:conclusion}.

In this paper we use the standard notation for symmetrized and antisymmetrized indices, namely $x^{(i} y^{j)}$ denotes symmetrization and $x^{[i} y^{j]}$ anti-symmetrization. We use a multi-index notation for products of vector components:
$x^{i j k} \equiv x^{i} x^{j} x^{k}$, and a capital letter superscript denotes a product of that dimensionality: $x^{L} \equiv x^{k_{1}} x^{k_{2}} \cdots x^{k_l}$.
\begin{figure*}
\centering
 \includegraphics[width=0.8\linewidth]{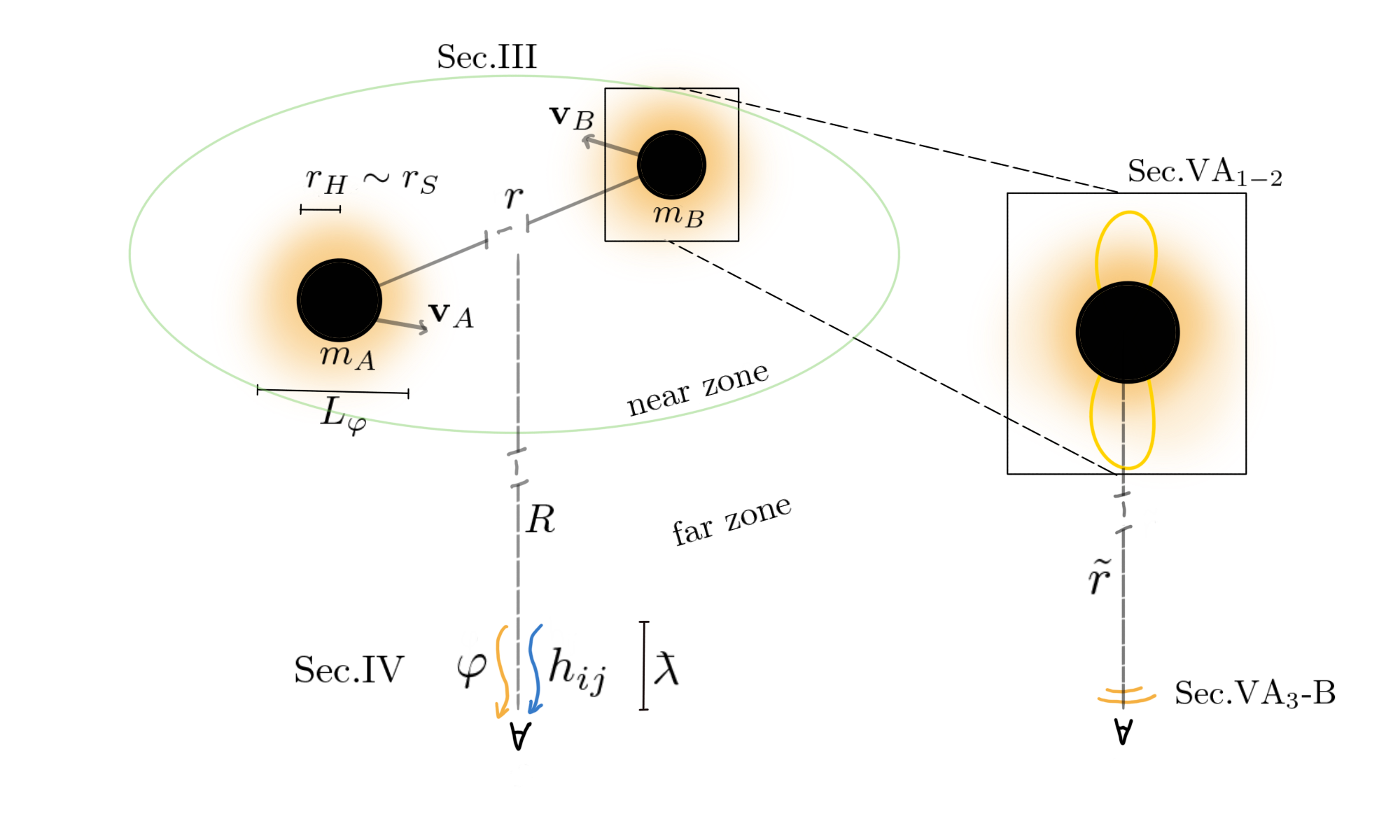}
\caption{Schematic figure of a binary system of the two BHs with masses and velocities $m_A$, $m_B$, $\mathbf{v}_A$ and $\mathbf{v}_B$ respectively. The scalar field configuration is shown in orange. The orbital near zone shows the characteristic length scales of the horizon radius $r_H$ and the scalar cloud $L_{\phi}$. At large distances R from the source the waveforms $h_{ij}$, $\varphi$ are found with characteristic wavelength $\lambdabar$. The inset shows the body zone around one BH at far distance $\tilde{r}$ from the observer. The sections for which the different zones are relevant are shown as well.}
\label{fig:scales}
\end{figure*}
\section{Scalar-Gauss-Bonnet Gravity}\label{sec:sGB}
\subsection{Action}
The action of sGB theory is given by:
\begin{equation}\label{action}
\begin{split}
    S=&\frac{c^3}{16\pi G}\int_{M} d^4x\sqrt{-g}\left[R-2 (\nabla\varphi)^2+\alpha f(\varphi)\mathcal{R}^2_{GB}\right]\\
    &+S_{m}[\Psi_m,\mathcal{A}^2(\varphi)g_{\mu\nu}]\,,
\end{split}
\end{equation}
where $R$ is the Ricci scalar on the four-dimensional manifold $M$ with the metric $g_{\mu \nu}$ and $\varphi$ is the canonical scalar field. The fundamental coupling constant of the theory is $\alpha$ and has dimensions of length squared; we discuss current bounds on it below. The function $f(\varphi)$ is a dimensionless coupling function, which we keep generic for most of the paper except for specific case studies in Sec.~\ref{sec:results}. The GB invariant $\mathcal{R}^2_{GB}$ is given by 
\begin{equation}
\mathcal{R}^2_{G B}=R^2-4 R^{\mu \nu} R_{\mu \nu}+R^{\mu \nu \rho \sigma} R_{\mu \nu \rho \sigma}.
\end{equation}
The quantity $S_{m}$ is the general matter action with $\Psi_m$ denoting the matter fields having a generic non-minimal coupling to the metric through a function ${\cal A}(\varphi)$.

There are two different interpretations of the action~\eqref{action}, which is expressed in the so-called Einstein frame. In this frame, the scalar field is not coupled to the Ricci scalar, so the GR limit is preserved even for large scalar fields. Using a conformal transformation, we can equivalently express the action in the Jordan frame, where the scalar field is also coupled to $R$ but couplings to standard-model fields are unaffected, which corresponds to $\mathcal{A}(\varphi)=1$ in $S_m$. For GB corrections derived from the low energy limit of quantum gravity theories such as string theories the direct result is in the Jordan frame, with~\eqref{action} representing a transformation of the theory with 
conformal factor $\mathcal{A}(\varphi)=e^{\varphi}$ for dilatonic couplings. A different perspective on the action~\eqref{action} is as an effective theory that reduces to GR in low-curvature regimes but includes higher-curvature corrections. In this case the Einstein frame action~\eqref{action} is the fundamental theory. 
In principle, the choice of frame is expected to leave observables unaffected, provided the calculations are carried out consistently to the final measurements.  
We also note that the Einstein and Jordan frame descriptions coincide in the weak-coupling limit, where $\varphi$ is small and the conformal factor can be approximated to be $1$. Having noted these subtleties, we will work with the action in the form~\eqref{action} throughout this paper.

We next discuss existing bounds on sGB theory. For the case of a dilatonic coupling function, current observational constraints on $\alpha$ are the following. Minimum constraints are set by solar system tests; the measurement of the Shapiro time delay by the Cassini spacecraft estimated $\sqrt{\alpha} < 8.9 \times 10^{6}$ km~\cite{Bertotti:2003rm}. Stronger constraints come from low-mass x-ray binary observations~\cite{Yagi:2012gp} and from Bayesian parameter estimation with GW detections~\cite{Nair:2019iur}. The latest stringent GW bounds are derived from the analysis of the O1-O3 datasets of the LIGO/Virgo and show
$\sqrt{\alpha} \lesssim 0.4-1.33$~\cite{Perkins:2021mhb,Wang:2021yll,Lyu:2022gdr,Saffer:2021gak} km. 
Cases with quadratic or Gaussian coupling functions have, so far, remained unconstrained from observations.

To proceed with calculations based on the action~\eqref{action} requires specifying the relevant matter action.  
To make the calculations analytically tractable we will use an adiabatic approximation for the system during the early inspiral. In this regime, the timescales associated with the source dynamics are much faster than the changes induced by GW losses. This enables an approximate description that separates between the dynamics of the source and the radiation, which are connected through energy and angular momentum balance laws. We further adapt this description to the hierarchy of lengthscales in the early inspiral, at large orbital separation.

\subsection{Hierarchy of scales in a binary inspiral}\label{scales}

The system considered in this work encompasses a binary of two non-spinning black holes with a nontrivial scalar configuration  at large separation $r$. We further assume that the black holes follow circular orbits that slowly decay due to GW losses. To calculate the GWs requires solving the equations of motion derived from the action~\eqref{action} for the dynamical spacetime and scalar field. This is a theoretical challenge. Thus, we focus here on the inspiral regime of the system, where the black holes are at large separation and a clear hierarchy of length- and time-scales emerges which makes the problem amenable to analytical approximation methods. Such methods rely on a tapestry of different perturbative expansions adapted to different patches of the spacetime. In the calculations below, we trace the flow of information between different parts of this system, from the consideration of a nearly isolated black hole experiencing scalar tidal perturbations to the orbital scales in the dynamics and ultimately to the waveforms at the GW detectors. These different regimes are characterized by different length scales that are important, as illustrated in Fig.~\ref{fig:scales}. 
We see that the smallest scale is the size of the BH, which is of order the Schwarzschild radius 
\begin{equation}
r_S= \frac{2GM}{c^2},\label{eq:rSdef}
\end{equation}
with $M$ being the mass of the BH. 
The typical scale associated to the scalar condensate surrounding the BH is $L_\varphi\sim 10 r_S$, as we show in Sec.~\ref{sec:Lovenumber} below. The orbital separation $r$ during the early inspiral is the second-largest scale, with the longest one being the reduced wavelength $\lambdabar$ of the GWs. Different physics dominates on these different scales, which makes it useful to divide the calculation into three zones: the body-zone near the BHs, the orbital (near)zone, and the far(wave)zone, the latter two we show explicitly in Fig. \ref{fig:scales}. In each zone we use a different approximation scheme, and match between them in  intermediate zones.

To establish suitable perturbation schemes in each zone requires identifying the relevant small dimensionless parameters to define different perturbation expansions. The first such parameter relevant here is the ratio of the characteristic size of each BH and scalar condensate to the orbital radius, which we denote by 
\begin{subequations}
\begin{equation}
\label{eq:epsilontidal}
\varepsilon_{\rm tid}\sim \frac{L_\varphi}{r}\ll 1.
\end{equation}
This assumption assures the validity of the skeletonization approximation which we introduce in the next section. We will comment further on the scale $L_\varphi$ and the range of validity of the skeletonized approximation in Sec.~\ref{sec:commentonLphi}. 
In addition, we use the PN weak-field, slow motion parameter characterizing relativistic corrections 
\begin{equation}
\varepsilon_{\rm PN}\sim \frac{GM}{rc^2}\sim \frac{v^2}{c^2}\ll 1,
\end{equation} 
 and the coupling strength of deviations from GR
\begin{equation}\label{eq:epsilon}
  \epsilon = \frac{c^4 \alpha}{G^2 m^2} \ll 1
  \end{equation}
where $m$ is the total mass of the binary system. We will also use a more restrictive version involving the GB coupling parameter that is based on each BH in the binary system, where 
\begin{equation}
\label{eq:alphahat}
\hat \alpha =\frac{\alpha}{r_S^2}\ll 1.
\end{equation}
\label{eq:smallparameters}
\end{subequations}
We treat all of these small parameters~\eqref{eq:smallparameters} as independent.\\

The organization of our calculation is that we start with the description of the system viewed on the orbital scale, where the BHs behave essentially as point particles with small corrections due to the scalar condensates encapsulated in a scalar-dependent mass and a tidally induced scalar dipole moment. For the computations in sections~\ref{sec:Binarydyn} and \ref{sec:WaveformPhasing} we expand perturbatively in the Post Newtonian parameter $\varepsilon_{\rm PN}$. This expansion is valid in the near zone in which the distance to a field point and the source is less than the characteristic wavelenght\cite{Pati:2000vt}. 

At larger distances, the retardation effects, which are considered small in the PN expansion, become important. Additionally in section \ref{sec:WaveformPhasing} we consider the waveforms at the detector, one can expand in large $R$, resulting in a multipolar treatment of the fields. In these sections we implicitly assume small $\varepsilon_{\rm tid}$ such that a skeletonized description applies but not making additional explicit approximations in this parameter. 

At this stage, our results are valid for generic values of the GB coupling. 
In Sec.~\ref{circularorbits} we specialize the binary calculations to quasi-circular orbits. To obtain closed-form expressions for quantities of interest requires expanding in $\epsilon$ and  $\varepsilon_{\rm tid}$, in addition to the expansion in $\varepsilon_{\rm PN}$. We obtain results for the GW phase in terms of global characteristics of each body, i.e. BH and scalar condensate,  when viewed from far away, such as a mass, sensitivity, and tidal deformability coefficient. To relate these parameters to the fundamental properties of the BH and scalar field configuration requires a different approximation adapted to studying the detailed behavior in the proximity of one BH, which we discuss in Sec.~\ref{sec:Lovenumber}. For the computations in this regime, again to obtain analytical closed-form results, we use a double perturbative expansion in $\hat{\alpha}$ and $\varepsilon_{\rm tid}$ to linear order in both parameters. With these results in hand we match to the parameters entering the description of the orbital dynamics and GWs at large distances $\tilde{r}$ from the black hole by extracting the multipole structure of the  configuration. 

\subsection{Skeletonized matter action with scalar-induced tidal effects}
\label{sec:tidalEFT}
The hierarchy of scales discussed above enables an approximate description of the orbital dynamics based on a worldline skeleton~\cite{Dixon:1970zza,Eardley}. In this skeletonization approach, a compact object, say body $A$, is reduced to a central worldline $z^\mu_A$ with tangent $u_A^\mu=dz^\mu_A/d\tau_A$, together with additional global characteristics such as a mass and higher multipole moments. As in other effective field theories, an effective action for the worldline dynamics, $S_m$, can be constructed as a functional of the fields $\varphi, \, g_{\mu\nu}$ expanded around their values on the central worldline for each body. By using field redefinitions and a derivative counting scheme one can truncate the expansion to a finite number of terms. For finite size effects, a more powerful approach than a direct derivative expansion that is capable of capturing additional internal dynamical effects of the object is to use the  multipole moments associated to the worldline as the fundamental dynamical degrees of freedom in the effective action. Symmetry considerations such as time-reversal and parity-invariance then dictate the kinds of couplings that can appear in such an action, and a multipole counting scheme truncates the number of relevant terms. The coupling coefficients in the effective action are fixed by matching to a full description of the body, which determines the information about the body they encode. The skeletonized action framework was initially developed to describe compact objects viewed on the orbital scale, however, it can also be used to describe objects surrounded by matter configurations~\cite{Eardley,Julie:2019sab,Bernard:2019yfz} provided that the matter configurations are sufficiently concentrated that the hierarchy of scales still applies. Here, we will use this formalism for including the effects of tidally induced scalar dipole moments, which are absent in GR, in addition to the previously established~\cite{Julie:2019sab} point-mass action. We comment on the regime of validity of this approach in Appendix~\ref{sec:apptidalEFT} and Sec.~\ref{sec:Lovenumber}.

Specifically, we consider the skeletonized matter action to be a sum of the point-particle action $S_{\rm pp}$ and the action for the scalar-induced dipolar tidal effects $S_{\rm tid}$:
\begin{equation}\label{skeletonization}
S_{m}[g_{\mu\nu},\varphi,x_{A}^{\mu}]= S_{\rm pp} +S_{\rm tid}\,.
\end{equation}
The point-particle effective action contains the center-of-mass contributions from the two objects $S_{\rm pp}=S_{\rm pp}^A+S_{\rm pp}^B$, which can be written in terms of a scalar-dependent mass $\bar M(\varphi)$ as
\begin{subequations}
\label{eq:totalSm}
\begin{equation}\label{eq:Sm}
S_{\rm pp}^A=-c\int \bar M_{A}(\varphi)ds_A\,,
\end{equation}
and similarly for body $B$, where $ds_A =\sqrt{-g_{\mu\nu}dx^{\mu}_{A}dx^{\nu}_{A}}$ is the differential along the worldline of particle A. Here, the scalar-dependent mass $\bar M_A$ is at this stage just a coupling coefficient, which accounts for the presence of a scalar condensate surrounding the body. However, as this mass term does not include any gradients of $\varphi$ it does not describe finite-size effects. We add such effects based on the action  describing linear tidal effects for the scalar dipole, which is derived in Appendix~\ref{sec:apptidalEFT} and was previously used in the context of scalar-tensor theories in~\cite{Damour:1998jk, Bernard:2019yfz}. For a body $A$ the tidal interactions involve a scalar dipole moment denoted by $Q^\mu_A$ and a scalar tidal field sourced by the companion ${\cal E}^\mu_A$. The tidal field felt by body $A$ and sourced by body $B$ is given by ${\cal E}^\mu_A=-\partial_\mu \varphi_B$, where an evaluation on the worldline of $A$ is implied; see Appendix~\ref{sec:apptidalEFT} for further discussion of the tidal field. The tidal action for scalar-induced dipolar effects consists of the coupling between the dipole and the tidal field, and a contribution from the internal dynamics of the dipole, and is given by $S^{\rm tid}_A=c\int ds_A {\cal L}^{\rm tid}_A$ with
\begin{equation}
\label{eq:StidwithQ}
{\cal L}^{\rm tid}_A=-Q^{\mu}_A {\cal E}_{\mu}^A+\frac{1}{2\lambda^A_s(\omega^A_s){}^2}\left( \dot Q^{\mu}_A\dot Q_{\mu}^A-(\omega^A_s){}^2Q_{{\mu}_A} Q^{\mu}_A\right),
\end{equation}
 where $\lambda_s^A$ and $\omega^A_s$ are 
 coupling coefficients already written in a suggestive form and we have chosen the normalization of the dipole such that the coupling coefficient for the first term is unity. Here, overdots denote derivatives with respect to proper time along the worldline. 
 The equations of motion for $Q_\mu^A$ derived from the action~\eqref{eq:StidwithQ} in the adiabatic limit, where we assume that variations in the tidal field are much slower than the internal timescales associated to the dipole, are
\begin{equation}
\label{eq:lambdasdef}
Q_\mu^A=-\lambda_s^A {\cal E}_\mu^A.
\end{equation}
Using~\eqref{eq:lambdasdef} to integrate out the dipole degrees of freedom from the adiabatic limit of~\eqref{eq:StidwithQ} leads to the effective tidal action
\begin{equation}\label{Stid}
S_{\rm tid}^A=c \int \mathrm{d} s_{A} \, \frac{\lambda_s^{A}}{2}\partial_{\mu} \varphi_{B}\partial^\mu \varphi_{B}\,.
\end{equation}
\end{subequations}
The coupling parameter is called the dipolar scalar tidal deformability parameter $\lambda_s^A$, which characterizes the response of A to a static scalar dipolar tidal field. The full definition of this parameter and its dependence on the fundamental properties of the theory and the BH can be found in Sec.~\ref{sec:calclovenumer}. Similar to the point-particle action, the total tidal action is the sum of the individual contributions $S_{\rm tid}=S_{\rm tid}^A+S_{\rm tid}^B$, with the results for body $B$ obtained by interchanging the body labels $(A\leftrightarrow B)$ in~\eqref{Stid}.

\subsection{Field Equations}
The field equations are found by varying the action~\eqref{action} and~\eqref{eq:Sm} with respect to the dynamical fields. Here we focus on the matter contribution from the point-particle action. The contribution from the tidal action is separately considered in Sec.~\ref{sec:2bodyL} at the level of the Lagrangian. As discussed in \cite{Bernard:2023eul} this equivalent to incorporating the tidal contribution at the level of the field equations at the order of the approximations we are considering. For the tensor field in the trace-reversed form, we obtain
\begin{equation}\label{FE1}
    \begin{aligned}
    R_{\mu \nu}=& 2 \nabla_\mu \varphi \nabla_\nu \varphi-4 \alpha\left(P_{\mu \alpha \nu \beta}-\frac{g_{\mu \nu}}{2} P_{\alpha \beta}\right) \nabla^\alpha \nabla^\beta f(\varphi)\\
    &+ \frac{8 \pi G}{c^4 }\left(T^{\rm pp}_{\mu\nu}-\frac{g_{\mu\nu}}{2}T^{\rm pp}\right),
    \end{aligned}
\end{equation}
where $P_{\mu \nu \rho \sigma}=R_{\mu \nu \rho \sigma}-2 g_{\mu[\rho} R_{\sigma] \nu}+2 g_{\nu[\rho} R_{\sigma] \mu}+g_{\mu[\rho} g_{\sigma] \nu} R$, and $P_{\mu \nu} \equiv P^\lambda{ }_{\mu \lambda \nu}$. The point-mass part of~\eqref{FE1} agrees with~\cite{Julie:2019sab}.
The distributional energy momentum tensors and their traces are given by
\begin{equation}
\label{EMtensors}
    T_{\mu\nu}^{\rm pp} = \frac{-2}{\sqrt{-g}}\frac{\delta S_{\rm pp}}{\delta g^{\mu\nu}},
    \end{equation}
    with the corresponding traces  $T^{\rm pp}=g^{\alpha\beta}T^{\rm pp}_{\alpha\beta}$ and similarly for the tidal part. 
Varying the action with respect to the scalar field leads to
\begin{equation}\label{FE2}
    \Box \varphi = -\frac{1}{4}\alpha f'(\varphi)R_{GB}^2- \frac{4\pi G}{c^4} (\bar{\delta} S_{\rm pp} ),
\end{equation}
with,
\begin{subequations}\label{deltaS}
    \begin{align}
    &\bar{\delta} S_{\rm pp} = \frac{1}{\sqrt{-g}}\frac{\delta S_{\rm pp}}{\delta \varphi},
    \end{align}
\end{subequations}
where $\square \equiv g^{\alpha \beta} \nabla_{\alpha} \nabla_{\beta}$ denotes the d'Alembertian operator. 

\section{Post-Newtonian, small coupling, and tidal approximations for the binary dynamics}\label{sec:Binarydyn}
\subsection{Expansions of the fields}
In order to find the contribution of the scalar tidal-effects to the conservative dynamics of compact bodies, we rely on an approximation scheme based on perturbative expansions in the PN approximation, where we work to 1PN order, as well as in the GB coupling and tidal effects, which we each consider to linear order. We treat these expansions as independent, though assume that the corrections are all small; more details are discussed in Sec.~\ref{subsec:commentexpansion}. For simplicity, we do not indicate the triple expansion on all the quantities. We also note that some of the results in this section have broader validity as, for instance, they do not require the small-coupling assumption. 

The PN framework restricts the dynamics to the region of weak-field and low-velocity for a gravitationally bound system, i.e., $Gm/rc^2\approx v^2/c^2\ll1$, where $m,r$, and $v$ are the characteristic mass, size, and velocity of the source. These are the relevant dimensionless expansion parameters that are used for the PN perturbative expansion. 
Here we keep track of factors of $1/c$, which are each counted as a half-PN order.

To solve the field equations \eqref{FE1} and \eqref{FE2} in the space-time region around the binary (i.e. the near-zone, see Fig.~\ref{fig:scales}), we expand the metric around Minkowski space-time and the scalar field around the background value $\varphi_{0}$~\cite{Damour:1990pi}
\begin{equation}
\label{eq:PNscalings}
    \begin{aligned}
    &g_{00} = -e^{-2U/c^2}+\mathcal{O}\left(c^{-6}\right)\,,\\
    &g_{0i} = -4 g_i/c^3+\mathcal{O}\left(c^{-5}\right)\,,\\
    &g_{ij} = \delta_{ij}e^{2U/c^2}+\mathcal{O}\left(c^{-4}\right)\,,\\
    &\varphi=\varphi_{0}+ c^{-2}\delta \varphi^{(1)}+\mathcal{O}\left(c^{-4}\right).
    \end{aligned}
\end{equation}

From the expansion for the scalar field~\eqref{eq:PNscalings} it follows that the scalar-dependent mass $\bar M_{A}(\varphi)$ has an expansion of the form
\begin{eqnarray} \label{eq:Mexpansion}
\bar M_{A}&=&m_{A}^0\left\{1+\frac{\alpha^0_A}{c^2}\delta \varphi^{(1)}+\frac{1}{c^4}\left(\alpha^0_A \delta\varphi^{(2)}\right.\right.\nonumber\\
&&\left.\left.+\left[\left(\alpha_A^0\right)^2+\beta^0_A\right]\delta \varphi^{(1)}\right)\right\} +\mathcal{O}(c^{-6})\,,
\end{eqnarray}
where $m_A^0$ denotes the leading-order coefficient, which we will match with the BH mass in Sec.~\ref{sec:Lovenumber}. The so-called \textit{sensitivity} parameter $\alpha_A^0$ measures the coupling of the skeletonized BH to the background scalar field and is defined by
\begin{equation}
\label{eq:alpha0def}
    \left.\alpha_{A}^{0}=\frac{d\, \ln[\bar M_{A}(\varphi)]}{d\varphi}\right|_{\varphi=\varphi_{0}},
    \qquad
    \left.\beta_{A}^{0}=\frac{d\, \alpha_{A}(\varphi)}{d\varphi}\right|_{\varphi=\varphi_{0}}.
\end{equation}
We will relate these parameters to the scalar charge in Sec.~\ref{sec:Lovenumber}.

With the above ansatz and approximations, we expand the equations of motion and solve for the expansion coefficients order by order in $1/c$. This results in solutions for the fields $U$, $g_i$ and $\varphi$ in the near-zone, where the PN expansion is valid, as discussed in Sec.~\ref{scales}. We reproduced the expressions for the near zone fields as found in \cite{Julie:2019sab, Julie:2019erratum} and we refer to their Appendix D for the explicit derivation. 
The solutions have the leading-order behavior
\begin{eqnarray}
\label{eq:UphiLO}
U&=& \frac{G m_A^0}{r}+(A\leftrightarrow B)+O(c^{-2}),\\
\delta \varphi^{(1)}&=& -\frac{G m_A^0 \alpha_A^0}{r} +(A\leftrightarrow B).
\end{eqnarray}
These solutions are valid in the region of the orbital scales, in the near zone~\cite{Pati:2000vt}. Using these solutions, Ref.~\cite{Julie:2019sab} constructed the effective PN Lagrangian for the dynamics of a binary system. Below, we discuss the modifications to the Lagrangian due to the tidal effects. 

\subsection{Two-body Lagrangian with dipolar tidal effects}\label{sec:2bodyL}
The Lagrangian of body $A$ in the field of body $B$ is given by 
\begin{equation}\label{deflag}
    \mathcal{L}_A = \frac{dS_{\rm pp}^A}{dt} + \frac{dS_{\rm tid}^A}{dt}.
\end{equation}
The term $dS_{\rm pp}^A/dt$ was calculated in~\cite{Julie:2019sab} and we do not repeat it here. Using similar methods for the tidal contribution leads to
\begin{equation}\label{tidalL}
\begin{aligned}
    \frac{dS_{\rm tid}^A}{dt} &= \frac{1}{2}c\lambda_A^{(s)}\sqrt{-g_{\alpha\beta}^A \frac{dx_A^{\alpha}}{dt}\frac{dx_A^{\beta}}{dt}}g^{\mu\nu}_B\partial_{\mu}\varphi_B\,\partial_{\nu}\varphi_B\\
    &=\frac{1}{2} c \lambda_A^{(s)}\sqrt{c^2\left(1-\frac{U_A}{c^2}- \frac{\mathbf{v}_A^2}{2c^2}\right)}\times \\
    &\qquad[-e^{2U_B/c^2}(\partial_0\varphi_B)^2 +e^{-2U_B/c^2}(\nabla\varphi_B)^2 ]+\ldots \\
    &=\frac{1}{2}\lambda_A^{(s)}c^2(\nabla \varphi_B)^2+O(c^{-4}).\end{aligned}
\end{equation}
Using the lowest order solution for $\varphi$ from \eqref{eq:UphiLO}, this can be written as
\begin{equation}\label{tidalLL}
\frac{dS_{\rm tid}}{dt}= \frac{1}{2}  \lambda_A^{(s)} \frac{G^2 (m_B^0 \alpha_B^0)^2}{c^2 r^4}\,+O(c^{-4}).
\end{equation}
Combining \eqref{tidalLL} with the point particle Lagrangian from~\cite{Julie:2019sab}, the overall two-body Lagrangian to 1PN order is given by
\begin{equation}\label{Lagrangian}
\begin{aligned}
&L_{A B} =-m_{A}^0c^2+\frac{1}{2} m^0_{A} \mathbf{v}_{A}^{2}+\frac{ G\bar\alpha\, m_{A}^0 m_{B}^0}{2r}+\frac{1}{8c^2} m^0_{A} \mathbf{v}_{A}^{4} \\
&\quad+\frac{G\bar\alpha\, m^0_{A} m^0_{B}}{rc^2}\left[-\frac{G\bar\alpha m^0_{A}}{2 r}\left(1+2 \bar{\beta}_{B}\right)+\frac{3}{2}\left(\mathbf{v}_{A}^{2}\right)\right.\\
&\quad\left.-\frac{7}{4}\left(\mathbf{v}_{A} \cdot \mathbf{v}_{B}\right)-\frac{1}{4}\left(\mathbf{n} \cdot \mathbf{v}_{A}\right)\left(\mathbf{n} \cdot \mathbf{v}_{B}\right)+\frac{\bar{\gamma}}{2}\left(\mathbf{v}_{A}-\mathbf{v}_{B}\right)^{2}\right] \\
&\quad+\frac{\alpha f^{\prime}\left(\varphi_{0}\right)}{r^{2}} \frac{G^2 m^0_{A} m^0_{B}}{r^{2} c^2}\left[m^0_{A}\left(\alpha_{B}^{0}+2 \alpha_{A}^{0}\right)\right]\\
&\quad +\frac{1}{4} \frac{G^2 \bar{\alpha}^2 \mu m}{r^4 c^2}\zeta +(A\leftrightarrow B)\,,
\end{aligned}
\end{equation}
where $m=m_A^0+m_B^0$ is the total mass, $\mu = m_A^0 m_B^0/m$ is the reduced mass, $r=\|\mathbf{r}\|=\|\mathbf{x}_{A}-\mathbf{x}_{B}\|$ is the relative separation vector and  $\mathbf{n} =\mathbf{r}/ r$ its unit vector. 
Following the notation of~\cite{Shiralilou:2021mfl}, we define the binary parameters:
\begin{equation}\label{EOMparams}
\begin{split}
&\bar{\alpha} \equiv \left(1+\alpha_{A}^{0} \alpha_{B}^{0}\right)\,,\quad
\bar{\gamma}\equiv-2 \frac{\alpha_{A}^{0} \alpha_{B}^{0}}{\bar{\alpha}}\,,\quad
\bar{\beta}_{A} \equiv \frac{1}{2} \frac{\beta_{A}^{0} (\alpha_{B}^{0})^{ 2}}{\bar{\alpha}^{2}}\,,\\
&\mathcal{S}_{\pm}  \equiv \frac{\alpha_{A}^{0} \pm \alpha_{B}^{0}}{2 \sqrt{\bar{\alpha}}}, \quad \beta_{\pm}  \equiv \frac{\bar{\beta}_{A} \pm \bar{\beta}_{B}}{2}\,,
\end{split}\end{equation}
with the coefficients $\alpha^0, \beta^0$ from~\eqref{eq:alpha0def}. We also define the quantity $\zeta$ to be the weighted averages of the tidal deformabilities
\begin{equation}\label{deftidbinding}
\zeta \equiv \lambda_{A}^{(s)} \frac{m_{B}^{0} \alpha_{B}^{0}{ }^{2}}{\bar{\alpha}^2 m_{A}^{0}}+\lambda_{B}^{(s)} \frac{m_{A}^{0} \alpha_{A}^{0}{ }^{2}}{\bar{\alpha}^2 m_{B}^{0}}\,.
\end{equation}
As discussed in~\cite{Shiralilou:2021mfl}, the scalar contributions to the dynamics have a similar effects as in scalar-tensor theories in that they re-normalize the gravitational constant to be $G \bar\alpha$, which also appears in the scalar finite-size terms but not in the additional GB effects from the higher curvature corrections. 

From the Lagrangian~\eqref{Lagrangian} we obtain the equations of motion using the Euler-Lagrange equations. After converting to the center-of-mass (CM) frame using the relations given in~\cite{Shiralilou:2021mfl} -- for which we have confirmed that the tidal corrections do not modify these equations at the considered order --
 we obtain the relative acceleration \begin{equation}\label{arel}
\begin{aligned}
&\mathbf{a}=-\frac{G \bar{\alpha} m }{r^{2}}\mathbf{n}+\frac{G \bar{\alpha} m}{c^2 r^{2}}\left\{\mathbf{n}\left[\frac{3}{2} \eta \dot{r}^{2}-(1+3 \eta+\bar{\gamma}) \mathbf{v}^{2}\right]\right.\\
&\left.+2 \mathbf{v} \dot{r}[2-\eta+\bar{\gamma}]+\frac{2 G \bar{\alpha} m \mathbf{n}}{r}\left[2+\eta+\bar{\gamma}+\beta_{+} - \frac{\Delta m}{m}\beta_{-}\right]\right.\\
&\left.+\frac{2 G \bar{\alpha} m \mathbf{n}}{r}\left[-\frac{2 \alpha f^{\prime}\left(\varphi_{0}\right)}{ \bar{\alpha}^{3/2} r^{2}}\left(3 S_{+} + \frac{\Delta m}{m} S_{-}\right) -\frac{\zeta}{m r^2}\right]\right\}\,,
\end{aligned}
\end{equation}
where $\Delta m=m_A^0 -m_B^0$ is the mass difference, $\eta=\mu/M$ is the symmetric mass ratio, $\mathbf{v}=\dot{\mathbf{r}}$ is the relative velocity.

\subsection{Discussion of the expansion scheme adopted here}
\label{subsec:commentexpansion}
 The results computed thus far were based only on the PN approximation, without any additional assumptions on the GB coupling. Within this approximation, the Lagrangian~\eqref{Lagrangian} contains terms at linear order in $\alpha$ and in $\lambda_s$, which first appear at $\mathcal{O}(1/c^2)$, i.e. the same order as the 1PN GR terms. However, they scale with the orbital separation as $1/r^4$, similar to 3PN GR terms, which arises because the coupling coefficients are dimensionful, for instance $\alpha/r^2$ and $G \zeta/(c^2 r^3)$ are dimensionless.
 Similar degeneracies of tidal and PN corrections can in fact be disentangled via analytic continuation \cite{Creci:2021rkz}.
In the calculations below, which will require perturbatively inverting various relations for circular orbits, we will explicitly expand to leading orders in the small coupling  $\epsilon$ and the tidal perturbations $\varepsilon_{\rm tid}$, as well as for small PN parameter, treating any cross-terms as negligible corrections for our purposes here. To avoid further notational complexity of the equations, we do not explicitly indicate this expansion on all the quantities, and also do not convert explicitly to appropriate dimensionless coefficients. We also note that adopting this triple expansion from the beginning leads to the same Lagrangian as~\eqref{Lagrangian}, which corroborates the consistency of our approach. Thus, given the above considerations, we do not use an explicit PN order counting for the tidal and GB contributions, but note that they are generally suppressed compared to 1PN GR terms. This is similar to the case of tensor tidal effects in GR, where an analogous situation arises and is addressed via a similar mathematical formalism~\cite{Flanagan:1997fn}. 

\subsection{Radius-frequency relation and binding energy for circular orbits}\label{circularorbits}
In the following, we will restrict our attention to the quasi-circular orbits with  $\dot{r}=\ddot{r}=0$. The orbital angular frequency $\omega$ is computed from $\omega^2 =-\mathbf{a}\cdot \mathbf{r}/r^2$ evaluated in the circular limit, which gives the generalized Kepler’s law: 
\begin{equation}\label{angfreq}
\begin{aligned}
    &\omega^2 = \frac{G \bar{\alpha} m }{r^{3}} \left\{1- \frac{G \bar{\alpha} m}{r c^2}\left[3 - \eta + \bar{\gamma} + 2\beta_{+} - 2\frac{\Delta m}{m}\beta_{-}\right.\right.\\
    &\left.\left.-\frac{4 \alpha f^{\prime}\left(\varphi_{0}\right)}{ \bar{\alpha}^{3/2} r^{2}}\left(3 S_{+} +  \frac{\Delta m}{m} S_{-}\right) -\frac{2 \zeta}{m r^2} \right]+\mathcal{O}(c^{-4})\right\}\,.
\end{aligned}
\end{equation}
Because the observable quantity is the frequency rather than the gauge-dependent orbital radius, we invert~\eqref{angfreq} to eliminate $r$ in favor of $\omega$. It is convenient to work with a frequency parameter
\begin{equation}
\label{eq:xdef}
  x=\left(\frac{G \bar{\alpha}m\omega}{c^3}\right)^{2/3}\,,
\end{equation}
which differs from the commonly-used analogous quantity in GR by a factor of $\bar{\alpha}$. We perturbatively invert~\eqref{angfreq} by expanding to  linear order in $\epsilon$ and $\varepsilon_{{\rm tid}}$ and truncate the remaining terms to 1PN order. This leads to the radius-frequency relationship
\begin{equation}\label{rtox}
\begin{aligned}
    r(x) &= \frac{G \bar{\alpha} m}{c^{2}x}\left[1-\frac{1}{3 }x\left(3 - \eta + \bar{\gamma} + 2\beta_{+} - 2\frac{\Delta m}{m}\beta_{-} \right)\right.\\
    &\left.+\frac{4 \alpha  f^{\prime}\left(\varphi_{0}\right) }{G m^2 \bar{\alpha}^{7/2} }c^4 x^3\left( S_{+} +  \frac{\Delta m}{3 m} S_{-}\right) + \frac{2 \zeta }{3 G^2\bar\alpha^2 m } c^4 x^3 \right]. \qquad
    \end{aligned}
\end{equation}
From the Lagrangian~\eqref{Lagrangian} we  derive the conserved binding energy of the system, again working perturbatively in the small parameters. In the CM frame, and after expressing the result in terms of $x$ using \eqref{rtox}, we obtain
\begin{subequations}
    \label{BE}
\begin{equation}
\begin{aligned}
    E(x) &= - \frac{\mu c^2 x}{2} \left[ 1+x \left( -\frac{3}{4} - \frac{\eta}{12} +E_{\rm s}\right)\right.\\
    &\left.+x^3c^4\left(E_{\rm GB}+E_{\rm tid}\right)\right],
    \end{aligned}
\end{equation}
with the sensitivity-dependent contributions
\begin{equation}
E_s=- \frac{2}{3} \bar{\gamma} +\frac{2}{3}\beta_{+} -\frac{2}{3}\frac{\Delta m}{m}\beta_{-},
\end{equation}
the GB contribution
\begin{equation}
\label{eq:GBbinding}
E_{\rm GB}=- \frac{10}{3} \frac{ \alpha f'(\varphi_0)}{G^2 \bar{\alpha}^{7/2} m^2}  \left(3 S_{+} + \frac{\Delta m}{m} S_{-}\right),
    \end{equation}
    and the tidal terms
    \begin{equation}
\label{eq:tidalbinding}
    E_{\rm tid}=- \frac{5}{3} \frac{ \zeta}{G^2 m^3 \bar{\alpha}^2} \,.
    \end{equation}
\end{subequations}
 In section~\ref{sec:results} we analyze the dependencies of the tidal and GB contributions to the binding energy on the parameters of the binary system. 

\section{Tidal effects in the scalar and  gravitational waveform}\label{sec:WaveformPhasing}
In this section we derive different contributions from the tidal terms as appearing in the scalar and tensor waveforms. Focusing on circular orbits, we also find tidal modifications to the scalar and tensor energy flux rates and the overall GW phasing in the Fourier domain.
\subsection{Tidal-Induced Dipolar Radiation}\label{diprad}
The scalar and tensor waveforms are constructed from radiative solutions to the equations of motion in the far-zone. For the scalar field we can rewrite \eqref{FE2} as 
\begin{equation}\label{EOMsourceterm}
\Box_{\eta}\frac{\delta\varphi}{c^2} = \Box_{\eta}\Phi = 4 \pi \mu_{s},
\end{equation}
with $\mu_{s}$ containing the source terms from~\eqref{FE2}.
For the waveform calculation, we are interested in the solutions to this equation at the position of a distant observer (such as the GW detectors), in contrast to the near-zone solutions discussed in Sec.~\ref{sec:Binarydyn}. This requires a different approximation scheme for the integration, as explaind in details in~\cite{Will:1996zj,Lang:2014osa}. 
The relevant part of the scalar field integral can be expressed as a multipole expansion of the form
\begin{equation}\label{phiNZgeneral}
\begin{aligned}
\Phi(x)&=\sum_{l=0}^{\infty} \Phi_{l}(x)\\
   &= - \sum_{l=0}^{\infty} \frac{(-1)^l}{l!} \partial_L \left(\frac{1}{R} I^L_s(\tau)\right),
   \end{aligned}
\end{equation}
with $R$ the distance from the source to the observer and $\tau=t-R/c$ is the retarded time. The scalar multipole moments are given by 
\begin{equation}\label{I}
    I^L_s(\tau) = \int_{\mathcal{M}} d^3 \mathbf{x'} \mu_s(\tau, \mathbf{x'}) \mathbf{x'}^L\,.
\end{equation}
We refer to \cite{Pati:2000vt, Shiralilou:2021mfl} for the details on this derivation following the DIRE approach~\cite{Pati:2000vt,Will:1996zj,Blanchet:1996pi}. 
The dominant radiative part of the scalar multipole fields at large distances is given by
\begin{equation}\label{poleswaveform}
 \Phi_{l}(x)=\frac{N_{L}}{R\, l !}\left(\frac{\partial}{c \partial t}\right)^{l} I_{s}^{L}+\mathcal{O}\left(R^{-2}\right),
\end{equation}
where $N=\mathbf{R}/R$ is the directional unit vector pointing from the source to the observer.
We are particularly interested in calculating the corrections to $\Phi_l$ due to the scalar tidal effects.

As discussed in Sec.~\ref{sec:tidalEFT}, the leading-order scalar tidal effects lead to induced dipole moments $Q_{\mu}^{(s)}$ of each body. From~\eqref{eq:UphiLO} and~\eqref{eq:lambdasdef} we obtain the leading-order induced dipole moment 
\begin{equation}\label{eq:Qi}
\begin{aligned}
    &Q^{(s)}_i = -\sum_{A\neq B}\frac{\lambda_A^{(s)} }{c^2} \partial_{i} \frac{G m_B^0 \alpha_B^0}{r}= \bar{\zeta} \frac{r_i}{c^2 r^3},
    \end{aligned}
\end{equation}
with 
\begin{equation}
\label{eq:barzetadef}
\bar{\zeta} =\lambda_A G m_B^0 \alpha_B^0 +\lambda_B G m^0_A \alpha_A^0.
\end{equation}
Note that this tidal quantity involves a different combination of parameters than the tidal contributions to the binding energy~\eqref{deftidbinding}. For instance, it depends linearly instead of quadratic on the sensitivities.
The contribution from this induced dipole moment to the overall scalar radiation is \footnote{The prefactor of $\frac{G}{c^2}$ results from matching the definition of the moments \eqref{I} which contain factors of $G$ and $c$ implicitly \eqref{differentIs}, and \eqref{eq:Qi}. Note there is also an overall factor of $c^2$ difference in the definitions of both moments which can be recovered from dimensional analysis.} 
\begin{eqnarray}
\label{tidaldeltaphi}
 \Phi_{i}^{\rm tid}(x)&=&\frac{G}{R c^{2}}\frac{1}{c}\frac{\,\,\,\,\,\partial Q_{i}^{(s)}}{ \partial t} N_{i}+\mathcal{O}(R^{-2})\\
 &=&\frac{ G \mu \sqrt{\bar{\alpha}}}{R c^{3}}
 \frac{\bar{\zeta} }{\bar{\alpha}^{3/2} G m^2 \eta r^2} 
 \left(\frac{G \bar{\alpha} m}{c^{2} r^{}}\right)\nonumber\\
 &&\quad \times
 \left[(\mathbf{N}\cdot\mathbf{v})-3\dot{r}\frac{(\mathbf{N} \cdot \mathbf{r})}{r} \right]\,. \qquad \qquad \nonumber~
\end{eqnarray}
 Together with the monopole, quadrupole and octupole scalar radiation of \cite{Shiralilou:2021mfl} we can construct the scalar waveform via \eqref{phiNZgeneral} up to relative 0.5PN order. We correct the waveform relative to \cite{Shiralilou:2021mfl} removing an overall factor of two and a different prefactor of the term proportional to the coupling constant. The full scalar waveform including the tidal contribution is given in Appendix~\ref{waveformphasing} via~\eqref{phitot}.

For the case of tensor waveforms, we have shown that the tidal contributions only modify the waveforms through their corrections to the acceleration term \eqref{arel} (i.e., they do not modify tensor multipole moments at the considered order). The full gravitational waveforms to 1PN order including these tidal terms is given in~\eqref{eq:tensorwaves}.

\subsection{Tidal Contributions to the Scalar Energy Flux}
The scalar energy flux is calculated from the scalar waveform via  
\begin{equation}\label{integralEsdot}
    {\cal F}_S =  \frac{c^3 R^2}{4\pi G} \oint \dot{\Phi}^2 \, d^2\Omega,
\end{equation}
where the angular integration over the products of unit vectors in $\dot{\Phi}^2$ is computed using the identities from \cite{Thorne:1980ru}.
Tidal contributions to the energy flux arise both from the tidal dipole radiation (proportional to $\bar{\zeta}$ defined in~\eqref{eq:barzetadef}) and from the tidal contribution in the relative acceleration (proportional to $\zeta$ defined in~\eqref{deftidbinding}). The latter comes in when differentiating the lowest order term of the scalar waveform. We find the total GB contributions to the scalar energy flux to be 
\begin{equation}\label{EsdotGB}
    \begin{aligned}
    &\mathcal{F}^{GB}_{S}=\frac{\eta^{2}}{G \bar{\alpha} c^{3}}\left(\frac{G \bar{\alpha} m}{r}\right)^{4}\Bigg[- \frac{8}{ c^{2}}\left(\frac{\alpha f^{\prime}\left(\varphi_{0}\right) \mathcal{S}_{-} \mathcal{S}_{+}}{\sqrt{\bar{\alpha}} r^{2}}\right)\\
    &\left(\mathcal{S}_{+}+\frac{\Delta m}{m} \mathcal{S}_{-}\right)\left[-3 \dot{r}^{2}+ v^{2}-\frac{2 G \bar{\alpha} m}{3 r}\right]+\frac{32 Gm\bar\alpha}{ 3rc^2}\\
&\left(\frac{\alpha f^{\prime}\left(\varphi_{0}\right) \mathcal{S}_{-}^2  }{\bar{\alpha}^{3/2} r^{2}}\right)\left(3S_+ + \frac{\Delta m}{m}S_-\right)+\mathcal{O}\left(c^{-3}\right)\Bigg],
    \end{aligned}
\end{equation}
and the dipolar tidal contribution
\begin{eqnarray}
{\cal F}^{\rm tid}_{S}&=&\frac{\eta^{2}}{G \bar{\alpha} c^{3}}\left(\frac{G \bar{\alpha} m}{r}\right)^{4} \Bigg[\frac{4 \bar{\zeta} S_-}{G m \mu \bar{\alpha}^{3/2} c^2 r^2 }(-3 \dot{r}^2 + v^2\nonumber\\
&&-\frac{2 G m \bar{\alpha}}{3 r}) + \frac{G m \bar{\alpha}}{r}\frac{16 \zeta S_{-}^2}{3 m c^2 r^2}  +\mathcal{O}\left(c^{-3}\right)\Bigg].
\label{Esdottid}
\end{eqnarray}
Both of these contributions enter at $0.5$PN order. Expressions for the total scalar energy flux including all other contributions up to that order is given in Appendix~\ref{waveformphasing}.

Specializing to circular orbits, and writing the expressions in terms of the dimensionless PN parameter $x$ defined in~\eqref{eq:xdef}, the scalar flux becomes
\begin{equation}\label{FSx}
    \mathcal{F}_S (x) = x^4 c^5  \left[S4+ x(S5 + S5_{\rm GB} x^2 c^4 + S5_{\rm tid} x^2 c^4)\right],
\end{equation}
with $S4$ and $S5$ given in~\eqref{eq:S4S5T5T6} and the  GB contribution given by
\begin{equation}
\begin{aligned}
\label{eq:S5GB}
&S5_{\rm{GB}} =  \left(\frac{4 \alpha f'(\varphi_0) \eta^2 S_-}{3 \bar{\alpha}^{7/2} G^3 m^2}\right) \left[ \frac{8 S_-}{3 \bar{\alpha}}(3 S_+ + \frac{\Delta m}{m} S_-)\right.\\
&\qquad \left.- 2S_+ (S_+ + \frac{\Delta m}{m} S_-)    \right],
\end{aligned}
\end{equation}
and the tidal contribution by
\begin{equation}\label{eq:S5tid}
    S5_{\rm{tid}} = \left(\frac{4 \eta S_-}{3 G^3 \bar{\alpha} m^3 }\right)\left(\frac{\bar{\zeta} }{G \bar{\alpha}^{3/2} m}  + \frac{4\eta S_- \zeta}{3} \right).
\end{equation}
Our result~\eqref{eq:S5GB} for $S5_{\rm GB}$   contains an additional term compared to~\cite{Shiralilou:2021mfl}, which results from substituting the relative acceleration in the lowest order term of the scalar waveform. 

\subsection{Tidal corrections to the Tensor Energy Flux}
The tensor energy flux is computed from the tensor waveform via 
\begin{equation}\label{integralEdotT}
{\cal F}_T=\frac{c^3 R^2}{32 \pi G} \oint \dot{h}_{\mathrm{TT}}^{i j} \dot{h}_{\mathrm{TT}}^{i j} d^2 \Omega.
\end{equation}
Here, tidal contributions enter only from the relative acceleration when differentiating the lowest order term in the tensor waveform. The resulting tidal correction is given by
\begin{equation}\label{ETdottid}
\begin{aligned}
{\cal F}^{\rm tid}_T &=-\frac{8}{15} \frac{\eta^2}{G \bar{\alpha}^2 c^5}\left(\frac{G \bar{\alpha} m}{r}\right)^4\frac{G \bar{\alpha} }{c^2 r^3} (56 \dot{r}^2 -48 v^2) \zeta.
\end{aligned}
\end{equation}
The GB contribution is
\begin{equation}\label{ETdotGB}
\begin{aligned}
&\mathcal{F}_T^{GB} =\frac{8}{15} \frac{\eta^2}{G \bar{\alpha}^2 c^5}\left(\frac{G \bar{\alpha} m}{r}\right)^4\Bigg[ \frac{f'(\varphi_0)}{7 c^2 \sqrt{\bar{\alpha}} r^2}\bigg((S_+(1+\frac{6}{25}\eta)\\
&+ \frac{\Delta m}{m}S_-)(1350 v^4 + 5100\dot{r}^4 + \frac{G \bar{\alpha} m}{r}(-750 v^2 + 950 \dot{r}^2)\\
&-900v^2 \dot{r}^2) + \frac{G m}{r}(3 S_+ \frac{\Delta m}{m} S_-)(672 v^2 - 784 \dot{r}) \bigg)\Bigg].
\end{aligned}
\end{equation}
Here we completed the result for the GB terms of \cite{Shiralilou:2021mfl}.
Again specializing to circular orbits in terms of PN parameter $x$ \eqref{eq:xdef}, the total tensor flux is
\begin{equation}\label{FTx}
    \mathcal{F}_T (x) = x^5 c^5\left[T5  + x(T6 + T6_{\rm GB} x^2 c^4 + T6_{\rm tid} x^2 c^4)\right],
\end{equation}
with $T5$ and $T6$ arising from the scalar and PN corrections and given explicitly in \eqref{eq:S4S5T5T6}. The GB contributions are 
\begin{eqnarray}
    T6_{GB}& =&  \left(\frac{128 \alpha f'(\varphi_0) \eta^2 }{5 \bar{\alpha}^{9/2} G^3 m^2}\right) \left[\frac{\Delta m}{m}S_- \left(\frac{25}{14}+ \frac{4}{3 \bar{\alpha}}\right) + \right.\nonumber\\
    &&\left.S_+\left(\frac{25}{14}+ \frac{4}{ \bar{\alpha}} + \frac{3}{7}\eta\right)\right],
    \end{eqnarray}
and the tidal terms are
\begin{equation}\label{eq:T6tid}
T6_{\rm tid} = \frac{256 \eta^2 \zeta}{15 G^3 \bar{\alpha}^4 m^3 }.
\end{equation}

The total flux is obtained from the sum of the tensor and scalar energy flux ${\cal F}={\cal F}_S + {\cal F}_T$ .

\subsection{Phase Evolution in Fourier Domain}\label{phaseevolution}
 To further analyse the effects of the tidal contribution on the gravitational wave signal we consider the tidal contributions to the phase evolution during an inspiral. In the adiabatic limit $\dot{\omega}/\omega^2\ll1$ assuming circular orbits, the radius slowly shrinks as determined by energy balance
\begin{equation}
\dot{E}(\omega) = - {\cal F}(\omega).
\end{equation}
  With the relation $\dot{\phi}=\omega$ for the evolution of the orbital phase we obtain the set of differential equations
\begin{equation}\label{diffeqphase}
\frac{d \phi}{d t}-\omega=0, \quad \frac{d \omega}{d t}+\frac{\mathcal{F}(\omega)}{E^{\prime}(\omega)}=0.
\end{equation}
The dominant GW mode due to the quadrupole oscillates at twice the orbital frequency $f=\omega/\pi$. There are different ways to solve the system~\eqref{diffeqphase} and obtain the GW phasing. For data analysis, it is useful to have closed-form expressions for the GW phasing in the Fourier domain. Using the stationary phase approximation (SPA) \cite{Droz:1999qx},  the Fourier transform of the GW signal is given by
\begin{equation}\label{eq:diffeqpsi}
\tilde{h}^{\mathrm{SPA}}(f)={\cal A} \sqrt{\frac{2 \pi}{m \dot{\omega}}} e^{-i\left[\psi\left(f\right)+\pi / 4\right]}.
\end{equation}
Here, ${\cal A}$ is the amplitude of the mode and the Fourier phase given by $\psi  \equiv 2 \phi(t(f))-2 \pi f t(f)$. With \eqref{diffeqphase} we can combine this to the second order differential equation for the Fourier phase 
\begin{equation}\label{psiDE}
\frac{d^2\psi(\omega)}{d\omega^2} = -2\frac{E'(\omega)}{\mathcal{F}(\omega)}.
\end{equation}
Using the parameter $x$ from~\eqref{eq:xdef}, the Fourier phase is calculated from
\begin{equation}\label{eq:psiv}
\psi = -\int \Bigg(\int \frac{E'(x)}{\mathcal{F}(x)} dx\Bigg) \frac{3 \sqrt{x} c^3}{G \bar{\alpha} m} dx+ \phi_{c}-2 \pi f t_{c}.
\end{equation}
Here the integration constants are expressed in terms of $t_c$ and $\phi_c$ and correspond to the choice of reference point in the evolution.

\begin{figure*}
\centering
 \includegraphics[width=0.8\linewidth]{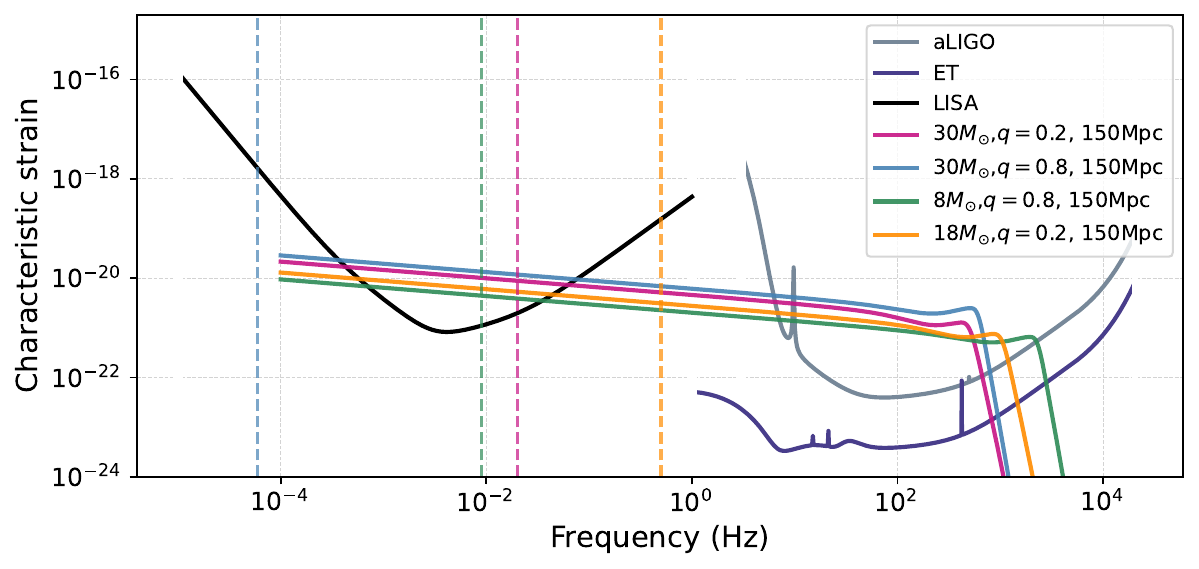}
\caption[]{The dimensionless characteristic strain $\sqrt{S_n\times f}$ of different GW detectors and the signal strain $h_c=f\times |\tilde{h}(f)|$ for different BH binary systems as a function of frequency $f$. The vertical lines mark the approximate transition frequency from DD to QD regime for each binary using~\eqref{eq:boundaryfreq} and assuming $\sqrt{\alpha}=1.7$~km.}
\label{fig:detect}
\end{figure*}

One can approach the expansion of the ratio in \eqref{eq:psiv} in different ways, here we use the so called Taylor~F2 approximant~\cite{Buonanno:2009zt}, where the ratio is perturbatively expanded using the explicit results for the fluxes~\eqref{ETdot} and~\eqref{Esdot} and the orbital energy~\eqref{BE}. Furthermore, we split the calculation of the Fourier phase in two domains: in the dipolar-driven (DD) regime where the scalar radiation dominates, and the quadrupolar-driven (QD) domain where the tensor radiation is dominant. The scalar dipolar dominated regime is relevant for
\begin{equation}\label{eq:boundaryfreq}
x^{\mathrm{DD}} \ll \frac{5 c^2 \mathcal{S}_{-}^2 \bar{\alpha}}{24} \; \text { or }\;  f^{\mathrm{DD}} \ll\left(\frac{5}{24}\right)^{3 / 2} \frac{c^3 \mathcal{S}_{-}^3 \sqrt{\bar{\alpha}}}{\pi G m},
\end{equation}
while at much higher frequencies the system is considered to be in the QD regime. Note that this condition is approximate and is based only on the leading-order contributions to the fluxes. We analyze the validity of this approximation in Sec.\ref{sec:results}.
In Fig.~\ref{fig:detect} we show the evolution of some example binary BH systems in the LISA~\cite{LISA:2017pwj}, advanced LIGO~\cite{LIGOScientific:2014pky} and ET~\cite{Hild:2010id} sensitivity bands and indicate the transition frequency for each case. The plot shows the characteristic strain $\sqrt{S_{n}}\times f$ and signal amplitude $h_c=f\times|\tilde{h}(f)|$ for four systems with average masses of $30M_{\odot}$ and $18M_{\odot}$, and a small total mass system with $8M_{\odot}$ all located at a distance of $d_{L}=150$~Mpc. Note that these amplitudes are found using the GR waveform template IMRPhenomXP~\citep{Pratten:2020ceb} which is sufficient for our purpose here. The GB and tidal corrections to the signal would result in accelerated mergers yet their contribution to amplitude changes is expected to be small. The vertical lines in Fig.~\ref{fig:detect} show the frequency based on \eqref{eq:boundaryfreq} for which the DD regime would shift to the QD regime for the different BH systems. In Sec.~\ref{sec:analysisfullphasing} we analyse these systems and their overall phasing behaviour in both the DD regime and the QD regime.

\subsubsection{Phase evolution in the Dipolar-Driven domain}\label{sec:psidd}
In the DD regime we factor out the leading order (-1)PN scalar contribution in the flux and we expand the ratio up to first order in $x$, which has the following structure
\begin{equation}\label{eq:EprimeFDDx}
    \frac{E'(x)}{\mathcal{F}^{\rm DD}(x)} = \frac{- 3 \bar{\alpha} G m}{8 \eta S_-^2 c^3 x^4}(1+(E'_0 - f_2^{\rm DD} )x),
\end{equation}
where the coefficients are given in~\eqref{eq:ratioDDcoeffs}. Performing the integration in \eqref{eq:psiv} we obtain for the Fourier phasing in the DD regime
\begin{widetext}
\begin{equation}\label{psiDD}
\begin{aligned}
\psi  &=\frac{1}{4 \eta \mathcal{S}_{-}^{2} x^{3/2}}\left\{1+\rho^{\rm D D} x+ c^4 x^3\left[(\rho_{\rm G B}^{\rm D D} +\rho_{\rm tid}^{\rm D D})\log \left(x\right) -\frac{1}{3}(\rho_{\rm G B}^{\rm D D} +\rho_{\rm tid}^{\rm D D})\right]\right\}+\phi_c-2 \pi f t_c\,,
\end{aligned}
\end{equation}
with
\begin{equation}\label{rhoDD}
\begin{aligned}
\rho^{\rm D D} &=-\frac{108}{5 \bar{\alpha} S_-^2}+\left(12-\frac{18 }{\bar{\gamma}}\right)\left(\beta _+ - \frac{\Delta m}{ m}\beta_-\right)+\frac{18}{\bar{\gamma}}\frac{ 
   S_+}{ S_-}\left(\frac{\Delta m}{m}\beta_+ - \beta_-\right)-3 \bar{\gamma}+\frac{21 \eta }{4}
   -\frac{18 S_+^2}{5 S_-^2}+\frac{117}{20}\\
\rho_{\rm G B}^{\rm D D} &= \frac{3 \alpha f^{\prime}\left(\varphi_{0}\right)}{G^2 m^{2} \bar{\alpha}^{5 / 2} S_-}\left[\frac{-16 S_- }{\bar{\alpha}}\left(\frac{\Delta m}{ m} S_- +3S_+\right)+2S_+\left(\frac{\Delta m }{m}S_-+S_+\right)\right],\\
\rho_{\rm tid}^{\rm D D} &=\frac{3 }{G^{2} \bar{\alpha}^2 m^3 \eta}\left[-\frac{\bar{\zeta}}{\bar{\alpha}^{3/2} G m S_- } - 8 \eta \zeta\right].
\end{aligned}
\end{equation}
\end{widetext}
Note that the GB and tidal contribution always come in the combination $(\rho_{\rm GB} + \rho_{\rm tid})$ hence have a degenerate scaling with the frequency. \\
From the prefactor of \eqref{rhoDD} we notice that in the limit of $S_-\xrightarrow{} 0$ and the limit $\alpha\xrightarrow{} 0$ corresponding to the equal mass system and GR limit respectively, the dipolar driven Fourier phase diverges. The inverse scaling with the mass ratio and coupling is coming from the inverse ratio of the change in binding energy and the total flux changing from the orbital phasing in the time domain to the Fourier domain as described by \eqref{diffeqphase} to \eqref{psiDE}. Therefore the interpretation of the scaling in the time domain is more intuitive; as the scalar flux vanishes for a vanishing coupling and the dipolar radiation scales linearly with $S_-$, the orbital phasing in the time domain vanishes in the GR limit and for equal mass systems as expected. In the Fourier domain this intuitive interpretation is absent and these limits only from \eqref{rhoDD} are not well defined. The total description does hold as the divergence coincides with a vanishing dipolar driven regime in both limits \eqref{eq:boundaryfreq}. We still analyse the parameter dependencies of the dipolar driven phasing in Sec. \ref{sec:results} outside these limits as the scalings in the Fourier domain are the relevant ones with respect to GW data analysis. 

\subsubsection{Phase evolution in the Quadrupolar-Driven domain}
For the QD phasing we repeat a similar analysis, but now we factor out the leading order tensor contribution in the total flux. We also split the QD flux in a part with and without dipolar terms. As the dipolar terms scale with $S_-$ we define the following decomposition
\begin{equation}\label{Fqdsplit}
\mathcal{F}^{\rm QD}=\mathcal{F}_{\text {non-dip }}+\mathcal{F}_{\text {dip }}, 
\end{equation}
with non-dipolar contribution defined by
\begin{equation}
\mathcal{F}_{\text {non-dip }} \equiv \lim _{\mathcal{S}_{-} \rightarrow 0} \mathcal{F}, 
\end{equation}
and the dipolar part being the remainder such that $\mathcal{F}_{\text {dip }} = \mathcal{F}-\mathcal{F}_{\text {non-dip }}.$
Explicitly we obtain when including up to 1PN fractional corrections
\begin{equation}\label{eq:FnondipFdip}
\begin{aligned}
&\mathcal{F}_{\text {non-dip }}=\frac{32 \eta^{2} \bar{\xi} c^{5}}{5 G \bar{\alpha}^{2}} x^5\left[1+f_{2}^{n d} x\right], \\
&\mathcal{F}_{\text {dip }}=\frac{4 \mathcal{S}_{-}^{2} \eta^{2} c^{5}}{3 G \bar{\alpha}} x^4\left[1+f_{2}^{d} x\right],
\end{aligned}
\end{equation}
with the explicit coefficients delegated to the Appendix, \eqref{eq:ratioQDcoeffs}.
With these definitions we can factor the non-dipolar contribution to the ratio in \eqref{eq:psiv} in the form
\begin{equation}
\label{eq:ratioQD}
\frac{E^{\prime}(x)}{\mathcal{F}(x)} \simeq \frac{E^{\prime}(x)}{\mathcal{F}_{\text {non-dip }}(x)}\left(1-\frac{\mathcal{F}_{\text {dip }}(x)}{\mathcal{F}_{\text {non-dip }}(x)}\right),
\end{equation}
which can be expressed as
\begin{equation}
\begin{aligned}
\frac{E^{\prime}(x)}{\mathcal{F}(x)} & \simeq -\frac{5 G m \bar{\alpha}^{2}}{64 c^{3} \eta \bar{\xi} x^5}\left[1+\left(E_{0}^{\prime}-f_{2}^{n d}\right) x\right] \\
&+\frac{25 G m \bar{\alpha}^{3} \mathcal{S}_{-}^{2}}{1536 c^{3} \bar{\xi}^{2} \eta x^5}\left[1+\left(E_{0}^{\prime}-2 f_{2}^{n d}+f^{d}\right) x\right],
\end{aligned}
\end{equation}
with 
\begin{equation}
    \bar{\xi}=\left(1+\mathcal{S}_{+}^2 \bar{\alpha} / 6\right).
\end{equation}
Integrating \eqref{eq:psiv} with~\eqref{eq:ratioQD} leads to 
\begin{equation}\label{psiQD}
\psi^{\rm QD}=\psi_{\text {non-dip }}+\psi_{\text {dip }}+\phi_c+2\pi f t_c,
\end{equation}
with \begin{subequations}
\label{rhoQD}
\begin{equation}
\begin{aligned}
&\psi_{\text {non-dip }}=\frac{6 \bar{\alpha}}{256\eta \bar{\xi} x^{5/2}}\left[1+ \rho^{n d} x+ (\rho_{\rm G B}^{n d}+\rho_{\rm tid}^{n d}) c^4 x^3\right], \\
&\psi_{\text {dip }}=-\frac{10 \mathcal{S}_{-}^{2} \bar{\alpha}^{2}}{3584\eta \bar{\xi}^{2} x^{7/2}}\left[1+ \rho^{d} x+ (\rho_{\rm G B}^{d}+ \rho_{\rm tid}^{d}) c^4 x^3\right].
\end{aligned}
\end{equation}
Here the non-dipolar coefficients are given by
\begin{eqnarray}
\rho^{n d}&=& \frac{6235}{756 \bar{\xi}} - \frac{10}{3} -(10-\frac{175}{\bar{\xi}})\frac{\eta}{27}+ \frac{80}{27}\left(\frac{1}{\bar{\xi}} -1\right)\bar{\gamma}  \nonumber\\
&&+ (\frac{80}{27} + \frac{160}{27\bar{\xi}})(\beta_+ - \frac{\Delta m}{m}\beta_-),\\
\rho_{\rm G B}^{n d}&=& \frac{f^{\prime}\left(\varphi_{0}\right) \alpha}{G^2m^{2} \bar{\alpha}^{5 / 2}}\left[-\frac{800 }{3\bar{\alpha}}(3 S_+ + \frac{\Delta m }{ m}S_-) \right.\nonumber\\
&&\left.+ \frac{S_+}{\bar{\xi}}(-\frac{1000}{7}-\frac{240}{7}\eta -\frac{320}{\bar{\alpha}})\right], \\
   \rho_{\rm tid}^{n d}&=& -\frac{\zeta}{G^{2} \bar{\alpha}^2 m^3 } \left[\frac{400}{3} + \frac{160}{3 \bar{\xi}}\right]\,,\qquad \qquad ~
\end{eqnarray}
and the dipolar parts are
\begin{eqnarray}
\rho^{d}&=&\frac{1247}{96 \bar{\xi}} - \frac{301}{40} + (\frac{245}{24 \bar{\xi}} - \frac{21}{8})\eta + (\frac{14}{3 \bar{\xi}}-\frac{7}{2})\bar{\gamma}\nonumber\\
&&+ \frac{7 S_+}{\bar{\gamma} S_-} (\beta_- - \frac{\Delta m}{m}\beta_+)\nonumber\\
&&+ \left(\frac{28}{3 \bar{\xi}}+\frac{7}{\bar{\gamma}}\right) (\beta_+ - \frac{\Delta m}{m}\beta_-), \\
\rho_{\rm G B}^{d}&=& \frac{f^{\prime}\left(\varphi_{0}\right) \alpha}{G^2m^{2} \bar{\alpha}^{5 / 2} S_-}\left[-35 S_+(S_+ + \frac{\Delta m}{m} S_-) \right.\nonumber\\
&&\left.- \frac{560 S_-}{3 \bar{\alpha}}(3 S_+ + \frac{\Delta m}{m} S_-) \right.\nonumber\\
&&\left.-\frac{S_- S_+}{\bar{\xi}}(250 + \frac{560}{\bar{\alpha}} +60\eta)\right],\\
   \rho_{\rm tid}^{d}&=& \frac{1}{G^{2} \bar{\alpha}^2 m^3 } \left[\frac{35\bar{\zeta}}{ 2 \bar{\alpha}^{3/2} G \mu S_-} - \frac{280 \zeta}{3}(1+\frac{1}{\bar{\xi}})\right].\qquad \qquad ~
   \end{eqnarray} 
   \end{subequations}
In the GR limit, obtained by setting $\bar \alpha = 1$ and all scalar parameters and the GB and tidal corrections to zero, the QD phase~\eqref{psiQD} agrees with known results for the 1PN TaylorF2 phasing, e.g.~\cite{Buonanno:2009zt}. This completes the expressions necessary to calculate the effects in the GW amplitude and phasing. However,
the results we have obtained are still parameterized in terms of the coupling coefficients in the skeletonized action such as the sensitivities $\alpha_0$ and tidal deformabilities $\lambda_s$. The next step is to perform a matching calculation that connects these parameters with the strong-field properties of the scalar condensate in the vicinity of the BH. We discuss this process in the next section, after briefly reviewing the case without tidal effects, which determines the mass parameters and sensitivities.
\section{Determining the skeleton parameters}
\label{sec:Lovenumber}
In this section, we discuss the information contained in the coupling coefficients that appear in the skeletonized action~\eqref{eq:Sm}, specifically the scalar charge $\alpha^0$ with the related parameter $\beta^0$ and the tidal deformability $\lambda_s$, which all depend on the fundamental properties of the BHs with scalar hair. Determining these coefficients requires calculations based on a fully relativistic approximation scheme. The starting points for this are the vacuum equations of motion~\eqref{FE1} and~\eqref{FE2}, omitting all the source terms that we used for the skeletonized description. We solve the equations of motion for equilibrium configurations to obtain the scalar charge and for their linearized perturbations to compute the Love number, then extract the asymptotic behavior at large distances from the center of the BH and match the coefficients in the skeletonized action. 
For these calculations, explicit expressions can only be obtained when specializing to the small-coupling approximation, which we will specialize to in this section. 

\subsection{Spacetime and scalar equation of motion to linear order in the coupling}
Solutions for BH spacetimes in sGB gravity have already been computed with both the analytical and numerical techniques~\cite{Guo:2008hf,Pani:2009wy,Yunes:2011we,Pani:2011gy,Sotiriou:2014pfa,Maselli:2015yva,Kleihaus:2015aje,Sullivan:2019vyi,Sullivan:2020zpf,Delgado:2020rev,Julie:2019sab}. Here, we are interested in analytical solutions, which rely on the assumption that the GB corrections are small; specifically that the dimensionless coupling defined in~\eqref{eq:alphahat} is small, $\hat \alpha \ll 1$.
From the equations of motion~\eqref{FE1} and~\eqref{FE2} we see that in the small-coupling limit, the leading order corrections to the BH metric appear at $\mathcal{O}(\hat{\alpha}^2)$, as the inhomogeneous solutions for $\varphi$ are $O(\alpha)$ and the source terms in the metric equations of motion are either quadratic in derivatives of $\varphi$ or involve an explicit factor of $\alpha$ together with a derivatives of $\varphi$. 
Therefore, the sGB static and spherically symmetric space-times reduce to the Schwarzschild solution at leading order in $\hat\alpha$~\cite{Yunes:2011we,Sotiriou:2014pfa,Julie:2019sab}. This sources the scalar field at $\mathcal{O}(\hat{\alpha})$ through the GB scalar $\mathcal{R}^2_{GB}=48M^2/\tilde r^6+\mathcal{O}(\hat{\alpha}^2)$, where $M$ and $\tilde r$ are the ADM mass and radial coordinate of the Schwarzschild spacetime. 
Corrections to the Schwarzschild metric enter only at subleading orders in $\hat\alpha$, which we neglect here. 
 In standard Schwarzschild coordinates $(t,\tilde r, \theta, \tilde \phi)$ centered on the BH, the line element is given by
\begin{equation}
d s_{S}^{2}=-(1-u) c^{2} d t^{2}+\frac{1}{1-u} d \tilde r^{2}+\tilde r^{2}d\Omega^2,
\end{equation}
with $d\Omega^2=\sin^2\theta \, d\theta\, d\tilde \phi$ the differential solid angle, $S$ shorthand for 'Schwarzschild', and $u$ the dimensionless inverse radial variable
\begin{equation}
\label{eq:defu}
u=\frac{r_{S}}{\tilde r},
\end{equation}
with $r_S$ defined in~\eqref{eq:rSdef}.

In this spacetime and to linear order in $\hat \alpha$ the scalar field equation of motion~\eqref{FE2} reduces to
\begin{equation}
\label{eq:eomphiSchw}
g^{\mu\nu}_{S} \nabla_\mu \nabla_\nu\varphi =-\frac{48 G^4 M^4 \hat{\alpha}}{c^8 \tilde r^6} f^{\prime}(\varphi).
\end{equation}

We look for solutions to~\eqref{eq:eomphiSchw} using an ansatz for the asymptotic expansion of the field for $\hat \alpha\to 0$ of the form 
\begin{equation}
\label{eq:phiexpand}
\varphi=\varphi^{(0)}+{\hat{\alpha}} \varphi^{(1)}+O\left({\hat{\alpha}}^{2}\right).
\end{equation}
Substituting~\eqref{eq:phiexpand} into \eqref{eq:eomphiSchw} and collecting orders of ${\hat{\alpha}}$ leads to the following set of equations
\begin{equation}
\label{eq:expandedeom}
\begin{aligned}
  \Box_{S} \varphi^{(0)}&=0,\\ \Box_{S}\varphi^{(1)}&=-\frac{48 G^{4} M^{4} }{c^{8} \tilde r^{6}} f^{\prime}\left(\varphi^{(0)}\right).
\end{aligned}
\end{equation}
We look for  static solutions $\varphi = \varphi(\tilde r,\theta,\tilde\phi)$ at each order. The angular functions that solve the homogeneous equations~\eqref{eq:expandedeom} are the spherical harmonics. This motivates the following separation of variables ansatz for the scalar field 
\begin{equation}\label{ansatz}
    \varphi =\sum_{\ell m} R^{\ell m}(\tilde r) Y_{\ell m}(\theta, \tilde \phi). 
\end{equation}

\subsubsection{Solution for the equilibrium scalar field configuration}
Substituting the ansatz~\eqref{ansatz} into~\eqref{eq:expandedeom}, using the property of the spherical harmonics $\tilde r^2\nabla^2 Y_{\ell m}=-\ell(\ell+1)Y_{\ell m}$
and converting from $\tilde r$ to $u$ using~\eqref{eq:defu} turns the equation at $\mathcal{O}({\hat{\alpha}}^0)$ into
\begin{equation}
\label{eq:radial00}
(1-u) R_{\ell m}^{(0) \prime \prime}(u)-R_{ \ell m}^{(0) \prime}(u)-\frac{\ell(\ell+1)}{u^{2}} R_{ \ell m}^{(0)}(u)=0.
\end{equation}
As we are interested in a background configuration that is spherically symmetric we focus on $\ell=0$, where the solution to~\eqref{eq:radial00} simplifies to
$R^{(0)}_{00}=c_{1}+c_{2} \log (1-u)$. 
Regularity at the horizon $u=1$ requires choosing $c_2=0$ so that only a constant remains, which we redefine such that the leading-order coefficient in~\eqref{ansatz} with $\ell=m=0$ is given by
\begin{equation}\label{sol00}
    \varphi^{(0)} = c_1 Y_{00} =\varphi_{\infty}.
\end{equation}
This makes explicit the dependence on the boundary conditions at infinity.

At order $\mathcal{O}({\hat{\alpha}}^1 )$ the radial equation with $\ell=0$ is given by
\begin{equation}
(1-u) R_{00}^{(1) \prime \prime}(u)-R_{00}^{(1) \prime}(u)=-6 \sqrt{\pi} u^{2} \hat{\alpha}f^{\prime}\left(\varphi_{\infty}\right),
\end{equation}
where we have used~\eqref{sol00} in the source term. 
Choosing the integration constant to enforce regularity at the horizon we obtain 
\begin{equation}
\label{sol10}
R^{(1)}_{00}=2\sqrt{\pi} f^{\prime}\left(\varphi_{\infty}\right)\left(u+\frac{1}{2} u^{2}+\frac{1}{3} u^{3}\right).
\end{equation}
The total background (B) equilibrium field configuration up to linear order in $\hat{\alpha}$ using~\eqref{eq:phiexpand} with~\eqref{sol00} and~\eqref{sol10} is given by
\begin{equation}
\label{eq:varphiBG}
\varphi_{\rm B} = \varphi_{\infty} + \hat{\alpha}f^{\prime}\left(\varphi_{\infty}\right)\left(u+\frac{1}{2} u^{2}+\frac{1}{3} u^{3}\right) +{\cal O}(\hat \alpha^2),
\end{equation}
which agrees with results in literature \cite{Julie:2019sab}.

\subsubsection{Characteristic scale for the scalar condensate}
\label{sec:commentonLphi}
The approximate results for the background field~\eqref{eq:varphiBG} enable more quantitative statements about the characteristic scale $L_\varphi$ introduced in Sec.~\ref{sec:tidalEFT}. As the falloff of the background scalar field configuration~\eqref{eq:varphiBG} is slow at large distances from the BH, a concern is that $L_\varphi$ could be very large such that the skeletonized description becomes inadequate early on in the inspiral. To quantify $L_\varphi$, we compute the ADM energy of the field given by \begin{equation}
E^\varphi=\int_{\cal S}\sqrt{\gamma}\,d^3x \, n^\alpha\, T_{\alpha\beta}\, t^\beta,
\end{equation}
where $t^\beta$ is the timelike Killing vector $\vec\partial/\partial t$ and ${\cal S}$ is a spatial hypersurface with induced metric $\gamma_{\alpha\beta}$ and a timelike unit normal $n^\alpha$. Using the energy-momentum tensor identified from the right hand side of the non-trace-reversed version of~\eqref{FE1} and the approximate field profile~\eqref{eq:varphiBG}, to second order in the coupling this becomes
$E^\varphi\sim 4\pi\, \hat\alpha^2 f^\prime(\varphi_{\infty})^2 \int dr (89r_S^5 + 7 r_S^4 r + 7 r_S^3 r^2 + 49 r_S^2 r^3 + r_S r^4 + r^5)/r^7$. 
Performing the integral we find that  $\gtrsim 95\%$ of the energy are contained within $\sim 5.5 r_S$. While there is no unique criterion for defining the bulk concentration of the field, we will 
consider here scales of 
\begin{equation}
L_\varphi\sim O(6\,r_S)
\end{equation}
as the characteristic size of the scalar condensate. The exact value of $L_\varphi$ is irrelevant for our purposes; the key point is that despite the slow falloff, the majority of scalar energy is nevertheless concentrated on relatively small scales $L_\varphi$, thus motivating a skeletonized approximation.  \\
Additionally we study the zero component of the energy momentum tensor related to the energy density $\rho$ of the scalar field. For an order of magnitude estimation we find the value of $\rho$ for a binary system consisting of $10 M_{\odot}$ and $20 M_{\odot}$, at a typical frequency of $100$Hz to be $\mathcal{O}(1) M_{\odot}/\textrm{pc}^3$. This is two to three orders of magnitude less than typical dark matter environments for which dynamical friction is relevant~\cite{Barausse:2014tra, Kavanagh:2020cfn}. Even though the dynamical friction effects contain more subtleties for massless scalar fields~\cite{Vicente:2022ivh}, this force in general scales linearly with the energy density hence this gives a rough estimation of the validity of neglecting dynamical friction effects. 

\subsubsection{Matching the sensitivity}
The sensitivity and its link to the characteristic parameters of GB theories and the BH was already discussed in detail in~\cite{Julie:2019sab,Julie:2022huo}. For completeness, we include here a brief overview, as we will extend this matching strategy to determine the tidal Love number coefficient in the skeletonized description. For a hairy BH, the scalar charge $D$ can be read off from the coefficient of the $1/\tilde r$ falloff in the asymptotic expansion of the scalar field near spatial infinity. Specifically, for $\tilde r\to \infty$ we have
\begin{subequations}
\label{eq:Schwarzschasymp}
\begin{equation}
\label{eq:Ddef}
\lim_{\tilde r \to \infty } \varphi_{\rm B}=\varphi_{\infty}+\frac{D}{\tilde r}+{\cal O}(\tilde r^{-2}).
\end{equation}
Comparing~\eqref{eq:Ddef} with the $O(\tilde r^{-1})$ coefficient in the asymptotic expansion of~\eqref{eq:varphiBG} near spatial infinity we read off that 
\begin{equation}
D=\hat \alpha \, r_S \, f^\prime(\varphi_{\infty}).
\end{equation}
Likewise, the ADM mass of the BH, which in the coordinates we are using coincides with the parameter $M$ in the metric potentials, can be read off from the asymptotic behavior at large distances, for example
\begin{equation}
\label{eq:Mfromasym}
\lim_{\tilde r \to \infty } g_{\tilde r\tilde r}^{\rm BG}=1+\frac{2GM}{\tilde rc^2}+{\cal O}(\tilde r^{-2}).
\end{equation}
\end{subequations}

The next step is to match these asymptotic results from the fully strong-field description above to the coefficients $m^0$ and $\alpha^0$ in the skeletonized action for a single body. This action is expressed in terms of the harmonic radial coordinate $r$, which differs from the Schwarzschild coordinate by the transformation $\tilde r=r+r_S/2$. Asymptotically near spatial infinity these coordinates coincide up to corrections of order $r^{-1}$, which do not contribute to the leading-order falloffs we are interested in. From the skeletonized description~\eqref{eq:UphiLO} we have for a single body
\begin{subequations}
\label{eq:harmonicasymp}
\begin{eqnarray}
g_{rr}&\sim &  1+\frac{2 G m_A^0}{r} +O(r^{-2})\\
\varphi &\sim & \varphi_{0}-\frac{Gm_A^0\alpha_A^0}{rc^2}+O(r^{-2}). 
\end{eqnarray}
\end{subequations}
Making the identification between~\eqref{eq:harmonicasymp} and~\eqref{eq:Schwarzschasymp} shows that 
\begin{equation}
m_A^0=M_A, 
\end{equation}
that the sensitivity is related to the scalar charge by 
\begin{equation}\label{eq:scalarcharge}
\alpha_A^0=-2 \hat \alpha  \, f^\prime(\varphi_{\infty})=-\frac{\alpha f^\prime(\varphi_{\infty}) c^4}{2 G^2 M_A^2},
\end{equation}
and that the asymptotic values of the field at spatial infinity are identified $\varphi_0=\varphi_\infty$.
The additional parameter $\beta^0_A$ can then be determined from~\eqref{eq:alpha0def}.

\subsection{Scalar tidal Love numbers}\label{sec:calclovenumer}
We next perform a similar analysis as above to derive the scalar tidal Love number in the small coupling approximation. In a patch of spacetime encompassing the BH and its vicinity but excluding the companion, the configuration is nearly that of an isolated BH and scalar condensate, with the effect of the distant companion being a small source-free tidal perturbation that causes the configuration to change away from its equilibrium state in isolation. Solving the equations of motion for the spacetime and scalar field for linearized, static perturbations to the isolated configuration and extracting the asymptotic behavior at large distances from the BH leads to the identification of the tidal Love numbers. We will focus here on the dipolar scalar tidal Love number, which can be computed by considering a scalar perturbation on a Schwarzschild background metric rather than having to solve the coupled system of metric and scalar perturbations.

\subsubsection{Definition of the scalar tidal Love number}
When considering linearized static perturbations of dipolar $\ell=1$ order of the scalar condensate away from its background configuration $\varphi_{\rm B}$ the asymptotic expansion of the scalar field at large distances $\tilde r$ can be written in the form 
\begin{equation}\label{Q1E1}
\begin{aligned}
\lim_{\tilde r\to\infty}(\varphi-\varphi_{\rm B}) 
& \sim  \sum_{m=-1}^{1} Y_{1 m}\left[\frac{G {Q}_{1 m}}{c^{2} \tilde r^{2}}+O\left(\frac{1}{\tilde r^{3}}\right)\right.\\
&\left. \quad -\tilde r {\mathcal{E}}_{1 m}+O\left(\tilde r^{2}\right)\right]+\ldots
\end{aligned}
\end{equation}
Here, the coefficients of $\tilde r^{-2}$ and $\tilde r$ correspond to the tidally induced dipole moment $Q_{1m}$ and the external tidal field ${\cal E}_{1m}$. The tidal deformability is defined to be the ratio 
\begin{equation}\label{QlE}
\lambda_s=-\frac{{Q}_{1m}^{s}}{ \mathcal{\tilde{E}}_{1m}^{s}} ,
\end{equation}
which holds for each $m$ for $\ell=1$.  
To compute the tidal deformability parameter~\eqref{QlE} requires solving the vacuum equations of motion for linearized perturbations to the background equilibrium configuration, and extracting the multipole moments from the asymptotic falloff behavior of the scalar field using~\eqref{Q1E1}.

\subsubsection{Tidal expansions of the small-coupling approximation}
We start by expanding each coefficient in~\eqref{eq:phiexpand} to linear order in the tidal perturbations. We track the tidal perturbations with a bookkeeping parameter $\varepsilon_{t}$\footnote{Note this parameter is not necessarily equal to $\varepsilon_{\rm tid}$ defined in Sec.\ref{scales} as now we are in the proximity of one BH experiencing small tidal perturbations from it's companion and not on the full orbital scale.} that we set to $1$ at the end of the calculation. This leads to the ansatz
\begin{equation}
\label{eq:expansionansatz}
\begin{aligned}
&\varphi^{(0)}=\varphi^{(0,0)}+ \varepsilon_{t} \varphi^{(0,1)}+\mathcal{O}(\varepsilon_{t}^2),\\
&\varphi^{(1)}= \varphi^{(1,0)}+\varepsilon_{t} \varphi^{(1,1)}+\mathcal{O}(\varepsilon_{t}^2),
\end{aligned}
\end{equation}
where the coefficients $\varphi^{(0,0)}$ and $\varphi^{(1,0)}$ given by~\eqref{sol00} and~\eqref{sol10} respectively. 
Substituting~\eqref{eq:expansionansatz} into ~\eqref{eq:eomphiSchw} and collecting orders of ${\hat{\alpha}}$ and $ \varepsilon_t $ leads to the following set of equations for the tidal perturbations
\begin{eqnarray}
\label{eq:expandedeom1}
  \Box_{S} \varphi^{(0,1)}&=&0, \\ 
 \Box_{S}\varphi^{(1,1)}&=&-\frac{48 G^{4} m^{4} }{c^{8} r^{6}} f^{\prime \prime}\left(\varphi^{(0,0)}\right) \varphi^{(0,1)}.
\end{eqnarray}

\subsubsection{Solutions at linear order in static tidal perturbations}
Substituting the separation of variables ansatz~\eqref{ansatz} into the equations of motion turns the $\mathcal{O}({\hat{\alpha}}^0 \varepsilon^1_{t})$ equation into
\begin{equation}
    (1-u)R_{lm}^{{(1,0)}''}(u) - R_{lm}^{{(1,0)}'}(u) - \frac{l(l+1)}{u^2}R_{lm}^{(1,0)}(u) =0.
\end{equation}
As we are interested in the solutions for dipolar perturbations we set $\ell=1$, which leads to the solution
\begin{equation}\label{R01}
R_{1 m}^{(0,1)}=\frac{(u-2)c_1 + 4c_2}{u} - \frac{(u-2 )c_2 \log(1-u)}{u}.
\end{equation}
Enforcing regularity at the horizon and renaming $c_1 = C$ gives for the full expansion coefficient
\begin{equation}\label{phi10}
    \varphi^{(1,0)} = \sum_{m=-1}^{1} R_{1 m}^{(0,1)} Y_{1m} = \sum_{m=-1}^{1} C\left(1-\frac{2}{u}\right)Y_{1m}.
\end{equation}
This solution has a growing behavior at infinity and thus lacks information about the tidal response. Therefore, we continue to work out results at higher orders in the $\hat\alpha$ expansion. 
At $\mathcal{O}({\hat{\alpha}}^1 \varepsilon^1_t)$ we have for the dipolar sector 
\begin{equation}
\begin{aligned}
&(1-u) R_{ 1 m}^{(1,1) \prime \prime}(u)-R_{ 1 m}^{(1,1) \prime}(u)-\frac{2}{u^{2}} R_{ 1 m}^{(1,1)}\\
&=-3 u^{2} f^{\prime \prime}\left(\varphi_{\infty}\right) R_{1 m}^{(0,1)}(u).
\end{aligned}
\end{equation}
Substituting the $\mathcal{O}({\hat{\alpha}}^0 \varepsilon^1_{t})$ solution and enforcing regularity at the horizon leads to
\begin{equation}
\label{eq:R11}
\begin{aligned}
R_{1 m}^{(1,1)}&=\left[c_{1}-14 C f^{\prime \prime}\left(\varphi_{\infty}\right)\right]\left(1-\frac{2}{u}\right)\\
&+\frac{1}{3} C f^{\prime \prime}\left(\varphi_{\infty}\right)\left(u^{3}-\frac{7}{2} u^{2}\right).
\end{aligned}
\end{equation}
Here the first term has the same structure as the $\mathcal{O}({\hat{\alpha}}^0\varepsilon_{t}^1)$ solution corresponding to an external field. We are interested here in the response to a generic external tidal perturbation, for which we do not assume any scaling with ${\hat{\alpha}}$ in this section. We thus focus on the solution~\eqref{phi10} for the external field and interpret the first term in~\eqref{eq:R11} as its $O(\hat \alpha)$ correction, which we consider to be subdominant. This leads to the solution for the scalar field expansion coefficient 
\begin{equation}
    \varphi^{(1,1)} = \sum_{m=-1}^{1} Y_{1m} \frac{1}{3} C  f^{\prime \prime}\left(\varphi_{\infty}\right)\left(u^{3}-\frac{7}{2} u^{2}\right).
\end{equation}
Thus, assembling the above results we obtain for the total perturbed scalar field to linear order in $\hat \alpha$ and the tidal perturbations and keeping only the terms of interest  
\begin{equation}\label{phitid}
\begin{aligned}
(\varphi-\varphi_{\rm BG})&=\sum_{m}Y_{1 m}C \left\{\left(1-\frac{2 }{u}\right)\right.\\
&\bigg. +\frac{1}{6} \hat{\alpha} f^{\prime \prime}\left(\varphi_{\infty}\right)u^2(2 u-7) \bigg\}.
\end{aligned}
\end{equation}

\subsubsection{Extracting the Love number}
\label{sec:Loveresult}
The asymptotic expansion of \eqref{phitid} is given by
\begin{equation}
\label{eq:asymptphitid}
\begin{aligned}
\lim _{\tilde r \rightarrow \infty} (\varphi-\varphi_{\rm B})&=\sum_{m=-1}^{1} Y_{1 m} C \left[-\frac{7 \hat{\alpha} r_S^2 f^{\prime \prime}\left(\varphi_{\infty}\right)}{6 \, \tilde r^{2}}\right.\\
&\left.+O(\tilde r^{-3})+\frac{c^{2}}{G M} \tilde r + O(\tilde r^{2})\right] .
\end{aligned}
\end{equation}
 Comparing~\eqref{eq:asymptphitid} with~\eqref{Q1E1} we can read off the induced dipole and tidal field as the coefficients of the $\tilde r^{-2}$ and $\tilde r$ terms
\begin{equation}
\tilde{Q}^{(s)}_{1 m}=-C\frac{14 \hat{\alpha} G^2 M^{2} f^{\prime \prime}\left(\varphi_{\infty}\right)}{3 c^{2}}, \quad \tilde{E}^{(s)}_{1 m}=\frac{c^{2}}{G M} C.
\end{equation}
This determines the tidal deformability~\eqref{QlE} to be
\begin{equation}\label{eq:lambda}
\lambda_{\ell=1}^{(s)}=\frac{7}{6} M \alpha f^{\prime \prime}\left(\varphi_{\infty}\right).
\end{equation}
Note that for coupling functions whose second derivative vanishes,  the scalar Love number and thus the dipolar tidal effects vanish.

\subsubsection{Matching to the skeletonized action}
The tidal deformability parameter appearing as a coupling coefficient in the effective action can be matched to the result~\eqref{eq:lambda} by considering the equations of motion~\eqref{eq:lambdasdef}. As discussed in more detail in the Appendix \ref{sec:apptidalEFT}, without loss of generality we can write~\eqref{eq:lambdasdef} in terms of the spatial pieces $Q_i=-\lambda_s {\cal E}_i$. Multiplying both sides by a general unit vector $n^i$ and using the identity~\cite{Thorne:1980ru}
\begin{equation}
Q_{i} n^{i}=\sum_{m=-1}^{1} {Q}_{1 m} Y_{1 m},
\end{equation}
and similarly for ${\cal E}_i$ we see that the parameter $\lambda_s$ in the skeletonized action has a one-to-one relation with the result~\eqref{eq:lambda}.

From the above calculations we now have an explicit connection between parameters in the action $m_A^0$, $\alpha^0_A$, $\beta_A^0$ and $\lambda_s^A$ and the characteristic properties of a full description of the BH and scalar condensate, i.e. the BH's ADM mass $M_A$, the scalar charge $D_A$, the GB coupling constant $\alpha$ and coupling function $f(\varphi)$ and the boundary values of the field at spatial infinity $\varphi_{\infty}$ in the small-coupling approximation. We next use these results to quantify the impact of tidal effects on the energy and GWs from binary inspirals.

\section{Results and case studies}\label{sec:results}
In this section, we present quantitative case studies on the impact of tidal effects on various gauge-invariant quantities. Ultimately, we are interested in assessing the signatures in the GWs, however, this involves various different contributions. To gain deeper insights, it is useful to separately consider effects in the binding energy and the fluxes, which we address below, before going on to the analysis of the GW phasing effects. For the case studies in this section we use the small-coupling results for the scalar charge~\eqref{eq:scalarcharge} and the tidal deformability~\eqref{eq:lambda} and consider an exponential coupling function $f(\varphi) = e^{2\varphi}/4$ as for a dilatonic field, also referred to as EdGB gravity.  

\subsection{Analysis of the tidal effects in the binding energy}
\begin{figure}
\centering
\begin{subfigure}{0.49\textwidth}
   \includegraphics[width=1\linewidth]{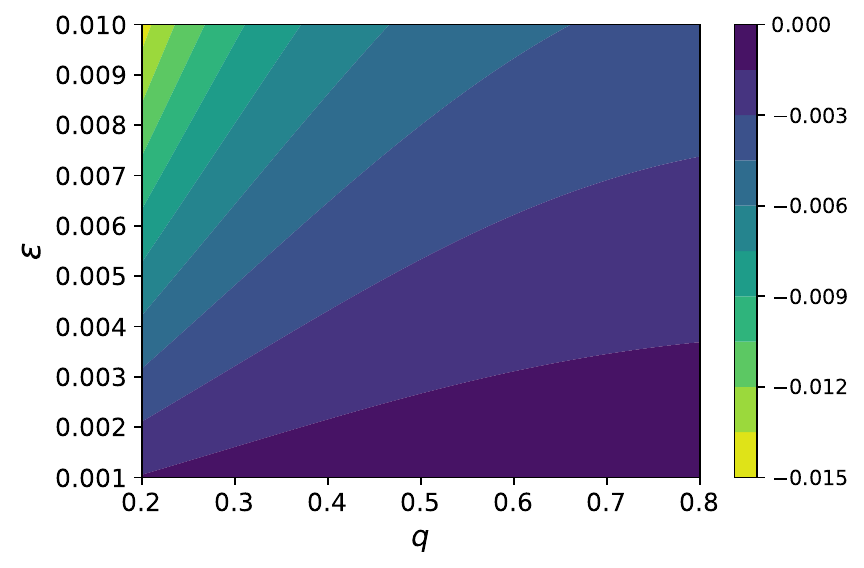}
       \centering 
\end{subfigure}

\caption[]{Contours of the ratio $E_{\rm tid}/E_{\rm GB}$ with respect to the mass ratio $q$ and the dimensionless coupling $\epsilon$ for a system of total mass $m = 15M_\odot$. The threshold for $\epsilon$ corresponds to $\sqrt{\alpha}=2.2$ km.}
\label{fig:contourBE}
\end{figure}

We start by analyzing the result for the binding energy given in~\eqref{BE}. As the dependence on the GB coupling constant $\alpha$ and the masses is not manifestly explicit due to the definitions of several composite parameters, we show in Fig.~\ref{fig:contourBE} the regimes of parameter space where tidal effects are most relevant. In particular, the plot shows the ratio between the contribution to the binding energy coming from tidal effects and the higher-curvature GB terms ($E_{\rm tid}/E_{\rm GB}$) from ~\eqref{eq:GBbinding} and ~\eqref{eq:tidalbinding}. This is given at a fixed total mass of $m=15M_{\odot}$, as a function of the mass ratio $q$ and the dimensionless coupling parameter $\epsilon$ defined in~\eqref{eq:epsilon}.
We analyse the parameter space for small values of the coupling. There is a theoretical bound on the coupling of $\epsilon=0.69$ from requiring regularity of the scalar field at the black hole horizon~\cite{Kanti:1995vq}, however from the current observational constraints the possible coupling parameter space is already reduced to around $\sqrt{\alpha}=1.7$km \cite{Perkins:2021mhb} which for this system corresponds to $\epsilon=0.006$. Latest analyses even constrain this value further down~\cite{Lyu:2022gdr, Wang:2021yll}. Therefore we limit our analysis to the small coupling regime.  The figure shows an overall negative ratio between the contributions, as the GB term contributes negative to the energy while the tidal contributions come in with an overall plus sign. This indicates 
decreased deviations from the GR terms as compared to when only GB terms were computed. Given that the ratio is always smaller than unity we conclude that the tidal contributions to the dynamics are subdominant as compared to the GB contributions. The relative importance of the tidal contributions to the binding energy is largest for large coupling and small mass ratios.   

\subsection{Tidal effects in the fluxes}
The tidal contributions to the fluxes differ qualitatively from those to the dynamics due to the fact that the tidally-induced dipole moment contributes together with the dipole moment associated with the orbital motion of the scalarized BHs. To gain insights into the consequences of this effect, we separate our analysis between the scalar and tensor fluxes. 
    \begin{figure*}
        \centering
        \begin{subfigure}[b]{0.47\textwidth}
            \centering
            \includegraphics[width=1\textwidth]{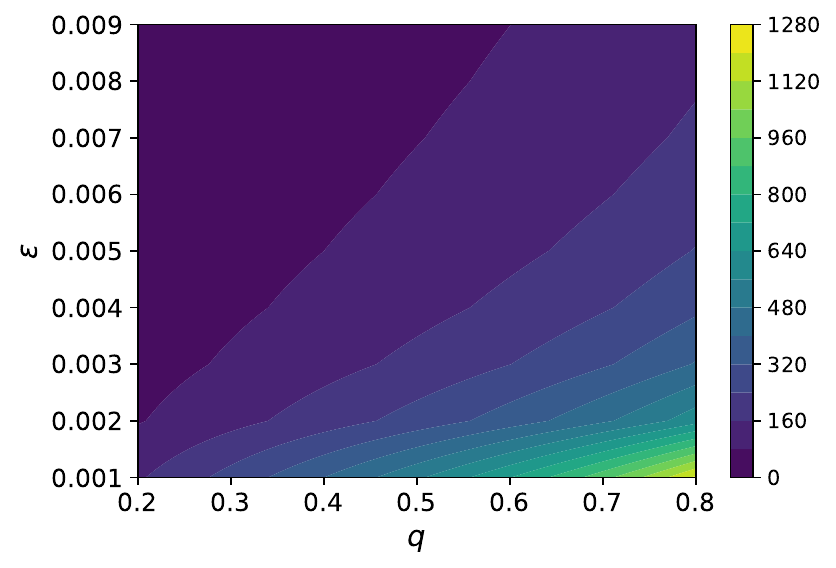}
        \end{subfigure}
        \begin{subfigure}[b]{0.49\textwidth}  
            \centering 
            \includegraphics[width=1\textwidth]{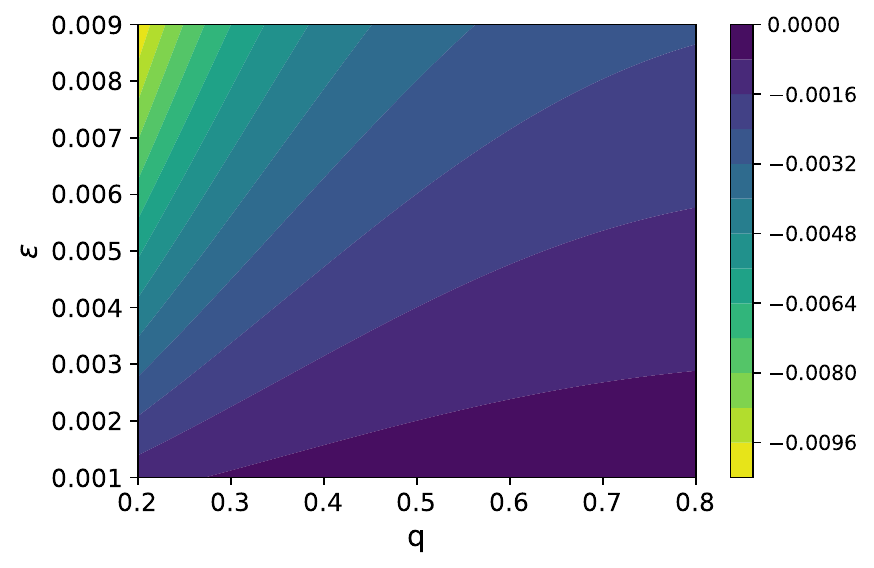}
        \end{subfigure}
        \caption{Contours of the ratio of the tidal and GB contributions to the fluxes for a system of total mass $m= 15M_\odot$ in the parameter space of the the dimensionless coupling $\epsilon$ and the mass ratio $q$. \emph{Left panel}: ratio of the contributions to the scalar flux $\cal{F}_S^{\rm tid}/\cal{F}_{S}^{\rm GB}$ and \emph{right panel}: tensor flux $\cal{F}_T^{\rm tid}/\cal{F}_{T}^{\rm GB}$. The upper end of the range in $\epsilon$ corresponds to $\sqrt{\alpha}=2.2$ km.   
        } 
       \label{fig:contourfluxes}
\end{figure*}

The left panel of Fig.~\ref{fig:contourfluxes} shows the ratio between the tidal terms of the scalar flux \eqref{Esdottid} and the GB terms of the scalar flux  \eqref{EsdotGB} with respect to the dimensionless coupling~\eqref{eq:epsilon} and the mass ratio. The right panel of Fig.~\ref{fig:contourfluxes} shows the ratio between the tidal \eqref{ETdottid} and GB terms~\eqref{ETdotGB} of the tensor-flux, where, similar to the binding energy, the tidal effects enter only through their contribution to the dynamics.\\

Comparing the values of the contours of both plots in Fig.~\ref{fig:contourfluxes} we see that the sign is positive for the scalar flux and negative for the tensor flux. The tidal terms in the scalar flux consist of the contribution related to the induced dipole moment and a contribution related to the dynamics, for which the former dominates in the here studied parameter space. Thus as for the scalar flux the tidal and GB contributions have the same sign they both enhance the difference with respect to GR while for the tensor flux the tidal term, consisting only of the contribution from the dynamics \eqref{eq:T6tid}, have the opposite sign to the GB contributions and therefore partly cancel the additional flux because of the higher curvature terms. \\
Furthermore, for the scalar flux, the ratio is much larger than unity, showing that tidal contributions to the scalar radiation dominate over the GB terms. By contrast to the trends in the binding energy shown in Fig.~\ref{fig:contourBE} and the tensor flux shown in the right panel of Fig.~\ref{fig:contourfluxes}, the relative size of the tidal contributions over the GB terms in the scalar flux becomes \emph{larger} for \emph{smaller} couplings, thus scaling in the opposite way. 
To analyse the origin of this behaviour we performed a small-coupling expansion of the tidal and GB contributions which also take into account the implicit dependencies on the coupling related to the scalar charge~\eqref{eq:scalarcharge}. We find that the contribution from the induced dipole terms to the scalar radiation is proportional to $\alpha^3$, while its contribution to the dynamics scales as $\alpha^5$. For comparison, the dominant scaling of the GB terms is $\sim \alpha^{4}$ in this limit. For small couplings, the ratio of tidal to GB contributions to the scalar flux thus scales as $1/\alpha$, while it is $\sim \alpha$ for the dynamics and tensor flux. This directly translates to the scaling with $\epsilon$ via \eqref{eq:epsilon} and corroborates the trends seen in the plots.

\subsection{Analysis of the Fourier phase evolution}\label{analysis}

From the expressions of the Fourier phase in both the dipolar and quadrupolar driven domains (\eqref{psiDD} and \eqref{psiQD}) we can see that the GB and tidal terms enter at the same PN order. Their similar scaling with the frequency is clear from the expressions however their dependence on the total mass of the BHs, mass ratio and the coupling is less straightforward. When analyzing these dependencies one compares different BH systems, and consequently comparisons at fixed dimensionful coupling parameter $\alpha$ differ from those at fixed dimensionless parameter $\epsilon$ defined in~\eqref{eq:epsilon}, which rescales the coupling by the total mass. For the analysis in this section we focus on comparisons based on the dimensionless coupling $\epsilon$ to better study the degeneracies with the mass terms.

\subsubsection{Analysis of the tidal and Gauss-Bonnet contributions}
    \begin{figure*}
        \centering
        \begin{subfigure}[b]{0.475\textwidth}
            \centering
            \includegraphics[width=0.9\textwidth]{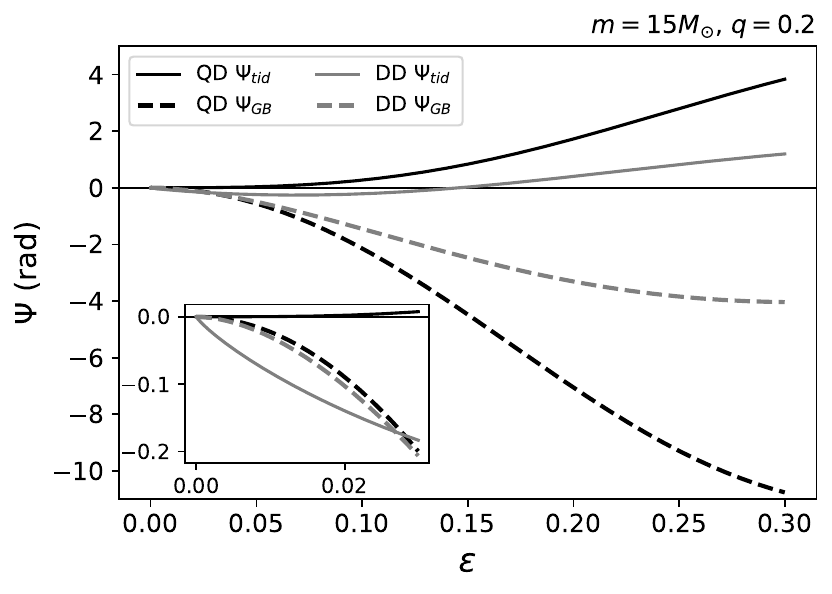}  
            \label{fig:alphadep1}
        \end{subfigure}
        \begin{subfigure}[b]{0.475\textwidth}  
            \centering 
            \includegraphics[width=0.9\textwidth]{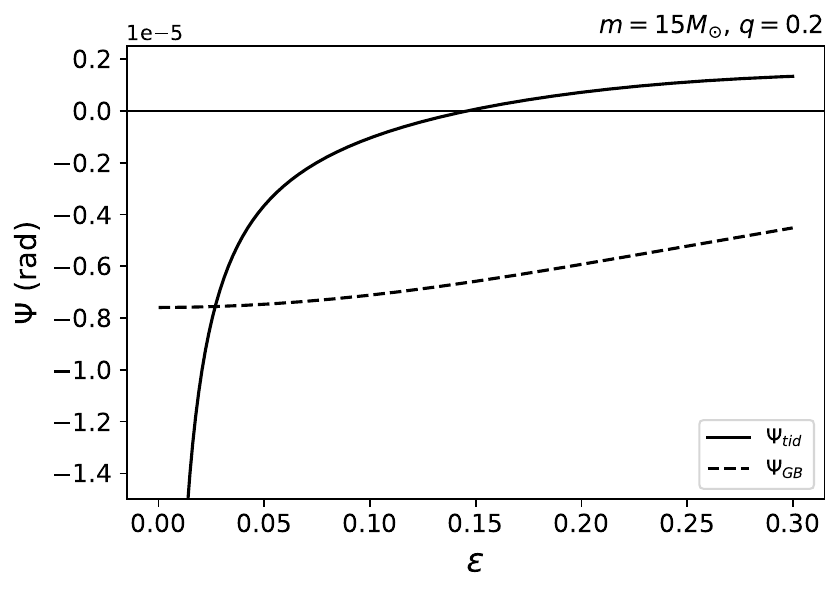}
        \end{subfigure}
        \vskip\baselineskip
        \begin{subfigure}[b]{0.475\textwidth}   
            \centering 
            \includegraphics[width=0.9\textwidth]{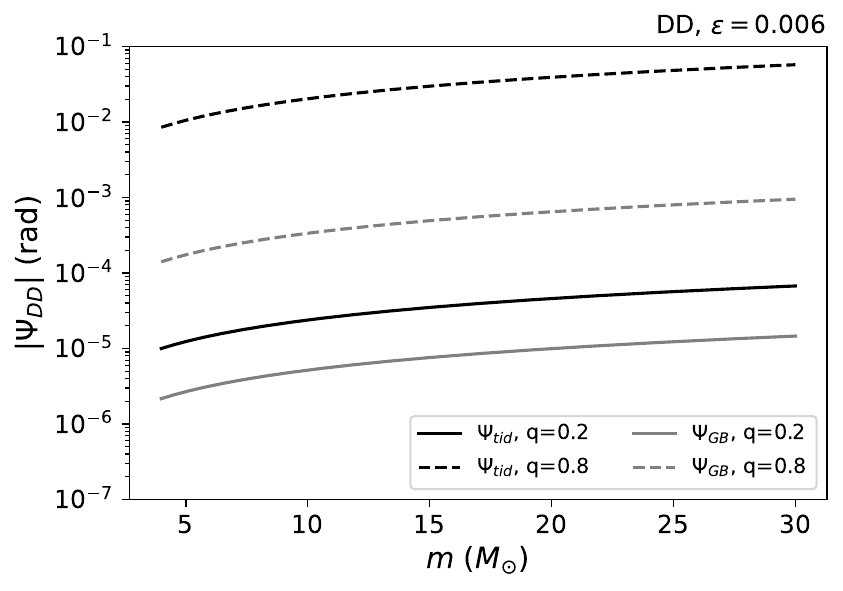}   
        \end{subfigure}
        \begin{subfigure}[b]{0.475\textwidth}   
            \centering 
            \includegraphics[width=0.9\textwidth]{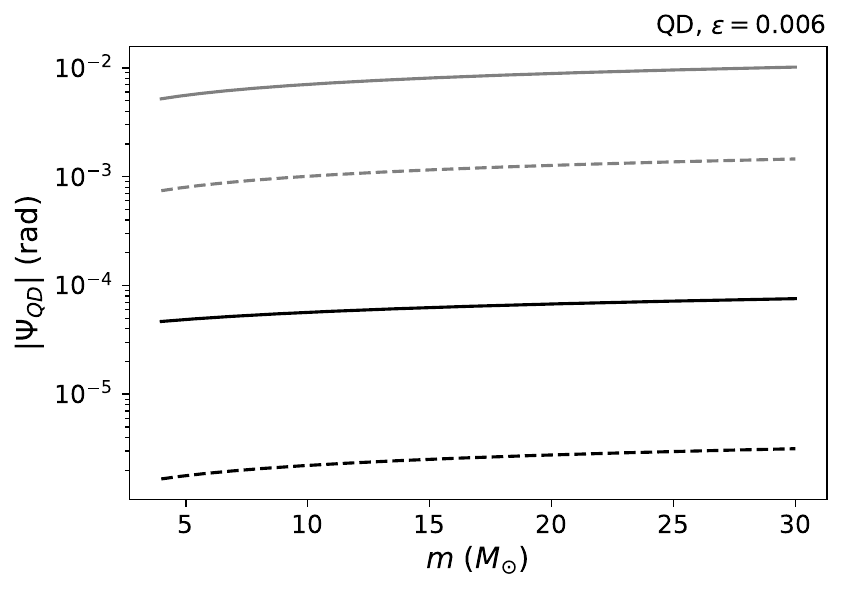} 
            \label{fig:mdep2}
        \end{subfigure}
        \caption{\emph{Top panels}: Dependence of the tidal (solid lines) and GB (dashed lines) phase contributions on the coupling parameter for a system of total mass $15 M_\odot$ and mass ratio $q=0.2$. Top left panel: Phase contributions in the DD regime at the transition~\eqref{eq:boundaryfreq} and  $586$Hz for the QD regime. Top right panel: DD phasing at fixed frequency of $10^{-4}$Hz. \emph{Bottom panels}: Absolute value of  the dependence of the tidal and GB phase contributions on the total mass for a fixed coupling of $\varepsilon=0.006$  for mass ratio $q=0.2$ (solid lines) and $q=0.8$ (dashed lines). Bottom left panel: DD phasing at fixed frequency of $10^{-4}$Hz, bottom right: QD regime at frequency $586$ Hz.} 
        \label{fig:tidGBcontr}
\end{figure*}
First we analyse the contribution of the tidal and GB terms in the phasing expressions of both the DD \eqref{psiDD} and QD \eqref{psiQD} regimes, at different values of $\epsilon$ in the top left plot of Fig.~\ref{fig:tidGBcontr}.

The tidal and higher curvature contributions to the phase is given for a system of $m=15M_{\odot}$ and $q=0.2$ at the boundary frequency \eqref{eq:boundaryfreq} for the DD regime and the ISCO frequency estimate $f=2(6^{2/3} c^3/G m\pi)\sim585$Hz for the QD regime. All contributions vanish in the zero coupling limit, in which GR is recovered. Note that the tidal contributions can change sign depending on the choice for in this case the coupling constant. 

Generally we find the GB contributions to be larger in magnitude than the tidal contributions, except in the DD regime for couplings below
 $\epsilon = 0.03$ shown in the inset, which is the relevant allowed regime of GB coupling strengths based on current bounds, when the tidal contribution becomes dominant over the GB contribution. To gain further insights into the behavior in the DD regime, 
the top right panel of Fig.~\ref{fig:tidGBcontr} shows the tidal contribution $\psi_{\rm tid}$ and the GB contribution $\psi_{\rm GB}$ to the DD phase evolution \eqref{psiDD} for a frequency of $f=10^{-4}$Hz, indicating that for small couplings the tidal phase contributions also have a strong inverse scaling with the coupling. This is also seen from a small-coupling expansion of the terms in~\eqref{psiDD}, which shows that the radiative tidal contributions scale as $\sim -\alpha^{-3}$, tidal contributions  to the dynamics as $\sim -\alpha^{-2}$, and the GB contributions as $\sim -\alpha^{0} + \alpha^2$. 
Note that the apparent divergence of the tidal contributions to the DD phasing in the zero coupling limit is counteracted by a vanishing extent of the DD regime, hence the phase contribution at the boundary frequency goes to zero in this limit, as expected. We also note that while the GB contributions scale as even powers of the coupling parameter and are thus unaffected by its sign, the tidal terms change sign for negative coupling. \\

The dependency of the tidal and GB contributions with respect to the total mass and for different mass ratios is depicted in the bottom plots of Fig.~\ref{fig:tidGBcontr}. Here we have fixed $\epsilon=0.006$ (which corresponds to $\sqrt{\alpha}=1.7$~km for a $m=$15$M_{\odot}$ system). When fixing the value of $\epsilon$ we find that the magnitude of the tidal and GB contributions to the phasing in the DD regimes both increase with increasing mass ratio as a consequence of the phasing analysis in the Fourier domain, as mentioned in Sec. \ref{sec:psidd}. This trend is the opposite in the QD regime, where both effects decrease with the mass ratio. Irrespective of the regime, the magnitude of both effects increases with increasing total mass. For the higher-curvature GB corrections, this may seem counter intuitive, as curvatures are higher for BHs with smaller mass. The opposite trends seen in Fig.~\ref{fig:tidGBcontr} are a consequence of fixing the dimensionless coupling parameter. At fixed $\epsilon$, a linear increase in the total mass induces a quadratic increase in the dimensionful coupling $\alpha$, which enhances the GB contributions. In Appendix~\ref{othercouplingplots} we show the dependencies of the QD phasing on the total mass for fixed $\alpha$, where the trends are consistent with expectations. 
From the bottom panels in Fig.~\ref{fig:tidGBcontr}, we also see that, regardless of the mass range and the mass ratios, the tidal contributions dominates over the GB contributions in the DD regime corresponding to the results from the top left plot of Fig. \ref{fig:tidGBcontr} for a small coupling of $\epsilon=0.006$. However this is reversed in the QD regime (black lines versus gray lines).\\

Lastly we highlight that whilst showing the absolute value of the phase contributions on a logarithmic scale, the contributions in the DD regime are all negative while in the QD regime the tidal contributions are positive and the GB contributions negative for this choice of the coupling constant. This corresponds to the curves shown in the inset of the left top panel in Fig.~\ref{fig:tidGBcontr} and the result of finding the same sign contribution for the scalar flux ratio and the opposite sign for the tensor flux ratio in Fig.~\ref{fig:contourfluxes}. 

The different trends and parameter dependencies in the DD versus QD regimes has important observational implications for GW tests of GR. As dipolar tidal effects significantly dominate over higher-curvature corrections in the DD regime, they must be included in waveforms for such systems to obtain more realistic constraints and avoid biases. The different parameter dependencies of the GWs in the DD versus QD regimes also open opportunities for breaking degeneracies and more accurately measuring the GB parameters if GWs from both regimes can be measured. For instance, as seen from~\eqref{eq:lambda} and the expressions in Sec.~\ref{sec:Binarydyn} and~\ref{sec:WaveformPhasing}, the GB effects depend on the first derivative of the coupling function, while the tidal terms involve the second derivative. A consequence is that tidal effects may be absent for some classes of coupling functions, which would also provide useful information about the theory. Additionally note that while our explicit results for the tidal GW signatures are specialized to GB theories, scalar tidal effects are likely to be an important phenomenon in broader classes of theories involving BHs with scalar hair, which motivates further studies of these effects, also going beyond the dipolar adiabatic tides considered here.

\subsubsection{Analysis of the full phasing expressions in the DD and QD domains}\label{sec:analysisfullphasing}
For the analysis of the overall Fourier phase evolution in both domains we focus specific binary BH systems. 
Based on the analysis above, it would be particularly interesting to detect both the DD and QD driven regimes of the early inspiral. This could be possible with multi-band detections of LISA and ground based detectors as illustrated in Fig.~\ref{fig:detect}, or with third-generation detectors such as ET.

\begin{figure}
\centering
\begin{subfigure}{0.49\textwidth}
   \includegraphics[width=0.9\linewidth]{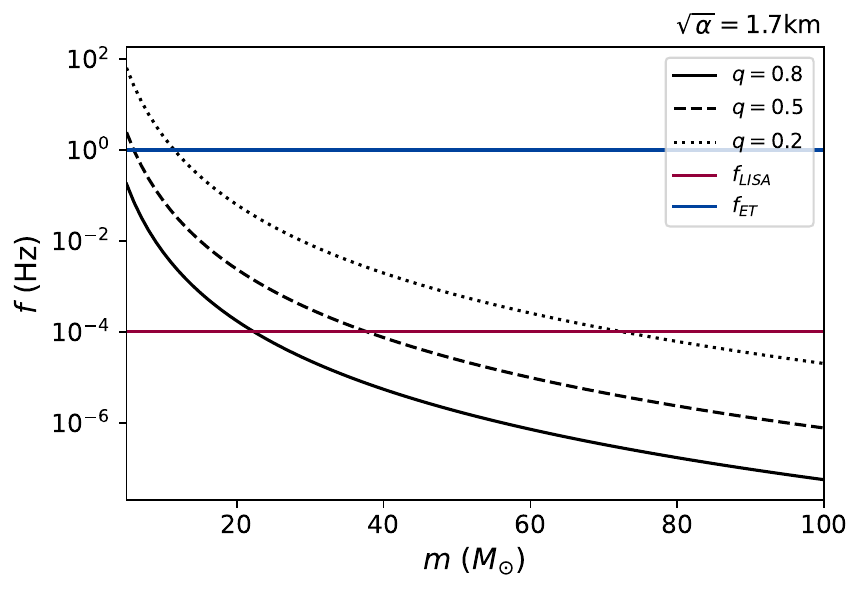}
       \centering
   \label{fig:boundaryfreq1} 
\end{subfigure}
\begin{subfigure}{0.49\textwidth}
    \includegraphics[width=0.9\linewidth]{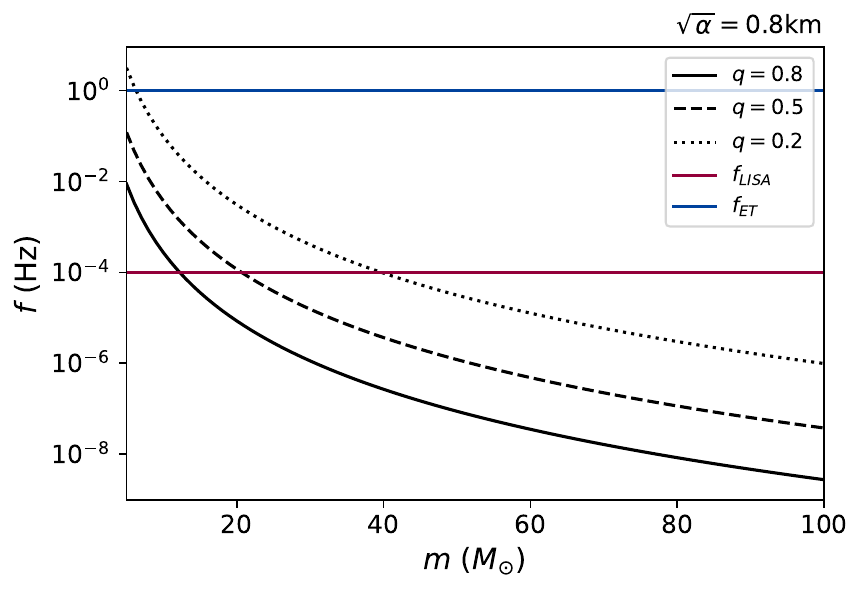}
    \label{fig:boundaryfreq2}
        \centering
 \end{subfigure}
\caption[]{Transition frequency~\eqref{eq:boundaryfreq} between the DD and QD regimes versus total mass for couplings of $\sqrt{\alpha}=1.7$km (top panel) and $\sqrt{\alpha}=0.8$km (bottom panel). The black curves with different dashing illustrate different mass ratios. The blue and red horizontal lines indicate the lower end of the sensitive frequency band of the ET and LISA detectors, respectively. 
}
\label{fig:boundaryfreq}
\end{figure}

In Fig.~\ref{fig:boundaryfreq} we show the boundary frequency \eqref{eq:boundaryfreq} for different total masses and mass ratios and a coupling of $\sqrt{\alpha}=1.7$km  and  $\sqrt{\alpha}=0.8$km also comparing to the lower bound frequencies of the ET and LISA sensitivity bands. In general, we see that the DD regime extends to higher frequencies for smaller mass ratios.
From the upper panel of Fig.~\ref{fig:boundaryfreq} we find that for the DD regime to fall within the ET frequency band the total mass should be below $11M_{\odot}$ and $q\sim 0.2$ which would make one of the BHs to have a mass below the astrophysical BH mass range. However for masses below $75M_{\odot}$ the DD regime falls in the LISA frequency band. For a smaller coupling depicted in the bottom panel of Fig.~\ref{fig:boundaryfreq}, this mass reduces to $42M_{\odot}$. Comparing to Fig.~\ref{fig:detect} these systems would be well within the LISA sensitivity band for a distance of around $150$ Mpc. The prospects for detecting the DD signal with multi-band observations rely significantly on the coupling parameter and the system properties, favoring small masses and mass ratios yet larger couplings for the DD regime to reach up to higher frequencies. However the smaller the total mass, the smaller the strain, which requires closer distances to the source for the possibility of detection. 

\begin{figure*}
        \centering
        \begin{subfigure}[b]{0.49\textwidth}
            \centering
            \includegraphics[width=1\textwidth]{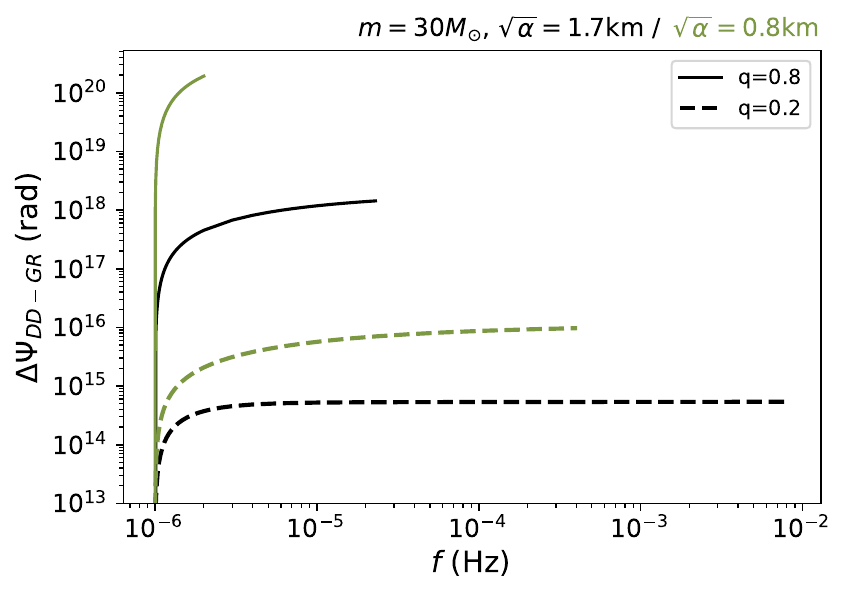}
        \end{subfigure}
        \begin{subfigure}[b]{0.49\textwidth}  
            \centering 
            \includegraphics[width=1\textwidth]{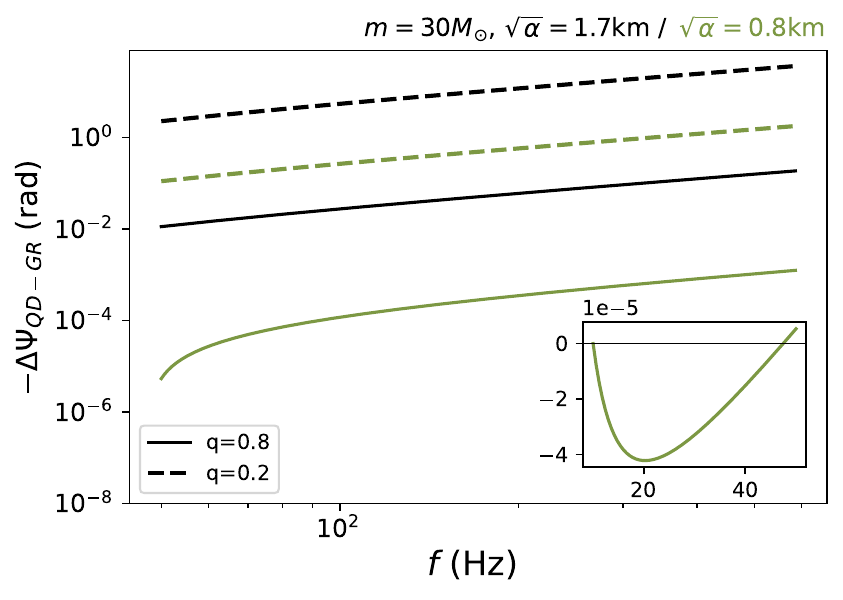}
        \end{subfigure}
        \vskip\baselineskip
        \begin{subfigure}[b]{0.49\textwidth}   
            \centering 
            \includegraphics[width=\textwidth]{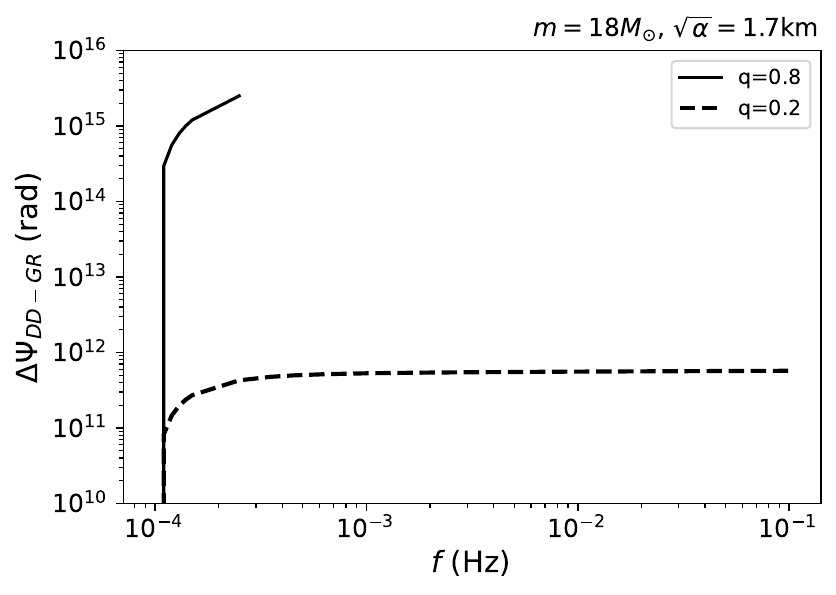}
        \end{subfigure}
        \begin{subfigure}[b]{0.49\textwidth}   
            \centering 
            \includegraphics[width=\textwidth]{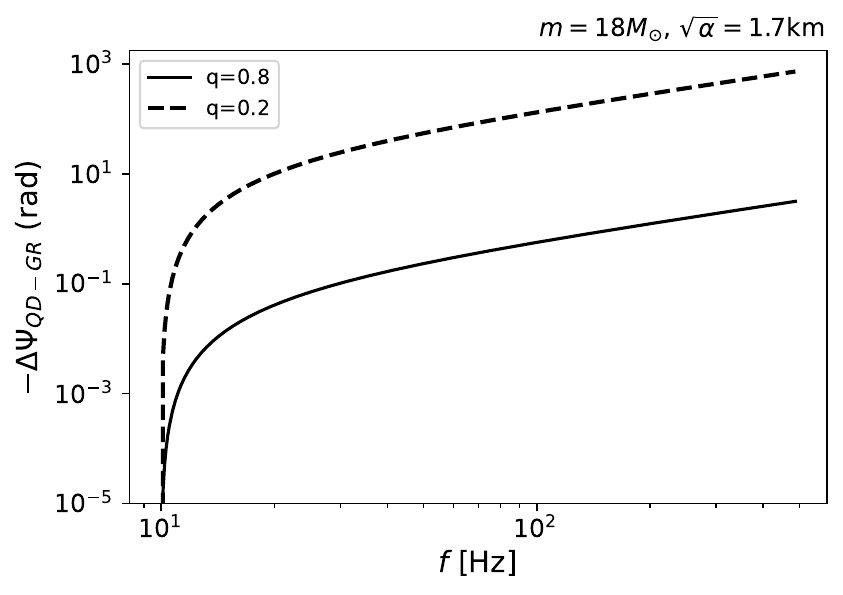}
        \end{subfigure}
        \caption{Phase difference between the DD (left panels) and QD (right panels) phasing from the 1PN GR case for different mass ratios. The curves in the DD regime terminate when the transition frequency~\eqref{eq:boundaryfreq} is reached.
        \emph{Top panels}:  systems of $m=30M_{\odot}$, with the black curves corresponding to a coupling of $
        \sqrt{\alpha}=1.7$km or $\epsilon=0.0015$ and the green curves to a coupling of  $
        \sqrt{\alpha}=0.8$km or $\epsilon=0.0003$. The inset in the top right panel shows the $\sqrt{\alpha}=0.8$km, $q=0.8$ case for the frequency range $10-50$Hz, note this inset does not have a logarithmic scale.
        \emph{Bottom panels}: systems of $m=18M_{\odot}$ and$
        \sqrt{\alpha}=1.7$km or $\epsilon=0.004$. }
        \label{fig:diffwGR}
\end{figure*}

Next, we consider systems with $m=30 M_{\odot}$ and $m=18 M_{\odot}$, for which the DD evolution spans a substantial frequency range in the sensitive band of LISA, while 
their evolution in the LIGO/Virgo/KAGRA band is fully QD. We study the large mass system for two different values of the coupling of $\sqrt{\alpha}=1.7$km and $\sqrt{\alpha}=0.8$km, and consider a coupling of $\sqrt{\alpha}=1.7$km for the smaller mass system. First we focus on the $m=30 M_{\odot}$ system. We evaluate the total Fourier phase evolution in the QD regime using~\eqref{psiQD} in the frequency band of current ground-based detectors, starting from $10$~Hz up to the estimate for the ISCO frequency of $293$~Hz. For the DD evolution, which partly overlaps with the LISA frequency band, we use~\eqref{psiDD} from a lower frequency of $10^{-6}$~Hz up to the transition frequency to the QD regime estimated from \eqref{eq:boundaryfreq}. As the transition frequency also depends on the mass ratio, we truncate the phase evolutions in the DD regime at their respective transitions in Fig.~\ref{fig:diffwGR}. We compare these phasings with the 1PN GR case. 
Note that when comparing different phasing evolutions, there is a remaining freedom to choose the reference constants $t_c$ and $\phi_c$. Here, we use this freedom to set the difference to zero at the starting frequencies and let their difference evolve to higher frequencies.

Figure~\ref{fig:diffwGR} shows the difference of the phasing in sGB including all effects we have calculated compared to  the 1PN GR case. 
For the DD evolution shown in the top left panel of Fig.~\ref{fig:diffwGR} the differences to GR are very large, as there is no dipolar radiation in GR. Additionally, the system stays at nearly the same frequency for a long time during the early inspiral. We find the green curves, corresponding to a coupling of $\epsilon=0.0003$, to lie above the black ones for $\epsilon=0.0015$, denoting larger deviations from GR for a smaller value of the coupling. In the limit of $q\xrightarrow{}1$, an equal mass system, the dipolar contribution in the QD phasing from \eqref{psiQD} vanishes and the DD phasing \eqref{psiDD} diverges as $S_-$ vanishes, as a consequence of working in the Fourier domain as described in Sec. \ref{sec:psidd}, however this is counteracted by a vanishing DD frequency regime from \eqref{eq:boundaryfreq}. Note that in the equal mass case there still is a tidal contribution proportional to $\zeta$ in the non dipolar part of \eqref{psiQD} related to the contribution from the dynamics.\\

The phase evolution in the QD regime is shown in the top right panel of Fig.~\ref{fig:diffwGR}. The phase evolution is very similar to the phase in GR where the higher curvature contributions make the QD phase smaller, therefore we show - minus - the phase difference. In this frequency regime the tidal contributions partly cancel the higher curvature terms. We found that for the small coupling $\epsilon=0.0003$ and large mass ratio $q=0.8$ case for frequencies below $50$Hz in the QD domain, the tidal and GB contributions are too small to counteract the enlarged difference of the QD versus GR phase evolution because of the lower order contributions given in \eqref{psiQD}. Therefore the phase difference for this system in this low frequency range as an opposite sign.  We show the curves for frequencies above $50$Hz and separately in the inset the $\epsilon=0.0003$ and $q=0.8$ curve for the frequency range $10-50$Hz. Note that the inset has no logaritmic scale while the total panel has. 
We see that in the QD regime the deviations from GR increase for smaller mass ratios, which corresponds with our analysis of the dependencies of the tidal and GB contributions in Fig. \ref{fig:tidGBcontr}. In the QD regime, the net phase differences from GR are small.  However for a coupling of $\epsilon=0.0015$ and mass ratio of $q=0.2$ the phase differences reach a maximum of $\mathcal{O}(10^{1})$ GW cycles, moving to the detectable range. 

We repeat this analysis for the smaller total mass system of $m=18M_{\odot}$ and $\sqrt{\alpha}=1.7$km, shown in the bottom right panel of Fig.\ref{fig:diffwGR}. The QD phasing differences for these systems are larger, becoming of $\mathcal{O}(10^1-10^2)$ GW cycles for mass ratios of order $q\sim 0.2$. 
Furthermore the DD differences are smaller compared to the $30 M_{\odot}$ system, which is due to the shift to higher frequencies of the DD regime and the positive relation with the total mass. 

\subsubsection{Analysis of the transition between DD and QD domains.}
Finally we analyse the accuracy of the phase evolution in the two domains and the regime at which the switch from DD to QD driven happens for two different total mass systems of $30M_{\odot}$ and $25M_{\odot}$ and two different mass ratios $q=0.2$ and $q=0.5$. We compare the QD and DD phase evolution by numerically solving \eqref{eq:psiv}, rather than using the Taylor~F2 approximation for phasings. Far in the QD regime this solution should approximate the QD evolution, ditto for the DD evolution for small frequencies. We expect the switch between the regimes to happen around \eqref{eq:boundaryfreq}.

\begin{figure}
\centering
\begin{subfigure}{0.48\textwidth}
   \includegraphics[width=0.9\linewidth]{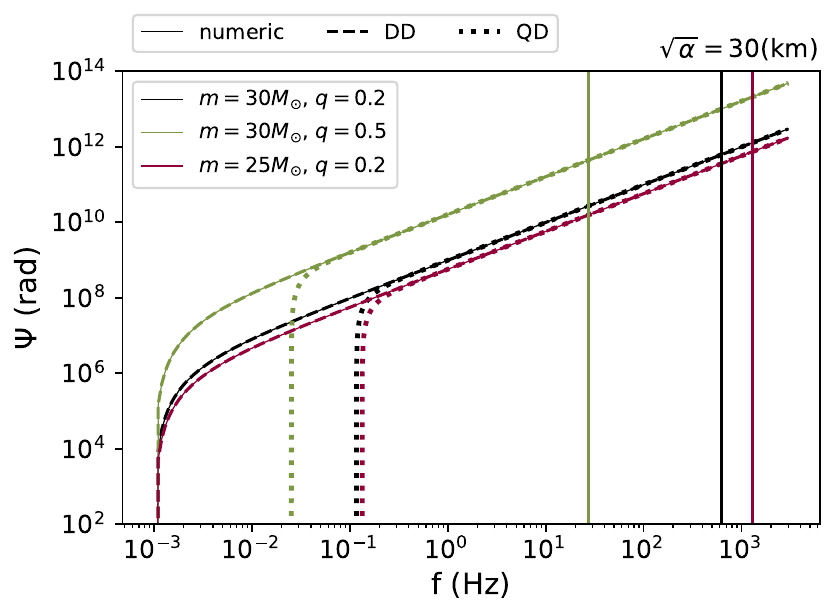}
       \centering
\end{subfigure}
\caption[]{Comparison of numerical and analytical phase evolution \eqref{psiDD} and \eqref{psiQD} for a coupling of $\sqrt{\alpha}=30 \textrm{km}$ for a system with $m=30M_\odot$ for $q=0.2$ and $q=0.5$ in black and green respectively and for a smaller mass $m=25M_\odot$ for $q=0.2$ in pink. Note the numerical curves overlap with the DD and QD lines. The vertical lines denote the analytical boundary frequencies \eqref{eq:boundaryfreq} of $f=27\textrm{Hz}$, $f=619\textrm{Hz}$ and $f=1289\textrm{Hz}$ in respectively green, black and pink. 
}
\label{fig:comparisonwithnumerical}
\end{figure}

For this comparison we match the analytic expressions with the numerical evolution. As this analysis is purely to see how well our analytic phase expressions evolve with respect to the numerical result we take a much larger value for the coupling than given by the current constraints. With this choice, the boundary frequency \eqref{eq:boundaryfreq} is much larger and therefore to consider a frequency far in the DD domain to match with the numerical result we do not have to go to very small values for the frequency where numerical inaccuracies need to be dealt with. However this analysis should take over to more realistic values of the coupling e.g. for a system of $25M_{\odot}$, $q=0.2$ with coupling $\sqrt{\alpha}=1.7\textrm{km}$ we were able to match the DD evolution with the numerical result starting at $f=10^{-8}\textrm{Hz}$. Our choice here for the large coupling has the advantage to be able to match up the DD evolution with the numerical evolution for all the systems at the same frequency of $f=10^{-3}$ which makes the comparison more clear. Then the analytic boundary frequency \eqref{eq:boundaryfreq} for a system of $30M_{\odot}$, $q=0.2$ is given by $f=619\textrm{Hz}$, for $30M_{\odot}$, $q=0.5$ by $f=27\textrm{Hz}$ and for $25M_{\odot}$, $q=0.2$ by $f=1289\textrm{Hz}$. As we have the freedom to set the integration constants $t_c$ and $\phi_c$ of \eqref{psiDD} and \eqref{psiQD}, we match the DD phasing with the numerical phase evolution around $10^{-3}\textrm{Hz}$, which is far in the DD regime for the three different systems we compare. For the QD phasing we match with the numerical phase evolution around $3000$Hz. Note that these frequencies are not realistic for a stellar mass binary inspiral because of the large choice for the coupling. The matching is done by integrating over the absolute difference between the two phasings and determine the constants $t_c$ and $\phi_c$ that minimize the integral. These phasings are shown in the top panel of Fig.~\ref{fig:comparisonwithnumerical}.\\

We find that the DD phasing, starting from $f=10^{-3}\textrm{Hz}$ follows the numerical phasing over the whole frequency domain as the curves for all the three systems overlap. The QD phasing follows the numerical evolution also well starting from $f=3000$Hz until far beyond the analytic estimates of the switch from QD to DD driven, which we denoted with the vertical lines in Fig.~\ref{fig:comparisonwithnumerical}. Where, as the boundary frequency becomes lower for a larger mass ratio or a larger mass, the frequency for which the QD evolution starts to differ also shifts to lower frequency. In the DD regime the QD evolution thus starts diverging strongly compared to the numerical result which makes sense as it is outside of the validity regime for quadrupolar radiation. The DD phasing keeps following the numerical evolution well far in the QD domain. The higher frequency where the DD description breaks down lies outside the analysed range. From this first analysis it seems that the QD and DD analytic expressions follow the total numerical result well in there respective domains and even beyond the boundary frequency estimate \eqref{eq:boundaryfreq}. 
Lastly we analyse the DD phase evolution with respect to the numerical result for a nearly equal mass system. As discussed in Sec.\ref{sec:psidd} the DD phase evolution in the Fourier domain is not well defined in the limit of $q\xrightarrow{}1$. By comparing it's evolution with the numerical result we try to find around which value for the mass ratio the expression becomes inaccurate. We analyse a range of values for the mass ratio in the limit of $q\xrightarrow{}1$ for a system of  $m=30M_{\odot}$ and $\sqrt{\alpha}=30 \textrm{km}$. For $q=0.999$ we find that after matching the DD phase to the numerical result the DD evolution quickly diverges after the starting frequency. This indicates that around this value for the mass ratio the DD analytic description becomes less accurate.

\section{Conclusion}\label{sec:conclusion}
In this article we studied signatures of scalar-induced adiabatic dipolar tidal effects in the dynamics and GWs from inspiraling BH binary systems in the sGB gravity. We used the PN approximation, augmented by independent perturbative expansions in the GB coupling and tidal effects, to compute the leading order tidal corrections to the scalar and tensor waves and obtained expressions for the GW phase evolution for circular orbits and nonspinning BHs. 

Our approach was based on an effective action for finite size effects adapted to the hierarchy of scales in the system. We recovered parameterized results in terms of the scalar tidal deformability from previous studies in the context of scalar-tensor theories~\cite{Bernard:2019yfz, Damour:1998jk} and, moreover, calculated the tidal deformability parameter in terms of the BH mass, the GB coupling strength, and the second derivative of the coupling function. We obtained the tidal corrections to the dynamics and to the fluxes, where the tidal contribution to the scalar waveform from the induced dipole moment introduces a difference dependence on the system parameters. We combined our results for the tidal effects with those due to higher curvature corrections from~\cite{Shiralilou:2021mfl}, which enter with the same scaling with GW frequency as the dipolar tidal effects, and derived a closed-form result for the GW phase in the Fourier domain in the stationary phase approximation. These results are tailored to two different regimes, the DD regime dominated by the scalar flux, which is important for low frequencies but also depends on the system parameters, and the QD regime dominated by the tensor flux. 

We performed several case studies of the implications of our results for GW signals for a dilatonic coupling function, which has nonvanishing derivatives at all orders when evaluated for a zero-field configuration. This is important since the GB and tidal corrections depend on the first and second such derivatives respectively, and thus may be absent for certain other choices of coupling function. By comparing the GB and tidal contributions to the energetics and phase evolution we found that for a positive value of the coupling around the current observational constraints the tidal contributions in the DD regime come with the same sign as the GB contributions, both enlarging the differences with GR, while in the QD regime the two contributions have opposite sign, partly cancelling each other. Our analysis showed that in the DD regime and for small couplings, the tidal effects dominate over GB contributions in magnitude. Moreover, the tidal effects in the GW phasing become larger for smaller couplings in this regime, though the cumulative contribution vanishes in the zero-coupling limit, as expected. We also found that at fixed dimensionless coupling, which rescales the fundamental GB coupling by the total mass, the size of the tidal and GB corrections increases for systems with larger total mass. This trend is counter intuitive for the GB corrections, which we expect to become larger for smaller masses corresponding to higher curvatures. We show that this is a consequence of fixing the dimensionless coupling and when considering a fixed fundamental coupling, the behavior is as expected. When including all effects of sGB gravity considered here and comparing with the GR case we found that, as expected, the differences in the DD regime are very large as there is no dipolar energy loss in GR. Additionally we highlighted that as a consequence of working in the Fourier domain there is an opposite scaling with respect to the mass ratio and coupling for the DD phase compared to the QD phase. In the QD regime, we found that for a coupling of $\sqrt{\alpha}=1.7$km, small mass ratio and systems with total mass around or below $30M_\odot$, the differences from GR are of the order of several GW cycles. For smaller values of the coupling the differences with GR in the QD regime decrease.

Based on estimates for the frequency regimes relevant for the DD and QD evolution respectively, we analyzed the prospects of measuring signals from the DD regime and the crossover to the QD regime with multiband observations with LISA and ground-based detectors. Our results indicate that this is a possibility for part of the parameter space of binary systems, and becomes more promising for smaller masses and mass ratios, as estimated for a coupling in agreement with the current observational constraints and a distance to the binary of $150 \rm{Mpc}$.
Measuring signals from both regimes would be useful for breaking degeneracies, as they involve different dependencies on the parameters. 

To gain further insights into the crossover between the DD and QD regimes we compare the analytical estimates with numerical studies of the phase evolution. Our analysis indicates that the accuracy of the DD and QD regimes spans beyond the analytic estimate for the crossover and depends on the system parameters. Repeating the same analysis for mass ratios close to an equal mass system gave an indication for the accuracy of the DD phase expressions in the limit to $q\xrightarrow{}1$. Overall we highlight that our analytical closed form results can be directly implemented in data analysis studies for improved tests of gravity.

As our work is an fist exploratory study of finite size effects for BH binaries in sGB gravity, it was limited in scope. Ongoing work focuses on including a scalar field mass to the sGB framework \cite{msGBfuturework}. Other pertinent advances are to include higher order corrections in our approximations and broadening our assumptions e.g. including higher order finite size, PN, coupling and spin effects and considering eccentric orbits. 
In addition it would be interesting to study different systems such as neutron star (NS) BH or NS binaries or other gravity theories. 

This work contributes an important step towards a more accurate modeling of the inspiral of a comparable-mass BH binary systems in sGB gravity. Given that the tidal signatures computed here enter in the GW signals with the same frequency scaling as the higher curvature contributions and, for couplings consistent with current constraints, even dominate over the higher curvature contributions in the DD regime, these effects will be important to incorporate in further analyses of constraints inferred from GW measurements. They will also be needed for future models of full inspiral-merger-ringdown GW templates for sGB models.

\textbf{\emph{Acknowledgments.}} We thank Laura Bernard, Simon Groot, Nan Jiang, Felix Julie, Samaya Nissanke, Nestor Ortiz, Justin Ripley, Stefan Vandoren, Helvi Witek, Kent Yagi and Nicolas Yunes and Vitor Cardoso for useful discussions. And Gastón Creci for pointing out an important error in the earlier version of this manuscript. This work was supported by the Netherlands Organization for Scientific Research (NWO). I. G. acknowledges the Dutch Black Hole Consortium (project NWA 1292.19.202) part of the National Research Agenda, B.S. the Projectruimte grant GW-EM NS (PI Nissanke), and T.H. the Beta Sectorplan and Cost Action CA18108. 
\appendix

\section{Construction of the effective  action for dipolar scalar tidal effects}\label{sec:apptidalEFT}
Several considerations dictate the kinds of terms that can enter an effective action description. For the degrees of freedom, we will work with the scalar multipole moments $Q_L^{(s)}$ and external fields ${\cal E}_L^{(s)}$ associated to the worldline skeleton. Here, $L$ denotes a string of $\ell$ spatial indices and in this section, we omit the body labels on all the quantities. We have written the moments in terms of three-dimensional spatial tensors in the rest frame, where $u^\mu=(1,0,0,0)$. We expect the description in the rest frame to contain the complete physical content of finite size effects. The tensors $Q_L^{(s)}$ and ${\cal E}_L^{(s)}$ are additionally symmetric and trace-free (STF), and transform irreducibly under rotations, which follows from the spherical symmetry of the unperturbed configuration. Specifically, the fields ${\cal E}_L^{(s)}$ are given by 
\begin{equation}
\label{eq:calEdef}
{\cal E}_L^{(s)}=-\underset{x\to z}{\rm FP}\partial_L \varphi ,
\end{equation}
where FP denotes the finite part. This is equivalent to using only the portion of the field $\varphi$ sourced by the companion. The covariant version of these quantities can be obtained by using the transverse projector
\begin{equation}
\label{eq:transverseP}
P_{\mu}^\alpha=\delta_{\mu}^\alpha +u_\mu u^\alpha,
\end{equation}
which has the property that $P_\mu^\alpha u^\mu=0$. 
Using this operator we can write the covariant version of ${\cal E}_I^s=-\underset{x\to z}{\rm FP}\partial_{I}\varphi$ as ${\cal E}_{\mu}^{(s)}=-\underset{x\to z}{\rm FP}P_{\mu}^\alpha \partial_{\alpha}\varphi$. The transformation to the covariant formulation can equivalently be accomplished using a tetrad, as discussed in Sec.III of Ref.~\cite{Gupta:2020lnv}.\\
We next consider the kinds of couplings involving these degrees of freedom that could appear in the effective action. For nonspinning configurations, whose unperturbed states are spherically symmetric and static, and non-dissipative effects, the couplings in the effective action should be invariant under time-reversal and parity transformations. The next consideration is about the truncation of such terms, which we base on considerations about the relevant characteristic scales discussed in section \ref{scales}.

Based on these small parameters introduced in this section, we truncate the terms in the effective action as follows. Firstly,
we include only terms up to second order in a derivative expansion of the fields, as higher order terms are suppressed by powers of $\varepsilon_{\rm tid}$ defined in~\eqref{eq:epsilontidal}. We also omit higher nonlinearities involving $Q_L^{(s)}$. Furthermore, we treat time derivatives of the external fields as suppressed compared to spatial gradients as is standard in the PN approximation on the orbital scale. We also work in the adiabatic limit, where internal relaxation timescales associated to the scalar condensate and thus $Q_L^{(s)}$, which are of order the quasi-normal mode frequency of a BH for $\hat \alpha\ll 1$ ~\cite{Yagi:2022vys}, are much faster than the variations of the external fields, which change on the orbital timescale, so that the configuration remains in equilibrium. Effective field theoretical approaches generally rely on integrating out degrees of freedom with frequencies higher than the scale under consideration -- here, the orbital scale -- and we will ultimately use this approach to express the finite size effects solely in terms of quantities defined in the orbital zone. 

The above considerations lead to the following leading-order terms describing finite size (FS) effects in the rest frame of one of the bodies as $S_{\rm FS}=c \int ds \, {\cal L}_{\rm FS}$ with\begin{equation}
\label{eq:SFSwithQ}
{\cal L}_{\rm FS}=c_1 Q_I^{(s)} {\cal E}^I_{(s)}+c_2 {\cal E}_I^{(s)}{\cal E}^I_{(s)}+ {\cal L}^{\rm int}(Q_I^{(s)},\dot Q_I^{(s)})+\ldots,
\end{equation}
where $\dot Q_I^{(s)}=DQ_I^{(s)}/d\tau=u^\mu \nabla_\mu Q_I^{(s)}$ and $\mathcal{L}^{\rm int}$ denotes the internal dynamics of the scalar dipole. We have omitted accelerations as they can be removed via field redefinitions. Within our approximations, the only possible internal terms are the quadratic combinations
\begin{equation}
{\cal L}^{\rm int}=-\frac{1}{2\lambda_s} Q^I_{(s)}Q_I^{(s)}+c_4 \dot Q^I_{(s)}\dot Q_I^{(s)},
\end{equation}
where we have chosen to label the first coefficient as $-1/(2\lambda_s)$ instead of $c_3$, which at this stage is merely a choice. The coefficients in~\eqref{eq:SFSwithQ} must still be matched to the properties of the BH and scalar condensate. One of the coefficients fixes the overall normalization of the quantities, which we choose such that the coupling of $Q_I^{(s)}$ with the external field scales as $-1/\ell!$. For the dipole $\ell=1$ case this implies that $c_1=1$. The equations of motion for $Q_I^{(s)}$ derived from the action~\eqref{eq:SFSwithQ} in the adiabatic limit, where we neglect any $\tau-$derivatives, are 
\begin{equation}
\label{eq:lambdasdef1}
Q_I^{(s)}=-\lambda_s {\cal E}_I^{(s)},
\end{equation}
which makes the physical interpretation of the coefficient $\lambda_s$ as the dipolar tidal Love number of the configuration explicit. The value of this coefficient cannot be determined from the effective description; it must instead be computed from a detailed study of the response of the scalar condensate to a relativistic scalar tidal perturbation, as we will analyze in Sec.~\ref{sec:Lovenumber}. Finally, using~\eqref{eq:lambdasdef1} to integrate out $Q_I^{(s)}$ leads to the reduced action
\begin{equation}
\label{eq:SFSwithoutQ}
S_{\rm FS}=c\int ds \, \left[\frac{\lambda_s}{2}{\cal E}_I^{(s)} {\cal E}^I_{(s)}+c_2 {\cal E}_I^{(s)}{\cal E}^I_{(s)}+\ldots\right].
\end{equation}
Here, we note that a contribution to the last term involving $c_2$ also arises from field-redefinitions of curvature quantities in the gravitational sector.
This term describes possible additional contributions not contained in the $\lambda_s$ terms, for example, from subdominant higher-frequency degrees of freedom.  We expect these contributions to be small and will omit them in our final analysis. 
Using the operator~\eqref{eq:transverseP} we can  write~\eqref{eq:SFSwithoutQ} in covariant form as
\begin{equation}
S_{\rm tid}=c\int ds \, \left[\frac{\lambda_s}{2}{\cal E}_\mu^{(s)} {\cal E}^\mu_{(s)}+\ldots\right],
\end{equation}
and using the definition~\eqref{eq:calEdef}, the fact that covariant derivatives of a scalar are equivalent to partial derivatives, and specializing to a binary system leads to the expression given in~\eqref{Stid}. This action describes the dipolar tidal effects on the orbital scale, with the coefficient $\lambda_s$ determined by matching to the strong-field vicinity of each object, as  discussed in Sec.~\ref{sec:Lovenumber}.

\section{Explicit expressions from waveform and phase evolution calculation}\label{waveformphasing}
In this appendix we show the full expressions coming from the scalar and part of the tensor waveform calculation. \\
As described in section \ref{diprad}, the scalar waveform can be obtained via the DIRE approach constructing the different scalar field moments from \eqref{poleswaveform}. The source term from \eqref{EOMsourceterm} and in the definition of the multipoles \eqref{I} is obtained from the full field equation \eqref{FE2} expanded to 1PN
\begin{equation}
\begin{aligned}
\mu_s&=-\frac{G}{c^4}\frac{\delta (S_m)}{\delta \varphi}-\frac{1}{16 \pi} \alpha f^{\prime}(\varphi) \mathcal{R}_{G B}^2 + \mathcal{O}(1/c^4)\\
&=\frac{G}{c^4}\sum_A \delta^{(3)}(\mathbf{x}-\mathbf{x_A}(t))\left[m_A^0\alpha_A^0\left(c^2 + (\alpha_A^0 + \frac{\beta_A^0}{\alpha_A^0 })\delta\varphi_c^{(1)}\right.\right. \\
&\left.\left.-  U^{(1)} - \frac{1}{2}  v_A^2\right) \right].
\end{aligned}
\end{equation}
Using in the last equality that the contributions of the $ \mathcal{R}_{G B}^2$ term vanishes in the integral of \eqref{I} as shown in \cite{Shiralilou:2021mfl}. The tidal contribution is considered in section \ref{diprad}. Substituting the lowest order field solutions \eqref{eq:UphiLO} and preforming the integration gives for the multipole moments
\begin{subequations}\label{differentIs}
\begin{eqnarray}
I_{s} &=&m_{A}^0 c^{-2} \alpha_{A}^{0}\left\{1-\frac{v_{A}^{2}}{2 c^{2}}-\frac{G m_{B}^0 \alpha_{A B}}{r c^{2}}\right.\nonumber\\
&&\left.+\mathcal{O}\left(c^{-4}\right)\right\}+(A \leftrightarrow B) \\
I_{s}^{i}& =&x_{A}^{i} m_{A}^0 c^{-2} \alpha_{A}^{0}\left\{1-\frac{v_{A}^{2}}{2 c^{2}}-\frac{G m_{B}^0 \alpha_{A B}}{r c^{2}}\right. \nonumber\\
&&\left.+\mathcal{O}\left(c^{-4}\right)\right\}+(A \leftrightarrow B),\\
I_{s}^{i j}& =&x_{A}^{i j} m_{A}^0 c^{-2} \alpha_{A}^{0}\left\{1+\mathcal{O}\left(c^{-2}\right)\right\}+(A \leftrightarrow B), \\
I_{s}^{i j k} &=&x_{A}^{i j k} m_{A}^0 c^{-2} \alpha_{A}^{0}\left\{1+\mathcal{O}\left(c^{-2}\right)\right\}+(A \leftrightarrow B).
\end{eqnarray}
\end{subequations}

Converting to the centre of mass frame with expressions (45) and (47) from \cite{Shiralilou:2021mfl} and substitution in \eqref{poleswaveform} gives for the expression of the scalar waveform up to a relative 0.5PN\footnote{using the convention of setting the lowest order GR waveform terms (can be extracted from the quadrupole formula) equal to 0PN, which is proportional to $1/c^4$.}
\begin{subequations}
    \label{phitot}
\begin{equation}\label{fullphi}
\Phi =\frac{ G \mu \sqrt{\bar{\alpha}}}{R c^{3}}\left\{P^{-1 / 2} \tilde{\Phi}+\frac{1}{c} \tilde{\Phi}+\frac{1}{c^{2}} P^{1 / 2} \tilde{\Phi}+\mathcal{O}\left(c^{-3}\right)\right\},
\end{equation}
with the components
\begin{widetext}
\begin{align}
&P^{-1 / 2} \tilde{\Phi}=2 \mathcal{S}_{-}(\mathbf{n} \cdot \mathbf{v}), \\
&\tilde{\Phi} =\begin{aligned}[t]&\left(\mathcal{S}_{+}-\frac{\Delta m}{m} \mathcal{S}_{-}\right)\left[-\frac{G \bar{\alpha} m}{r}\left(\frac{\mathbf{n} \cdot \mathbf{r}}{r}\right)^{2}+(\mathbf{n} \cdot \mathbf{v})^{2}-\frac{1}{2} v^{2}\right] \\
&+\frac{G \bar{\alpha} m}{r}\left[-2 \mathcal{S}_{+}+\frac{8}{\bar{\gamma}}\left(\mathcal{S}_{+} \beta_{+}+\mathcal{S}_{-} \beta_{-}\right)\right] ,\end{aligned}\\
&P^{1 / 2} \tilde{\Phi}=\begin{aligned}[t]&\left(-\frac{\Delta m}{m} \mathcal{S}_{+}+(1-2 \eta) \mathcal{S}_{-}\right)\left[\frac{3}{2} \frac{G \bar{\alpha} m}{r^{4}} \dot{r}(\mathbf{n} \cdot \mathbf{r})^{3}-\frac{7}{2} \frac{G \bar{\alpha} m}{r^{3}}(\mathbf{n} \cdot \mathbf{v})(\mathbf{n} \cdot \mathbf{r})^{2}+(\mathbf{n} \cdot \mathbf{v})^{3}\right] \\
&+(\mathbf{n} \cdot \mathbf{v})\left\{\left(\frac{\Delta m}{m} \mathcal{S}_{+}-\eta \mathcal{S}_{-}\right) v^{2}+\frac{G \bar{\alpha} m}{r}\left[\frac{1}{2} \frac{\Delta m}{m} \mathcal{S}_{+}+\left(2 \eta-\frac{3}{2}\right) \mathcal{S}_{-}\right.\right.\\
&\left.\left.\quad-\frac{4}{\bar{\gamma}} \frac{\Delta m}{m}\left(\mathcal{S}_{+} \beta_{+}+\mathcal{S}_{-} \beta_{-}\right)+\frac{4}{\bar{\gamma}}\left(\mathcal{S}_{-} \beta_{+}+\mathcal{S}_{+} \beta_{-}\right)\right]\right\} \\
&+\frac{G \bar{\alpha} m}{r^{2}} \dot{r}(\mathbf{n} \cdot \mathbf{r})\left[\frac{3}{2} \mathcal{S}_{-}-\frac{5}{2} \frac{\Delta m}{m} \mathcal{S}_{+}+\frac{4}{\bar{\gamma}} \frac{\Delta m}{m}\left(\mathcal{S}_{+} \beta_{+}+\mathcal{S}_{-} \beta_{-}\right)-\frac{4}{\bar{\gamma}}\left(\mathcal{S}_{-} \beta_{+}+\mathcal{S}_{+} \beta_{-}\right)\right] \\
&+2 \frac{G \bar{\alpha} m}{r} \frac{\alpha f^{\prime}\left(\varphi_{0}\right)}{\sqrt{\bar{\alpha}} r^{2}} \mathcal{S}_{+}\left(\mathcal{S}_{+}+\frac{\Delta m}{m} \mathcal{S}_{-}\right)\left[3 \frac{\dot{r}}{r}(\mathbf{n} \cdot \mathbf{r})-(\mathbf{n} \cdot \mathbf{v})\right] \\
&- \frac{\bar{\zeta} }{\sqrt{\bar{\alpha}} \mu r^3}\left[3 \frac{\dot{r}}{r}(\mathbf{n} \cdot \mathbf{r})-(\mathbf{n} \cdot \mathbf{v})\right].
\end{aligned}
\end{align}
\end{widetext}
\end{subequations}
Compared to (55) of \cite{Shiralilou:2021mfl} we find an overall factor of $2$ difference in \eqref{fullphi} and a slightly modified prefactor of the term proportional to $\alpha$.
Integrating the square of the differentiated scalar wavefunction in \eqref{integralEsdot} using the identities from \cite{Thorne:1980ru} gives for the scalar energy flux up to a relative 0PN
\begin{widetext}
\begin{eqnarray}
\label{Esdot}
\dot{E}_{S}&=&\frac{\eta^{2}}{G \bar{\alpha} c^{3}}\left(\frac{G \bar{\alpha} m}{r}\right)^{4}\left[\frac{4}{3} \mathcal{S}_{-}^{2}+\frac{8}{15 c^{2}}\left(\frac { G \overline { \alpha } m } { r } \left[\left(-23+\eta-10 \bar{\gamma}-10 \beta_{+}+10 \frac{\Delta m}{m} \beta_{-}\right) \mathcal{S}_{-}^{2}\right.\right.\right. \nonumber\\
&&\left.-2 \frac{\Delta m}{m} \mathcal{S}_{+} \mathcal{S}_{-}\right]+v^{2}\left[+2 \mathcal{S}_{+}^{2}+2 \frac{\Delta m}{m} \mathcal{S}_{+} \mathcal{S}_{-}+(6-\eta+5 \bar{\gamma}) \mathcal{S}_{-}^{2}-\frac{10}{\bar{\gamma}} \frac{\Delta m}{m} \mathcal{S}_{-}\left(\mathcal{S}_{+} \beta_{+}+\mathcal{S}_{-} \beta_{-}\right)\right. \nonumber\\
&&\left.+\frac{10}{\bar{\gamma}} \mathcal{S}_{-}\left(\mathcal{S}_{-} \beta_{+}+\mathcal{S}_{+} \beta_{-}\right)\right]+\dot{r}^{2}\left[+\frac{23}{2}\mathcal{S}_{+}^{2}\right.\nonumber\\
&&-8 \frac{\Delta m}{m} \mathcal{S}_{+} \mathcal{S}_{-}+\left(9 \eta-\frac{37}{2}-10 \bar{\gamma}\right) \mathcal{S}_{-}^{2}-\frac{80}{\bar{\gamma}} \mathcal{S}_{+}\left(\mathcal{S}_{+} \beta_{+}\right. \nonumber\\
&&\left.\left.\left.+\mathcal{S}_{-} \beta_{-}\right)+\frac{30}{\bar{\gamma}} \frac{\Delta m}{m} \mathcal{S}_{-}\left(\mathcal{S}_{+} \beta_{+}+\mathcal{S}_{-} \beta_{-}\right)-\frac{10}{\bar{\gamma}} \mathcal{S}_{-}\left(\mathcal{S}_{-} \beta_{+}+\mathcal{S}_{+} \beta_{-}\right)+\frac{120}{\bar{\gamma}^{2}}\left(\mathcal{S}_{+} \beta_{+}+\mathcal{S}_{-} \beta_{-}\right)^{2}\right]\right)\nonumber\\
&&\left.- \frac{8}{ c^{2}}\left(\frac{\alpha f^{\prime}\left(\delta \varphi_{0}\right) \mathcal{S}_{-} \mathcal{S}_{+}}{\sqrt{\bar{\alpha}} r^{2}}\right)\left(\mathcal{S}_{+}+\frac{\Delta m}{m} \mathcal{S}_{-}\right)\left[-3 \dot{r}^{2}+ v^{2}-\frac{2 G \bar{\alpha} m}{3 r}\right]\right.\nonumber\\
&&\left.-\frac{16}{ c^2}\left(\frac{\alpha f^{\prime}\left(\delta \varphi_{0}\right) \mathcal{S}_{-}^2  }{\bar{\alpha}^{3/2} r^{2}}\right)(3S_+ + \frac{\Delta m}{m}S_-)(-\frac{ 2 G m \bar{\alpha} }{3 r})\right.\\
&&\left. + \frac{4 \bar{\zeta} S_-}{G m \mu \bar{\alpha}^{3/2} c^2 r^2 }[-3 \dot{r}^2 + v^2 -\frac{2 G m \bar{\alpha}}{3 r}] + \frac{8 \zeta S_-^2}{m c^2 r^2}(+\frac{2 G m \bar{\alpha}}{3 r}) +\mathcal{O}\left(c^{-3}\right)\right].
\end{eqnarray}
\end{widetext}
For the scalar flux compared to (61) of \cite{Shiralilou:2021mfl} we find an additional factor 2 in the 6th line and a second term proportional to the coupling. This term is sourced from the relative acceleration \eqref{arel}, coming in from the product of the -0.5 and 0.5PN terms of \eqref{fullphi}.\\

For the calculation of the tensor waveform we do not repeat the calculation of the DIRE integral from the field equations \eqref{FE1} from scratch. Instead we refer to \cite{Shiralilou:2021mfl} for the definitions of the Epstein-Wagoner moments constructing the multipole expansion of the near zone tensor waveform. We start from their result of the tensor waveform (21). The second order time derivative results in the tensor waveform up to a relative 1PN
\begin{widetext}
\begin{subequations}
\label{eq:tensorwaves}
\begin{equation}
\begin{aligned}
&h_{T T}^{i j}=\frac{2 G \mu}{R c^4}\left\{Q^{i j}\right\}_{T T}=\frac{2 G \mu}{R c^4}\left\{\tilde{Q}^{i j}+\frac{1}{c} P^{1 / 2} \tilde{Q}^{i j}+\frac{1}{c^2}\left(P \tilde{Q}^{i j}+P \tilde{Q}_{G B}^{i j} + P \tilde{Q}^{i j}_{tid}\right)+\mathcal{O}\left(c^{-3}\right)\right\}_{T T}, \\
&\tilde{Q}^{i j}=2\left[v^{i j}-\frac{G m \bar{\alpha} r^{i j}}{r^3}\right],\\
&P^{1 / 2} \tilde{Q}^{i j}=\frac{\Delta m}{m}\left[3 \frac{G m \bar{\alpha}}{r^3}(\hat{\mathbf{n}} \cdot \mathbf{r})\left(2 v^{\left(i r^j\right)}-\frac{\dot{r} r^{i j}}{r}\right)-(\hat{\mathbf{n}} \cdot \mathbf{v})\left(2 v^{i j}-\frac{G m \bar{\alpha} r^{i j}}{r^3}\right)\right], \\
&P \tilde{Q}^{i j}=\frac{1-3 \eta}{3}\left\{(\mathbf{r} \cdot \hat{\mathbf{n}})^2 \frac{G \bar{\alpha} m}{r^3}\left[\left(6 \bar{E}-15 \dot{r}^2+13 \frac{G \bar{\alpha} m}{r}\right) \frac{r^{i j}}{r^2}+30 \dot{r} \frac{r^{(i} v^{j)}}{r}-14 v^{i j}\right]\right. \\
&\left.\quad+(\hat{\mathbf{n}} \cdot \mathbf{v})^2\left[6 v^{i j}-2 \frac{G \bar{\alpha} m}{r^3} r^{i j}\right]+\frac{1}{2}(\mathbf{r} \cdot \hat{\mathbf{n}})(\hat{\mathbf{n}} \cdot \mathbf{v}) \frac{G \bar{\alpha} m}{r^2}\left[12 \frac{\dot{r} r^{i j}}{r^2}-32 \frac{r^{(i} v^{j)}}{r}\right]\right\} \\
&\quad+\frac{1}{3}\left\{\left[3(1-3 \eta) v^2-2(2-3 \eta) \frac{G \bar{\alpha} m}{r}\right] v^{i j}+4 \frac{G \bar{\alpha} m}{r}(5+3 \eta+3 \bar{\gamma}) \frac{\dot{r}}{r} r^{(i,} v^{j)}\right. \\
&\left.\quad+\frac{G \bar{\alpha} m}{r^3} r^{i j}\left[3(1-3 \eta) \dot{r}^2-(10+3 \eta+6 \bar{\gamma}) v^2+\left(29+12 \bar{\gamma}+12 \beta_{+}-12 \frac{\Delta m}{m} \beta_{-}\right) \frac{G \bar{\alpha} m}{r}\right]\right\}
\end{aligned}
\end{equation}

\begin{equation}
\begin{aligned}
&P \tilde{Q}_{G B}^{i j}=4 \frac{G \bar{\alpha} m}{r} \frac{\alpha f^{\prime}\left(\phi_0\right)}{\sqrt{\bar{\alpha}} r^2}\left[-\left(\mathcal{S}_{+}+\frac{\Delta m}{m} \mathcal{S}_{-}\right)\left(\frac{12 \dot{r} r^{(i} v^{j)}}{r}+\frac{r^{i j}}{r^2}\left(6 \tilde{E}+\frac{5 G \bar{\alpha} m}{r}-15 \dot{r}^2\right)-2 v^{i j}\right)\right. \\
&\left.-2 \frac{G m}{r} \frac{r^{i j}}{r^2}\left(3 \mathcal{S}_{+}+\frac{\Delta m}{m} \mathcal{S}_{-}\right)\right]+3 \frac{G \bar{\alpha} m}{r} \frac{\alpha f^{\prime}\left(\phi_0\right)}{\sqrt{\bar{\alpha}} r^4}\left(\mathcal{S}_{+}(1-2 \eta)+\mathcal{S}_{-} \frac{\Delta m}{m}\right) \times \\
&\left\{(\mathbf{r} \cdot \hat{\mathbf{n}})^2\left[\frac{r^{i j}}{r^2}\left(10 \tilde{E}-35 \dot{r}^2+\frac{9 G \bar{\alpha} m}{r}\right)+\frac{\left.10 \dot{r}  2 r^{(i} v^j\right)}{r}-2 v^{i j}\right]-2(\hat{\mathbf{n}} \cdot \mathbf{v})^2 r^{i j}+\frac{1}{2}(\mathbf{r} \cdot \hat{\mathbf{n}})(\hat{\mathbf{n}} \cdot \mathbf{v})\left[20 \frac{\dot{r} r^{i j}}{r}-4r^{(i} v^j) \right]\right\} ,
\end{aligned}
\end{equation}

\begin{equation}
    P \tilde{Q}^{i j}_{tid} = -\frac{4 G^2 \bar{\alpha}^2 m}{r^6} \zeta r^{ij}.
\end{equation}
\end{subequations}
\end{widetext}
We correct additional factors 2 in front of the terms in the symmetric brackets notation, two factors $1/2$ in front of the $(\mathbf{r}\cdot\hat{\mathbf{n}})(\mathbf{v}\cdot\hat{\mathbf{n}})$ terms and a factor of $\bar{\alpha}$ in the first term of the second line in the $P\tilde{Q}^{ij}_{GB}$ expression. In our case we obtain a tidal contribution from the tidal term in the relative acceleration \eqref{arel}.
We integrate this expression over the angular dependencies in \eqref{integralEdotT}, again using the identities in \cite{Thorne:1980ru}. With this we obtain the tensor energy loss up to a a relative 1PN.
\begin{widetext}
\begin{equation}\label{ETdot}
\begin{aligned}
\dot{E}_T &=\frac{8}{15} \frac{\eta^2}{G \bar{\alpha}^2 c^5}\left(\frac{G \bar{\alpha} m}{r}\right)^4\left\{\left(12 v^2-11 \dot{r}^2\right)\right.\\
&+\frac{1}{28 c^2}\left[-16\left(170-10 \eta+63 \bar{\gamma}+84 \beta_{+}-84 \frac{\Delta m}{m} \beta_{-}\right) v^2 \frac{G \bar{\alpha} m}{r}\right.\\
&+(785-852 \eta+336 \bar{\gamma}) v^4-2(1487-1392 \eta+616 \bar{\gamma}) v^2 \dot{r}^2+3(687-620 \eta+280 \bar{\gamma}) \dot{r}^4 \\
&\left.+8\left(367-15 \eta+140 \bar{\gamma}+168 \beta_{+}-168 \frac{\Delta m}{m} \beta_{-}\right) \dot{r}^2 \frac{G \bar{\alpha} m}{r}+16(1-4 \eta)\left(\frac{G \bar{\alpha} m}{r}\right)^2\right] \\
&\alpha\frac{f'(\varphi_0)}{7 c^2 \sqrt{\bar{\alpha}} r^2}\Bigg((S_+(1+\frac{6}{25}\eta) + \frac{\Delta m}{m}S_-)(1350 v^4 + 5100\dot{r}^4 + \frac{G \bar{\alpha} m}{r}(-750 v^2 + 950 \dot{r}^2) -900v^2 \dot{r}^2) + \frac{G m}{r}(3 S_+\\
&+ \frac{\Delta m}{m} S_-)(672 v^2 - 784 \dot{r}) \Bigg)-\frac{G \bar{\alpha} }{c^2 r^3} (56 \dot{r}^2 -48 v^2) \zeta\}.
\end{aligned}
\end{equation}
\end{widetext}
We find our terms proportional to $\alpha$ to differ with the expression (59) in \cite{Shiralilou:2021mfl}. In section \ref{phaseevolution} the scalar \eqref{Esdot} en tensor \eqref{ETdot} energy loss are used to obtain the total energy flux and phase evolution. For deriving the Fourier phase evolution we specialize the flux expressions to circular orbits and rewriting the terms to PN parameter $x$ \eqref{eq:xdef} as presented in \eqref{FSx} and \eqref{FTx}. The missing $S4$, $S5$ ,$T5$ and $T6$ contributions are
\begin{widetext}
\begin{equation}\label{eq:S4S5T5T6}
    \begin{aligned}
    &S4 = \frac{4  \eta^2 S_-^2}{3 \bar{\alpha} G},\\
    &S5 = \left(\frac{8  \eta^2 S_-}{45 \bar{\alpha} G}\right)\Bigg(-\frac{30 \Delta m (   S_-\beta_- +   S_+\beta_+  )}{ \bar{\gamma} m}+\frac{30 (S_-\beta_+ +S_+\beta_- )}{ \bar{\gamma}}+10\frac{ \Delta m}{m} S_-\beta_-\\
    &- S_-(5 \bar{\gamma}+10\beta_++10 \eta +21)+\frac{6  S_+^2}{ S_-}\Bigg),\\
    &T5 = \frac{32  \eta^2 }{5 \bar{\alpha}^2 G},\\
    &T6 = \left(\frac{2  \eta^2 }{105 \bar{\alpha}^2 G}\right)\left(-1247 - 448 \bar{\gamma} + 896 \frac{\Delta m}{m} \beta_-- 896 \beta_+ - 980 \eta\right).
    \end{aligned}
\end{equation}
\end{widetext}
Then for the coefficients in the expansion $E^\prime/{\cal F}$ in the DD regime \eqref{eq:EprimeFDDx} we find\\
\begin{widetext}
\begin{subequations}
\label{eq:ratioDDcoeffs}
\begin{equation}
    \begin{aligned}
    &E'_0 =-\frac{3}{2}-\frac{\eta}{6}-\frac{4 \bar{\gamma}}{3}+\frac{4}{3}\left(\beta_{+}-\frac{\Delta m}{m} \beta_{-}\right)-\frac{40 c^{4}}{3 G^{2}} \frac{\alpha f^{\prime}\left(\delta \varphi_{0}\right)}{m^{2} \bar{\alpha}^{7 / 2}}\bar{v}^{4}\left(3 \mathcal{S}_{+}+\frac{\Delta m}{m} \mathcal{S}_{-}\right) - \frac{20 c^4}{3 \bar{\alpha}^2 G^2 m^3} \zeta \bar{v}^4, 
    \end{aligned}
\end{equation}
  \begin{equation}
   \begin{aligned}
   f_2^{DD} &=\frac{24}{5 \bar{\alpha} S_-^2}+\frac{4 S_+^2}{5 S_-^2} -\frac{4 \beta _+}{3}+\frac{4 \beta _-   \Delta m}{3 m}-\frac{14}{5}-\frac{4 \eta }{3} -\frac{2 \bar{\gamma}}{3}  +\frac{4 \beta _- S_+}{\bar{\gamma} S_-} -\frac{4 \beta _+   \Delta m S_+}{\bar{\gamma} m S_-} +\frac{4 \beta _+}{\bar{\gamma}}-\frac{4 \beta _- \Delta m}{\bar{\gamma} m} \\
    &+\frac{\alpha f'[\varphi_0] c^4 \bar{v}^4}{\bar{\alpha}^{5/2}  G^2 m^2 S_-}\left(\frac{8 S_- }{3 \bar{\alpha}} ( 3S_+ + \frac{\Delta m}{m} S_-) - 2S_+ (S_+ + \frac{\Delta m}{m} S_-) \right)\\
   &+\frac{2 \bar{\zeta} c^4 \bar{v}^4}{2 \bar{\alpha}^{7/2}  \mu G^3 m^3 S_-}
   +\frac{4
   \zeta c^4 \bar{v}^4}{3 \bar{\alpha}^2  G^2 m^3}.
   \end{aligned}
\end{equation}
\end{subequations}
\end{widetext}
And in the QD regime splitting the contributions in the non-dipolar and dipolar parts \eqref{eq:FnondipFdip}
\begin{widetext}
\begin{subequations}
\label{eq:ratioQDcoeffs}
 \begin{equation}
 \begin{aligned}
     &f_2^{nd} =\frac{1}{\bar{\xi}}\Bigg( -\frac{8 \beta _+}{3}-\frac{4 \bar{\gamma}}{3}-\frac{35 \eta }{12}+\frac{8 \beta _- \Delta m}{3 m}-\frac{1247}{336}+\frac{\alpha f'[\varphi_0] c^4 \bar{v}^4}{\bar{\alpha}^{5/2} G^2 m^2 }\left(S_+ (1+ \frac{6}{25}\eta)\frac{50}{7} + S_+\frac{16}{ \bar{\alpha}} \right)+\frac{8
   \zeta c^4 \bar{v}^4}{3 \bar{\alpha}^2  G^2 m^3}\Bigg),\\
     &f^{d} =  -\frac{4 \beta _+}{3}+\frac{4 \beta _-   \Delta m}{3 m}-\frac{14}{5}-\frac{4 \eta }{3} -\frac{2 \bar{\gamma}}{3}  +\frac{4 \beta _- S_+}{\bar{\gamma} S_-}-\frac{4 \beta _+   \Delta m S_+}{\bar{\gamma} m S_-}+\frac{4 \beta _+}{\bar{\gamma}}-\frac{4 \beta _- \Delta m}{\bar{\gamma} m} +\frac{\alpha f'[\varphi_0] c^4 \bar{v}^4}{\bar{\alpha}^{5/2}  G^2 m^2 S_-}\\
     &\left(\frac{8 S_- }{3 \bar{\alpha}} ( 3S_+ + \frac{\Delta m}{m} S_-) - 2S_+ (S_+ + \frac{\Delta m}{m} S_-) \right)+\frac{2 \bar{\zeta} c^4 \bar{v}^4}{2 \bar{\alpha}^{7/2}  \mu G^3 m^3 S_-}
+\frac{4
   \zeta c^4 \bar{v}^4}{3 \bar{\alpha}^2  G^2 m^3}.\\
\end{aligned}
 \end{equation}
 \end{subequations}
   \end{widetext}
The expressions in this appendix supplement the discussion in Sec. \ref{sec:WaveformPhasing}.
\begin{figure*}[]
        \centering
        \begin{subfigure}[b]{0.475\textwidth}
            \centering
            \includegraphics[width=0.9\textwidth]{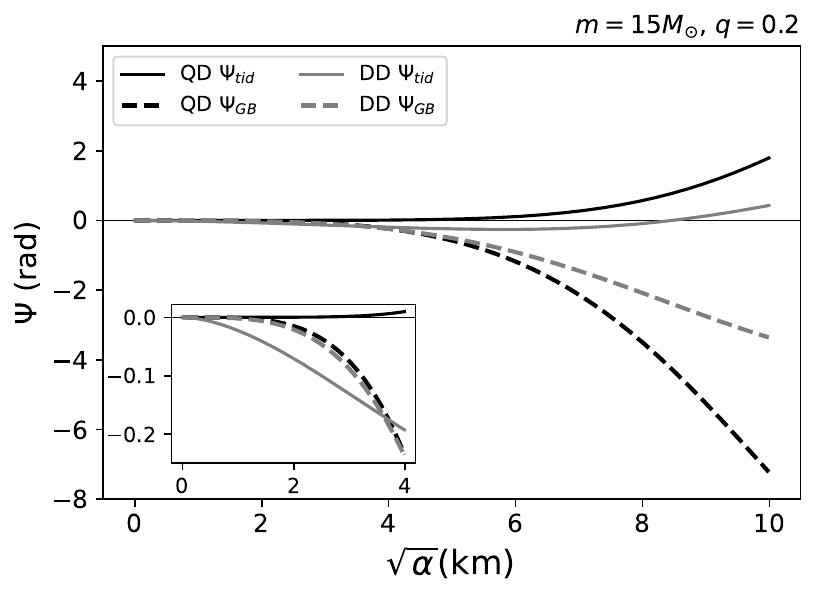}
        \end{subfigure}
        \begin{subfigure}[b]{0.475\textwidth}  
            \centering 
            \includegraphics[width=0.9\textwidth]{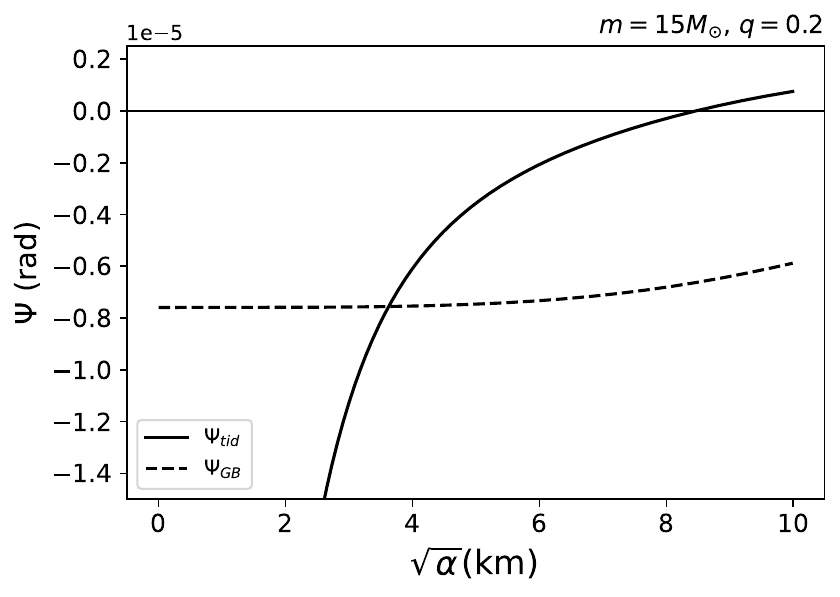}
            \label{fig:alphadep2}
        \end{subfigure}
        \vskip\baselineskip
        \begin{subfigure}[b]{0.475\textwidth}   
            \centering 
            \includegraphics[width=0.9\textwidth]{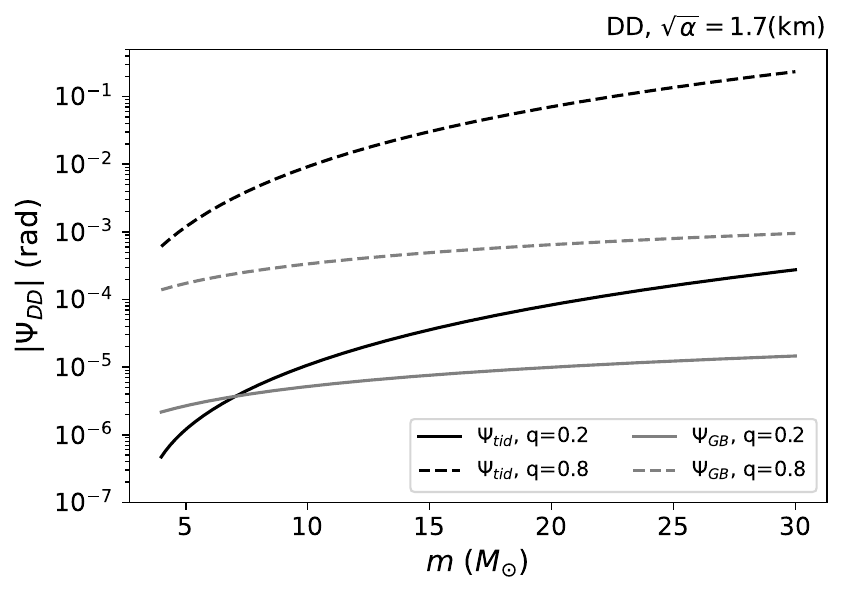}  
            \label{fig:mdep1}
        \end{subfigure}
        \begin{subfigure}[b]{0.475\textwidth}   
            \centering 
            \includegraphics[width=0.9\textwidth]{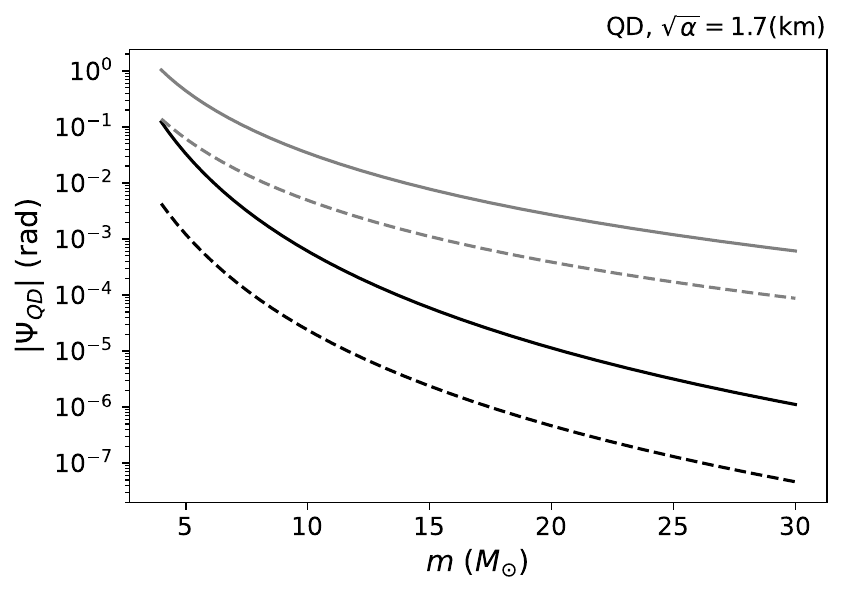}  
        \end{subfigure}
        \caption{\emph{Top panels}: Tidal (solid lines) and GB (dashed lines) phase contributions with respect to the dimensionfull coupling for a system of $15 M_{solar}$ and $q=0.2$. Top left: Phase contribution up to \eqref{eq:boundaryfreq} for the DD regime and $586$Hz for the QD regime. Top right panel: DD contributions at fixed frequency of $10^{-4}$Hz. 
        \emph{Bottom panels}: Absolute value of the tidal and GB phase contributions with respect to the total mass and $q=0.2$ (solid lines) respectively $q=0.8$ (dashed lines) for $\sqrt{\alpha}=1.7$km. Bottom left panel: DD regime at $10^{-4}$Hz, bottom right: QD regime at $586$Hz. } 
        \label{fig:tidGBcontralpha}
\end{figure*}
\section{Analysis tidal and Gauss Bonnet phase contributions with respect to the dimensionfull coupling}\label{othercouplingplots}
In this Appendix we present the plots related to Fig. \ref{fig:tidGBcontr} but regarding the dimensionfull coupling $\alpha$. 
For the dependencies of the contributions with respect to varying the coupling we find the same characteristics as a dominating tidal contribution in the DD regime for small coupling and a negative scaling of the tidal contributions. Only the evolution is different because of the rescaling via $\epsilon = \frac{\alpha c^4}{G^2 m^2}$.
The contributions for the different total masses significantly differ comparing the bottom plots of Fig. \ref{fig:tidGBcontralpha} with Fig. \ref{fig:tidGBcontr}. In the figures \ref{fig:tidGBcontralpha} we see purely the change in phase from varying the mass, the coupling input is fixed. We find still a positive scaling with the total mass in the DD regime but a negative scaling in the QD regime, which is expected as for smaller separations, the higher order curvature terms in the GB contributions become dominant and curvature effects are larger for smaller BH mass.
\clearpage

\bibliography{bib.bib}
\end{document}